\documentclass[12pt]{article}
\pdfoutput=1
\usepackage{comment}
\usepackage{jheppub}
\usepackage{todonotes}
\usepackage{smartdiagram}
\newcommand{\CommaPunct}{\mathpunct{\raisebox{0.1ex}{,}}}

\author[a]{Arash$\,$Arabi$\,$Ardehali,}
\author[b,c]{Mykola$\,$Dedushenko,}
\author[d]{Dongmin$\,$Gang,}
\author[e]{Mikhail$\,$Litvinov}

\affiliation{${}^a\,$Physics Department, Sharif University of Technology,\\
${}\,$\ Azadi Avenue, Tehran, Iran}

\affiliation{${}^b\,$Center for Mathematics and Interdisciplinary Sciences, Fudan University, Shanghai, China}

\affiliation{${}^c\,$Shanghai Institute for Mathematics and Interdisciplinary Sciences, Shanghai, China}

\affiliation{${}^d\,$Department of Physics and Astronomy \& Center for Theoretical Physics,\\
${}\,$\ Seoul National University, 1 Gwanak-ro, Seoul 08826, Korea}

\affiliation{${}^e\,$C.N. Yang Institute for Theoretical Physics, Stony Brook University,\\ ${}\,$\ Stony Brook, NY 11794, USA}

\emailAdd{a.a.ardehali@gmail.com}\emailAdd{dedushenko@gmail.com}\emailAdd{arima275@snu.ac.kr}\emailAdd{mikhail.litvinov@stonybrook.edu}

\title{Bridging 4D QFTs and 2D VOAs\\ via 3D high-temperature EFTs}

\date{}

\begin{document}

\abstract{The high-temperature limit of the superconformal index, especially on higher sheets, often captures useful universal information about a theory. In 4d $\mathcal{N}=2$ superconformal field theories with fractional r-charges, there exists a special notion of high-temperature limit on higher sheets that captures data of three-dimensional topological quantum field theories arising from r-twisted circle reduction. These TQFTs are closely tied with the VOA of the 4d SCFT. We study such high-temperature limits. More specifically, we apply Di~Pietro-Komargodski type supersymmetric effective field theory techniques
to r-twisted circle reductions of $(A_1,A_{2n})$ Argyres-Douglas theories, leveraging their Maruyoshi-Song Lagrangian with manifest  $\mathcal{N}=1$ supersymmetry. The result on the second sheet is the Gang-Kim-Stubbs family of 3d $\mathcal{N}=2$ SUSY enhancing rank-$0$ theories with monopole superpotentials, whose boundary supports the Virasoro minimal model VOAs $M(2,2n+3)$. Upon topological twist, they give non-unitary TQFTs controlled by the $M(2,2n+3)$ modular tensor category (MTC). The high-temperature limit on other sheets yields their unitary or non-unitary Galois conjugates. This opens up the prospect of a broader four-supercharge perspective on the celebrated correspondence between 4d $\mathcal{N}=2$ SCFTs and 2d VOAs via interpolating 3d EFTs. Several byproducts follow, including a systematic approach to 3d SUSY enhancement from 4d SUSY enhancement, and a 3d QFT handle on Galois orbits of various MTCs associated with 4d $\mathcal{N}=2$ SCFTs.}

\maketitle

\section{Introduction and summary}\label{sec:intro}
The tools of two-dimensional Conformal Field Theory (CFT) \cite{Belavin:1984vu} and three-dimensional Topological Quantum Field Theory (TQFT) \cite{Witten:1988hf} permeate physics in diverse dimensions. 
Vertex operator algebras (VOAs) and their representation categories play a special role here, both controlling the structure of 2d CFT and 3d TQFT, and appearing in other dimensions.
In particular, a far-reaching correspondence was discovered a decade ago between a subsector of 4d $\mathcal{N}=2$ superconformal field theories (SCFTs) and 2d vertex operator (super)algebras \cite{Beem:2013sza}.
We refer to it as the SCFT/VOA correspondence in this paper.
Recently, it has been put in a broader context \cite{Dedushenko:2023cvd}, relating it with the 3d TQFT and 2d CFT in a natural way.
Exploring this link in greater detail holds promise of putting the world of (super)conformal field theories under better control.
The SCFT/VOA correspondence inspired a plethora of developments in recent years.
It has guided, on the one hand, explorations of the landscape of $\mathcal{N}=2$ superconformal field theories, and on the other hand, foundational developments in logarithmic vertex operator algebras. A sample of works in the first direction is 
\cite{Liendo:2015ofa,Lemos:2015orc,Buican:2016arp,Beem:2017ooy,Beem:2018duj,Kaidi:2022sng,Buican:2023efi,Beem:2024fom} and in the second \cite{Arakawa:2016hkg,creutzig2017logarithmic,Creutzig:2018lbc,Auger:2019gts,Arakawa:2023cki,Arakawa:2024ejd}. See the surveys \cite{Lemos:2020pqv,Argyres:2022mnu,Arakawa:2017aon,Arakawa:2017fdq} for more context and related directions of research.

In this paper, building on \cite{Dedushenko:2023cvd,Dedushenko:2018bpp} on the one hand, and  \cite{DiPietro:2014bca,ArabiArdehali:2015ybk,DiPietro:2016ond,ArabiArdehali:2019zac,Cassani:2021fyv,ArabiArdehali:2021nsx,Ardehali:2021irq} on the other, we present a 3d effective field theory (EFT) bridge between the two sides that sheds significant light on the correspondence in the rational case.\footnote{The high-temperature expansions in \cite{ArabiArdehali:2023bpq} suggest that a single 3d EFT might fall short of capturing logarithmic modules of the VOA, and a direct sum of multiple 3d EFTs may be needed in that case. This possibility is currently being studied.} In particular, while in the original 4d/2d context, non-vacuum modules of the VOA correspond to surface defects in 4d \cite{Cordova:2017mhb,Pan:2017zie,Nishinaka:2018zwq,Dedushenko:2019yiw,Zheng:2022zkm,Pan:2024bne}, whose fusion category is rather poorly understood, in the 3d EFT, they correspond to line operators, which are under much better control (\emph{cf.}~\cite{Costello:2018swh,Dimofte:2019zzj,Ballin:2023rmt}).

The 3d bridge arises via an $R$-twisted circle reduction \cite{Cecotti:2010fi,Dedushenko:2018bpp} of the 4d SCFT as follows. We begin with the holomorphic-topological (HT) twist \cite{Kapustin:2006hi} of the 4d SCFT on a (holomorphic) Riemann surface $\Sigma$ times a (topological) cigar $C$. This yields a commutative vertex Poisson algebra on $\Sigma$ \cite{Oh:2019mcg}. We then quantize the latter into a non-commutative VOA by localizing the transverse excitations to the tip of the cigar via $\Omega$-deformation \cite{Nekrasov:2002qd,Nekrasov:2010ka} along $C$ \cite{Oh:2019bgz,Jeong:2019pzg}. Reduction to 3d is now possible along the angular direction of the cigar, where fields acquire $U(1)_r$-twisted boundary conditions due to the unit $U(1)_r$ flux involved in the topological twist \cite{Dedushenko:2023cvd}.

We restrict the discussion here to what may be considered the simplest setting of the 4d/2d correspondence, where the VOA is just a Virasoro minimal model. On the 4d side, we have the $(A_1,A_{2n})$ series of Argyres-Douglas theories \cite{Argyres:1995jj,Eguchi:1996ds,Gaiotto:2010jf,Cecotti:2010fi}, and on the 2d side the $M(2,2n+3)$ series of minimal models \cite{Cordova:2015nma,Beem:2017ooy}. By applying the $R$-twisted reduction to $(A_1,A_{2n})$, we will derive as the 3d bridge the $\mathcal{T}_n$ series of Gang-Kim-Stubbs \cite{Gang:2018huc,Gang:2023rei} rank-$0$ 3d $\mathcal{N}=4$ superconformal theories. The category of Wilson lines in these 3d theories is under good control (via systems of polynomial Bethe equations in the 3d A-model \cite{Kapustin:2013hpk,Nekrasov:2014xaa,Closset:2016arn}, for instance), and known to be in correspondence with that of the modules in $M(2,2n+3)$ \cite{Gang:2023rei,Ferrari:2023fez}. Upon the 3d topological A-twist (the mirror of B-twist, also known as Rozansky-Witten or Blau-Thompson twist) \cite{Blau:1996bx,Rozansky:1996bq,Kapustin:2010ag}, these theories lead to semi-simple non-unitary 3d TQFTs. Such TQFTs are captured by some modular tensor categories (MTC) \cite{Moore:1989yh,Turaev+2016}, and in our case, not surprisingly, these are the categories of modules for $M(2,2n+3)$.

\subsection*{$R$-twisted circle reduction}

The $R$-symmetry group of the 4d $\mathcal{N}=2$ superconformal group is $\mathrm{SU}(2)_R\times U(1)_r\,.$ For our purposes, the main result of \cite{Dedushenko:2023cvd} can be summarized as follows: the 4d $\mathcal{N}=2$ SCFT reduced on a circle of length $\beta$, with $U(1)_r$-twisted (supersymmetric) boundary conditions around the circle on all fields~$\phi$:
\begin{equation}
    \phi(x+\beta)=\, e^{i\pi (F+2r)}\, \phi(x),\label{eq:R_twisted_bc}
\end{equation}
yields a 3d $\mathcal{N}=4$ SCFT, whose topological A-twist supports the desired 2d VOA on its holomorphic boundary. The quantum numbers $F$ and $r$ in \eqref{eq:R_twisted_bc} are the fermion number and the $U(1)_r$ charge of the field $\phi$.

Through this device, the 4d/2d correspondence reduces to the more familiar realm of 3d/2d correspondences between TQFTs and their boundary VOAs, which have longer history and are far more developed since their inception in \cite{Witten:1988hf,Moore:1988qv,Moore:1989yh,Elitzur:1989nr}. Among other things, the emergence of infinite-dimensional chiral symmetry is now largely clarified.

More precisely, starting from a generic 4d $\mathcal{N}=2$ SCFT, the twisted reduction followed by the topological twist may yield a TQFT with local operators. This happens when the 3d theory has Higgs and/or Coulomb branches of vacua \cite{Costello:2018swh}. Such TQFTs constitute a more general type of bulk/boundary correspondence \cite{Costello:2018swh,Creutzig:2021ext,Garner:2022vds}. (See also \cite{Coman:2023xcq,Creutzig:2024abs}). In our setting, however, the $(A_1, A_{2n})$ theories have no 4d Higgs branch to begin with, and the 4d Coulomb branch is lifted via the $R$-twisting, so our TQFTs will have no local operators (even in the presence of line defects), and we will be in the standard territory.

It remains to identify the 3d $\mathcal{N}=4$ SCFT arising from the $R$-twisted reduction, or better, its A-twisted version supporting the VOA on its boundary. The main result of the present paper is that this remaining challenge is overcome efficiently via the EFT tools forged in the context of Cardy limits of the 4d superconformal index \cite{DiPietro:2014bca,ArabiArdehali:2015iow,ArabiArdehali:2015ybk,DiPietro:2016ond,ArabiArdehali:2019zac,GonzalezLezcano:2020yeb,Cassani:2021fyv,ArabiArdehali:2021nsx,Ardehali:2021irq,Cabo-Bizet:2021plf}.

\subsection*{SUSY index on the second sheet}

Consider the 4d $\mathcal{N}=2$ superconformal index defined as \cite{Kinney:2005ej,Romelsberger:2005eg,Gadde:2011uv}:
\begin{equation}
    \mathcal{I}_t(p,q,t):=\mathrm{Tr}^{}_{S^3}(-1)^F\, p^{j_1+j_2+r}q^{j_1-j_2+r}t^{I_3-r},
\end{equation}
where $I_3$ is the generator of the Cartan of $\mathrm{SU}(2)_R$, and $j_{1},j_2$ are the Lorentz spins.

Going to the \emph{$(\gamma+1)$st sheet} of the index via $p\to p\, e^{2\pi i\gamma}$, we get (\emph{cf.} \cite{Kim:2019yrz,Cabo-Bizet:2019osg,Cassani:2021fyv}):
\begin{equation}
    \begin{split}
    \mathcal{I}^{\gamma}_t(p,q,t):=\mathcal{I}_t(p\, e^{2\pi i\gamma},q,t)&=\mathrm{Tr}\,(-1)^{F}\, e^{2\pi i \gamma (j_1+j_2+r)}\,  p^{j_1+j_2+r}q^{j_1-j_2+r}t^{I_3-r}\\
    &=\mathrm{Tr}\,(-1)^F\,  e^{i\pi  \gamma(F+2r)}\,  p^{j_1+j_2+r}q^{j_1-j_2+r}t^{I_3-r},
    \end{split}
\end{equation}
where on going to the second line we used $(-1)^{F+2(j_1+j_2)}=1$.

The index can be computed via localization of the path-integral on a Hopf surface with topology $S^3\times S^1$ \cite{Assel:2014paa}. Writing $p=e^{2\pi i\sigma},\,q=e^{2\pi i\tau},$ the complex-structure moduli $\sigma,\tau$ of the Hopf surface encode the length $\beta$ of the circle, the squashing parameter $b$ of the unit three-sphere, and two angular twist parameters $\Omega_{1,2}$ as (\emph{cf.}~\cite{Cassani:2021fyv})
\begin{equation}
    \sigma=i\frac{\beta}{2\pi}(b+i\,\Omega_1)\,,\qquad \tau=i\frac{\beta}{2\pi}(b^{-1}+i\,\Omega_2)\,.\label{eq:HopfCSMs}
\end{equation}
The insertion of $e^{i\pi \gamma(F+2r)}$ inside the trace corresponds to the 4d fields having twisted boundary conditions around the $S^1$ (\emph{cf.}~Eq.(1.6) of \cite{Cassani:2021fyv}):
\begin{equation}
    \phi(x+\beta)=\, e^{i\pi \gamma(F+2r)}\, \phi(x).\label{eq:gamma_twisted_bc}
\end{equation}
For $\gamma=1$, corresponding to the \emph{2nd sheet}, this is exactly our desired $U(1)_r$-twist in~\eqref{eq:R_twisted_bc}. This twist appeared in \cite{Dedushenko:2023cvd}, within the $\Omega$-deformation approach to the SCFT/VOA correspondence, as a combination of the usual $(-1)^F$ (due to the NS spin structure arising from folding the topological plane of the HT twist into a cigar) with an $e^{2\pi i\,r}$ arising from the unit $U(1)_r$ flux involved in the topological twist. Higher sheets, as will be explained below, allow making contact with the Galois orbits discussed in \cite{Dedushenko:2018bpp}.

In the body of the paper we focus on the $\mathcal{N}=1$ limit of $\mathcal{I}_t^\gamma(p,q,t),$ which we denote by $\mathcal{I}^\gamma(p,q)$. This is the index that has been better understood from an EFT perspective in \cite{Ardehali:2021irq}; other choices will be discussed in Section~\ref{subsec:sense4dBackground}. The $\mathcal{N}=1$ limit corresponds to setting $t=(pq)^{2/3},$ and thus our index of interest is\footnote{Note that the 2nd sheet index of \cite{Cassani:2021fyv} is instead $\mathcal{I}_\text{there}=\mathcal{I}_t(p\, e^{2\pi i},q,(pq)^{2/3}e^{4\pi i/3})$. Different notions of ``2nd sheet'' correspond to different subgroups of the $R$-symmetry used in the twisting.}
\begin{equation}
    \mathcal{I}^{\gamma=1}(p,q):=\mathcal{I}_t(p\, e^{2\pi i},q,(pq)^{2/3}).\label{eq:theRtwistedIndex}
\end{equation}
In fact, since we are primarily interested in topological structures descending from 4d to 3d, we further set $b=1,$ $\Omega_{1,2}=0,$ implying $p=q\in\mathbb{R}.$ The relevant index is hence
\begin{equation}
    \mathcal{I}^{\gamma=1}(q):=\mathcal{I}^{\gamma=1}(q,q)=\mathcal{I}_t(q\, e^{2\pi i},q,q^{4/3}).\label{eq:theRtwistedIndex}
\end{equation}

With the $R$-twisting implemented as above, we can now perform the reduction via the Cardy limit $q\to1$ of the index $\mathcal{I}^{\gamma=1}(q)$.

\subsection*{EFT data from the Cardy limit}

The tools and formulas developed for analyzing the Cardy limit of the 4d $\mathcal{N}=1$ index in \cite{DiPietro:2014bca,ArabiArdehali:2015ybk,DiPietro:2016ond,ArabiArdehali:2019zac,Cassani:2021fyv,ArabiArdehali:2021nsx,Ardehali:2021irq} will be upgraded here to an efficient procedure for finding the data of the 3d EFT arising from the $R$-twisted reduction. The procedure requires as an input a 4d $\mathcal{N}=1$ Lagrangian description of the 4d $\mathcal{N}=2$ SCFT, which for the $(A_1,A_{2n})$ series is available thanks to the work of Maruyoshi and Song \cite{Maruyoshi:2016tqk,Maruyoshi:2016aim}.

There are three steps to the procedure:
\begin{itemize}
    \item \emph{EFT field content} is found via a (patch-wise) saddle-point variant of the real-analytic methods used in \cite{Rains:2006dfy,ArabiArdehali:2015ybk,Ardehali:2021irq}. This variant, as explained in Section~\ref{sec:Cardy}, solves Problem~1 in \cite{ArabiArdehali:2015ybk}. The resulting holonomy saddles \cite{Hwang:2018riu} of $\mathcal{I}^{\gamma=1}$ yield abelian EFTs from the non-abelian Maruyoshi-Song starting points for $(A_1,A_{2n})$.
    \item \emph{EFT Chern-Simons couplings} are found via formulas from \cite{Ardehali:2021irq}. For $(A_1,A_2)$ the Chern-Simons (CS) coupling derived as such, and the field content obtained as above, match with those of the Gang-Yamazaki theory \cite{Gang:2018huc}, confirming the conjecture in \cite{Dedushenko:2023cvd}. Moreover, as explained in Section~\ref{sec:A1A2}, the theory obtained from 4d in this way comes equipped with a gravitational CS coupling that resolves via inflow the 't~Hooft anomaly matching puzzle raised in \cite{Dedushenko:2023cvd}.  
    \item \emph{EFT monopole superpotentials} are found via an upgraded version of the technique used in \cite{ArabiArdehali:2019zac} for diagnosing (multi-) monopole superpotentials from the periodic polynomials \eqref{eq:QhDef} and \eqref{eq:LhDef} (the former vanished in \cite{ArabiArdehali:2019zac}, and the latter was referred to as the Rains function there). For $(A_1,A_4)$ the monopole superpotential derived as such, together with the field content and CS couplings obtained as above, match those of the Gang-Kim-Stubbs $\mathcal{T}_2$ theory \cite{Gang:2023rei}, confirming the conjecture in \cite{Dedushenko:2023cvd}. Moreover, our 4d derivation provides us with a microscopic mechanism for dynamical generation of the monopole superpotential \`{a} la Affleck-Harvey-Witten \cite{Affleck:1982as} in the compactified Maruyoshi-Song theory. As explained in Section~\ref{sec:A1A4}, more generally, the monopole superpotentials of all $\mathcal{T}_n$ theories can be thought of as arising from the Affleck-Harvey-Witten type mechanism in the parent Maruyoshi-Song theory. See \cite{ArabiArdehali:2024vli} for more details.
\end{itemize}

We thus establish the Gang-Kim-Stubbs family $\mathcal{T}_n$ of 3d $\mathcal{N}=4$ SCFTs (including the Gang-Yamazaki minimal theory \cite{Gang:2018huc} for $n=1$) as the 3d bridge in the 4d/2d correspondence between $(A_1,A_{2n})$ and $M(2,2n+3).$ Indeed, the A-twisted $\mathcal{T}_n$ theory is known via the half-index calculations \cite{Gang:2023rei} to host $M(2,2n+3)$ on its boundary. Our approach to the family, however, brings to light additional background CS couplings supplied by the 4d UV completion, which resolve anomaly mismatches of the kind noticed in~\cite{Dedushenko:2023cvd}.

\subsection*{Higher sheets and Galois orbits}


Our approach also brings to light, following \cite{Dedushenko:2018bpp}, a whole collection of other three-dimensional QFTs closely related to the $\mathcal{T}_n$ family.

These arise from non-minimal $R$-twists in \eqref{eq:gamma_twisted_bc} corresponding to $1<\gamma<2n+3$, with the upper bound present because $\gamma\in\mathbb{Z}_{2n+3}$ \cite{Dedushenko:2023cvd}. The associated EFT data can be obtained from the Cardy limit $q\to1$ of the higher-sheet indices
\begin{equation}
    \mathcal{I}^{\gamma}(q):=\mathcal{I}_t(q\, e^{2\pi i\gamma},q,q^{4/3}).\label{eq:gamma_twisted_index}
\end{equation}

The 3d EFTs we find are either gapped and flow to unitary TQFTs, or enjoy a $U(1)$ flavor symmetry with respect to which the $F$-maximization gives $\mathcal{N}=4$ SCFTs whose A-twist yields non-unitary TQFTs. These higher-sheet 3d TQFTs (some of which are unitary and some are not) are labeled by $\gamma\in\{1,\dots,2n+2\}$, or equivalently, by the roots of unity $e^{2\pi i \frac\gamma{2n+3}}$, and denoted TQFT$^\gamma$. It is natural to expect that they are permuted by the Galois group associated to the extension $\mathbb{Q}(\zeta)$, where $\zeta= e^{\frac{2\pi i}{2n+3}}$ is a simple root of unity. Such a group is known to be the group of units in $\mathbb{Z}_{2n+3}$:
\begin{equation}
\label{Gal}
    {\rm Gal}(\mathbb{Q}(\zeta)/\mathbb{Q}) = \mathbb{Z}_{2n+3}^\times,
\end{equation}
where $\mathbb{Z}^\times_N$ is a multiplicative group of integers modulo $N$.
This suggests a tantalizing connection to the Galois group action on the TQFT data \cite{Rowell:2007dge}, as was noticed in
\cite{Dedushenko:2018bpp}. More specifically, when a TQFT is determined by an underlying modular tensor category, the MTC data obeys polynomial equations, and one may consider various Galois groups acting on MTCs \cite{Rowell:2007dge}. The most relevant for us is the \emph{Galois group of the modular data}, acting on the modular $S$ and $T$ matrices. It is the Galois group of the extension of $\mathbb{Q}$ by the matrix elements of $S$ and $T$. This group is usually bigger than ours, given by $\mathbb{Z}_{N}^\times$, where $N$ is called the \emph{conductor} \cite{Rowell:2007dge} (see also \cite{Harvey:2018rdc,Harvey:2019qzs}). If we, however, choose the normalization $T_{00}=1$ (rather than $e^{-2\pi i \frac{c}{24}}$), then the Galois group acting on the modular data of our 3d theories TQFT$^\gamma$ is precisely ${\rm Gal}(\mathbb{Q}(\zeta)/\mathbb{Q})$ given in \eqref{Gal}. 

Note that when $2n+3$ is prime, we have $\mathbb{Z}_{2n+3}^\times \cong \mathbb{Z}_{2n+2}$, and the possible values $\gamma\in\{1,\dots,2n+2\}$ constitute the group's single orbit. For non-prime $2n+3$, the set $\{1,\dots,2n+2\}$ breaks into multiple orbits of $\mathbb{Z}_{2n+3}^\times$. The orbit of our interest in this paper is the one containing $\gamma=1$, whose members are labeled by $\gamma$ coprime to $2n+3$. For such $\gamma$, the Coulomb branch is fully lifted in the 3d limit. The other orbits would require dealing with unlifted Coulomb branches which are beyond the scope of this work.

We thus obtain from a single 4d theory, by considering different values of $\gamma$ relatively prime to $2n+3$, a sequence of 3d TQFTs:
\vspace{.1cm}
\begin{equation*}
    \text{TQFT}^{\gamma=1}_\text{non-uni}\,,\quad \text{TQFT}^{\gamma=2}_\text{(non-)uni}\,,\quad \cdots,\quad  \text{TQFT}^{\gamma=2n+1}_\text{(non-)uni}\,,\quad  \text{TQFT}^{\gamma=2n+2}_\text{non-uni}\,
\end{equation*}
whose modular $S$ and $T$ matrices are related via the Galois/Frobenius conjugation (\emph{cf.}~\cite{Coste:1993af,Harvey:2019qzs}). The $\gamma=1$ TQFT is of course the non-unitary A-twisted bridge in the 4d/2d correspondence \cite{Dedushenko:2023cvd}.

Our evaluation of the TQFT $S$ and $T$ matrices is based on the $\mathcal{N}=2$ field content and CS levels (see \emph{e.g.}~\eqref{eq:LY}), using Nekrasov-Shatashvili type Bethe root techniques \cite{Nekrasov:2014xaa,Closset:2019hyt}, via the handle-gluing and fibering operators used in the BPS surgery (see Appendix~\ref{app:Bethe_techniques}). At its current stage of development, this approach yields results up to sign ambiguities in entries of $S$, and an overall phase ambiguity in $T$. Fixing the sign ambiguities in the entries of $S$ by demanding compatibility with an MTC structure, and fixing the normalization of $T$ by $T_{00}=1$ allows one to unambiguously talk about the Galois group action. The physical $T_{00}$, however, is some phase: our ignorance of this phase reflects the possibility of multiplying a TQFT by an invertible TQFT \cite{Freed:2004yc,Freed:2016rqq}.

Like in \cite{Dedushenko:2018bpp}, our main example here is $(A_1,A_2)$ (although we also consider $(A_1, A_4)$). On the second sheet $\gamma=1$, we of course find the Lee-Yang (LY) MTC. In this case, $2n+3=5$, the relevant Galois group is $\mathbb{Z}_5^\times = \mathbb{Z}_4$, and the Galois orbit also contains, in addition to LY, conjugate Fibonacci ($\overline{\text{Fib}}$), Fibonacci (Fib), and conjugate Lee-Yang ($\overline{\text{LY}}$) modular tensor categories: 

\begin{center}
\usetikzlibrary{shapes.geometric, positioning}
\smartdiagramset{
font=\footnotesize,
module minimum width=1.7cm,
module minimum height=.7cm,
text width=1cm,
circular distance=2cm, 
arrow tip=to
}
\smartdiagram[circular diagram:clockwise]{LY, $\overline{\text{Fib}}$, Fib, $\overline{\text{LY}}$}
\end{center}
The natural expectation is to see this orbit on our four sheets, for $\gamma=1$ to $4$, respectively. Computations of $S, T$ in the normalization $T_{00}=1$ indeed corroborate this expectation. However, once we try to account for the phase of $T_{00}$, things start to slightly deviate from \cite{Dedushenko:2018bpp}.

Our concrete EFT data enable us to take a further step here and access the boundary VOAs of these TQFTs via half-index calculations as in \cite{Gadde:2013wq,Yoshida:2014ssa,Dimofte:2017tpi,Okazaki:2019bok,Costello:2020ndc}. In the $(A_1,A_2)$ case, we obtain the ${\rm LY}\cong M(2,5)$ characters \cite{Gang:2023rei} on the second sheet (\emph{i.e.}~$\gamma=1$), as expected from the SCFT/VOA correspondence. Similarly, on the fifth sheet (\emph{i.e.}~$\gamma=4$) we see evidence of $\overline{\rm LY}$, thus realizing half of the Galois/Hecke orbit at the level of VOA characters (\emph{cf.}~\cite{DeBoer:1990em,Coste:1993af,Bantay:2001ni,Harvey:2018rdc,Harvey:2019qzs}). On the third and fourth sheet, however, where the 3d EFT is gapped, the half-index computations indicate a spin-TQFT. Since we study the low-energy regimes of supersymmetric gauge theories, this finding should not be too surprising. We perform a half-index calculation for the TQFT$^{\gamma=2}_\text{uni}$ on the third sheet of $(A_1,A_2)$, with Dirichlet boundary conditions in Section~\ref{subsec:FibBar}. The resulting characters are those of a fermionic VOA, namely the fermionized tricritical Ising (known as the supersymmetric minimal model $\mathrm{SM}(3,5)$) times a free Majorana fermion (often denoted as SO$(1)_1$). On the fourth sheet, we find evidence that TQFT$^{\gamma=3}_\text{uni}$ is the conjugate of the third sheet spin-TQFT. Thus, we propose:

\begin{center}
\usetikzlibrary{shapes.geometric, positioning}
\smartdiagramset{
font=\footnotesize,
module minimum width=2.2cm,
module minimum height=.7cm,
text width=1cm,
circular distance=2.4cm, 
arrow tip=to
}
\smartdiagram[circular diagram:clockwise]{$\text{M(2,5)}$, $\text{SM(3,5)}\otimes\text{SO(1)}_1$, $\overline{\text{SM(3,5)}}\otimes\overline{\text{SO(1)}_1}$, $\overline{\text{M(2,5)}}$}
\begin{tikzpicture}[overlay, remember picture]
    \node at ([xshift=-3.8cm,yshift=3.1cm]current bounding box.center) {\large$(A_1, A_2)$};
\end{tikzpicture}
\end{center}
At first, this looks very disappointing, as it differs from the Galois orbit shown earlier. However, as we explain in Section \ref{sec:A1A2}, $\text{SM(3,5)}\otimes\text{SO(1)}_1$ actually agrees with $\overline{\rm Fib}\cong(F_4)^{}_1$ up to multiplication by an invertible spin-TQFT. Likewise, $\overline{\text{SM(3,5)}\otimes\text{SO(1)}_1}$ agrees with ${\rm Fib}\cong(G_2)^{}_1$ up to an invertible factor. Therefore, our findings still agree with the Galois orbit proposal formulated up to invertible factors:
\begin{equation}
    \boxed{\text{Higher-sheet TQFTs form a Galois orbit, up to invertible spin-TQFT}}
\end{equation}
We test this proposal by computing half-indices of various natural boundary conditions and identifying them as characters of the corresponding VOAs. For the third sheet theory, besides the ${\text{SM(3,5)}\otimes\text{SO(1)}_1}$ characters on the right Dirichlet boundary, we also find the $(G_2)_1 \cong {\rm Fib}$ characters on the left enriched Neumann boundary, as explained in Section \ref{subsec:FibBar}. On the fourth sheet, we find the same, with the left and right boundaries swapped. This implies a level-rank type duality between ${\text{SM(3,5)}\otimes\text{SO(1)}_1}$ and $(G_2)_1$, analogous to \cite{Ferrari:2023fez}. Notice also that up to invertible factors, we have ${\rm Fib}$ and $\overline{\rm Fib}$ on the opposite boundaries, which is expected.
We also note in passing that the ${\rm LY}$ and $\overline{\rm LY}$ theories are mirror duals of each other.
It would be interesting to realize the $(F_4)^{}_1\cong \overline{\rm Fib}$ \cite{Mukhi:2022bte} characters as well.

We do not explore the Galois orbit of $(A_1,A_4)$ in as much detail, but we find that besides the Gang-Kim-Stubbs $\mathcal{T}_2$ theory arising on the 2nd sheet, quite interestingly, its mirror arises on the 3rd sheet, suggesting that more generally, the mirror is always on the Galois orbit of $\mathcal{T}_n$. We also uncover a fascinating unitary TQFT with a non-abelian gauge group and fractional monopoles on the fourth sheet of $(A_1,A_4)$, which may be seen as a harbinger of richer possibilities at larger $n.$

We hope this concrete 3d perspective on the Galois orbits of 4d $\mathcal{N}=2$ SCFTs finds further synthesis with related results in \cite{Buican:2017rya,Buican:2019huq}.

\subsection*{Outline of the paper}

In Section~\ref{sec:Cardy} we outline how the twisted reduction of 4d $\mathcal{N}=1$ gauge theories can be tracked via EFT tools developed in the context of Cardy-like limits of the 4d superconformal index. While most of Section~\ref{sec:Cardy} is review of known results, the extremely simple relations in \eqref{eq:kij&kjR_from_Q&L} are new. We use them here to sharpen the twisted reduction procedure, but they have wider implications that will be explored elsewhere. In Section~\ref{sec:A1A2} we apply the $R$-twisted reduction procedure to $(A_1,A_2),$ reproducing via the minimal twist the Gang-Yamazaki theory, properly dressed with a gravitational CS coupling that resolves the anomaly mismatch puzzle of \cite{Dedushenko:2023cvd}. With non-minimal $R$-twists, we obtain the Galois orbit of the Gang-Yamazaki theory, up to invertible spin-TQFT factors. In particular, on the third (resp. fourth) sheet we obtain a spin-TQFT with $S$ and $T^2$ matrices matching those of the conjugate Fibonacci (resp. Fibonacci) MTC, up to an overall phase of $T$, that supports ${\text{SM(3,5)}\otimes\text{SO(1)}_1}$ characters on a Dirichlet boundary (resp.  $(G_2)^{}_1$ characters on a Neumann boundary).
In Section~\ref{sec:A1A4} we explore part of the Galois orbit arising from the $R$-twisted  reduction of $(A_1,A_4)$, making contact with the $\mathcal{T}_2$ theory of Gang-Kim-Stubbs on the second sheet, its mirror on the third sheet, and an intriguing TQFT with non-abelian gauge group on the fourth sheet. Some remaining open questions are discussed in Section~\ref{sec:openQs}, and various technical calculations are outlined in Appendices~\ref{app:I4toZ3} and \ref{app:effective_CS}. Appendix~\ref{app:3d_toolkit} summarizes our toolkit for studying 3d $\mathcal{N}=2$ gauge theories: the 3d superconformal index, which we use as a diagnostic tool to verify that our (possibly twisted) EFTs flow to TQFTs without local operators; the squashed three-sphere partition function, which arises in the Cardy limit of the 4d $\mathcal{N}=1$ index as explained in Appendix~\ref{app:I4toZ3}; the Bethe root techniques, which we use to determine $S$ and $T^2$ matrices from the EFT data; half-indices with Neumann or Dirichlet boundary conditions, which we use to find characters of the VOAs on holomorphic boundaries of our TQFTs.

\section{Cardy limit of the 4d index and 3d monopoles}\label{sec:Cardy}

The superconformal indices $\mathcal{I}^\gamma(p,q)$ of our interest in this work are made out of $q$-Pochhammer symbol and elliptic gamma function building blocks:
\begin{equation}
    (z;q):=\prod_{k=0}^{\infty}(1-zq^k),\qquad \Gamma(z;p,q):=\prod_{j,k\ge 0}\frac{1-z^{-1}p^{j+1}q^{k+1}}{1-z
    p^{j}q^{k}}.
\end{equation}
We often denote $\Gamma(z;p,q)$ by $\Gamma_e(z)$ for simplicity, keeping the dependence on $p,q$ implicit. Note that
\begin{equation}
    \frac{1}{\Gamma(z;p,q)}=\Gamma(z^{-1}pq;p,q).\label{eq:invEllGam}
\end{equation}

For a general 4d $\mathcal{N}=1$ gauge theory with a $U(1)_R$ symmetry, with a semi-simple gauge group $G$ and flavor group $F$, the index takes the form (\emph{cf.}~\cite{Dolan:2008qi}):
\begin{equation}
\mathcal{I}^\gamma(q)=(q;q)^{2r_G} \frac{1}{|W|}\int_{\mathfrak{h}_{\text{cl}}} \prod_{j=1}^{r_G} \mathrm{d} x_j\ \frac{\prod_\chi\prod_{\rho^\chi}\Gamma_e\big(e^{2\pi i\rho^\chi\cdot\boldsymbol{x}}\,q^{r_\chi}\, e^{2\pi i\,q^\chi\cdot\boldsymbol{\xi}}\big)}{\prod_{\alpha_+}\Gamma_e(e^{2\pi i\alpha_+\cdot\boldsymbol{x}})\, \Gamma_e(e^{-2\pi i\alpha_+\cdot\boldsymbol{x}})},\label{eq:indexIntegralGeneral}
\end{equation}
where $|W|$ is the order of the Weyl group and $r_G$ the rank of $G$. The parameters $x_j$ will be referred to as the \emph{holonomies}. They parametrize a moduli space denoted $\mathfrak{h}_{\text{cl}}$, which we write explicitly as
\begin{equation}
\mathfrak{h}_\text{cl}=\left(-\frac{1}{2},\frac{1}{2}\right]^{r_G}\!.    
\end{equation}
The $r_G$-tuple $x_1,\dots,x_{r_G}$ is denoted $\boldsymbol{x}$, and the $r_F$-tuple $\xi_1,\dots,\xi_{r_F}$ is denoted $\boldsymbol{\xi}$. The positive roots of $G$ are denoted $\alpha_+$, and the weights of the gauge group representation of the chiral multiplet $\chi$ are denoted $\rho^\chi$. 
We will sometimes use the notation 
$z_j:=e^{2\pi i x_j}$ below.
The $r_F$-tuple flavor charge of $\chi$ is denoted $q^\chi$, and its $U(1)_R$ charge $r_\chi$. We assume $r_\chi\in(0,2)$ for all $\chi.$
This guarantees, among other things \cite{Gerchkovitz:2013zra}, a universal large-$\beta$ (or ``low-temperature'') limit \cite{ArabiArdehali:2015ybk}.

In our application to $R$-twisted reductions below, we will encounter phases of the form $e^{2\pi i\,q^\chi\cdot\boldsymbol{\xi}}$ in the arguments of the gamma functions arising from the replacement $p\to q\,e^{2\pi i\gamma}$ as in \eqref{eq:gamma_twisted_index}. See for example the phases in~\eqref{eq:I_4 for n=1}. We do not consider actual flavor fugacities in this work (except in Appendix~\ref{app:flavor}, but there we will take $\xi$ to be  $\mathcal{O}(\beta)$, rather than $\mathcal{O}(\beta^0)$ as in the formal discussion of the present section).

Defining $\beta$ via $q=e^{-\beta}$, we refer to the $q\to1^{-}$ limit of the index as the Cardy limit. The most interesting part of \eqref{eq:indexIntegralGeneral} for our purposes in this limit is the integrand. But since we want to obtain the Cardy-like growth of the index as well, let us begin with the asymptotics of the Pochhammer prefactor in \eqref{eq:indexIntegralGeneral}, which simplifies in the limit as
\begin{equation}
    \log\,(q;q)=
-\frac{\pi^2}{6\beta}-\frac{1}{2}\log\beta+\mathcal{O}(\beta^0).\label{eq:PochAsy}
\end{equation}

\noindent\textbf{Decomposition of the moduli space and the outer patch.} The index \eqref{eq:indexIntegralGeneral} can be thought of as a supersymmetric partition function on $S^3\times S^1$, with $\mathfrak{h}_\text{cl}/W$ the (classical) moduli-space of flat connections. From a Kaluza-Klein perspective,  $\mathfrak{h}_\text{cl}/W$ coincides with a middle-dimensional section of the classical Coulomb branch of the 3d field theory living on $S^3$ (the other half of the Coulomb branch is parameterized by the dual photons). A subset $\mathcal{S}$ of $\mathfrak{h}_\text{cl}$ referred to as the ``singular set'',
\begin{equation}
\begin{split}
\mathcal{S}_g:=\bigcup_{\alpha_+}\{\boldsymbol{x}\in
\mathfrak{h}_\text{cl}|\, \alpha_+\cdot
\boldsymbol{x}\in&\, \mathbb{Z}\},\quad\quad\mathcal{S}_\chi:=\bigcup_{\rho^\chi\neq0}\{\boldsymbol{x}\in\mathfrak{h}_\text{cl}|\, \rho^{\chi}\cdot
\boldsymbol{x}+q^\chi\cdot\boldsymbol{\xi}\in\mathbb{Z}\},\\
&\mathcal{S}:=\bigcup_{\chi}\mathcal{S}_\chi\cup
\mathcal{S}_g,\label{eq:SSsDef}
\end{split}
\end{equation}
supports additional massless modes compared to generic $\boldsymbol{x}$. We excise an $\epsilon$ neighborhood $\mathcal{S}_\epsilon$ of the singular set and refer to the rest of $\mathfrak{h}_\text{cl}$ as the \emph{outer patch}. The details of the excision scheme are largely irrelevant for our purposes here, but an important point is that for small enough $\epsilon$ the outer patch consists of multiple disconnected components, denoted as out$_n$. These are separated from each other by $\mathcal{S}_\epsilon$. One can also decompose $\mathcal{S}_\epsilon$ into various \emph{inner patches} denoted in$_n$. We refer the interested reader to \cite{Ardehali:2021irq} for precise definitions, and here just illustrate the ideas with a couple of simple examples as in Figures~\ref{fig:patchesGamma=1} and~\ref{fig:patchesGamma=2}. 

The leading asymptotics of the index $\mathcal{I}^\gamma(q)$ in this limit can be found using the estimate~\cite{ArabiArdehali:2015ybk} (\emph{cf.}~\cite{Rains:2006dfy}):
\begin{equation}
    \begin{split}
\log\Gamma_e(z\,q^r)&= i\,\frac{8\pi^3}{\beta^2}\frac{\kappa(x)}{12}
-\frac{4\pi^2}{\beta}\frac{1-r}{2}\vartheta(x)-\frac{\pi^2}{3\beta}(r-1)+O(\beta^0),
    \end{split}\label{eq:GammaAsy}
\end{equation}
where 
\begin{equation}
    \begin{split}
        \vartheta(x)&:=\{x\}(1-\{x\}),\\
        \kappa(x)&:=\{x\}(1-\{x\})(1-2\{x\}),
    \end{split}
\end{equation}
with $\{x\}:=x-\lfloor x\rfloor.$

Using \eqref{eq:PochAsy} and \eqref{eq:GammaAsy}, we find in the $\beta\to0$ limit  (\emph{cf.} Eq.~(3.9) of \cite{ArabiArdehali:2015ybk}):
\begin{equation}
    \mathcal{I}^\gamma(q)\approx e^{-\frac{\pi^2}{3\beta}\mathrm{Tr}R} \left(\frac{2\pi}{\beta}\right)^{r_G}\int_{\mathfrak{h}_\text{cl}}
\mathrm{d}^{r_G}x\
e^{-\frac{4\pi^2}{\beta}L_h(\mathbf{x})+i\frac{8\pi^3}{\beta^2}Q_h(\mathbf{x})},\label{eq:simplifiedIntegral}
\end{equation}
where the two functions in the exponent\footnote{The role that $Q_h,L_h$ play in the 4d$\,\to\,$3d reduction of four-supercharge gauge theories is in some ways analogous to that played by $W,\Omega$ in the 3d$\,\to\,$2d reduction~\cite{Nekrasov:2014xaa}; see Appendices~\ref{app:I4toZ3} and \ref{app:Bethe_techniques}. This is seen most clearly from the 4d A-model perspective \cite{Nekrasov:2009uh,Closset:2017bse,Ardehali:upcoming}.  In particular, $Q_h$ (resp. $W$) encodes black hole entropy in AdS$_5$/CFT$_4$ (resp. AdS$_4$/CFT$_3$) via Cardy limit of the 4d (resp. 3d) superconformal index \cite{Choi:2018hmj,Honda:2019cio,ArabiArdehali:2019tdm,Kim:2019yrz,Cabo-Bizet:2019osg,Amariti:2019mgp,Amariti:2021ubd,Choi:2019dfu,Choi:2019zpz}. Alternatively, $Q_h,L_h$ may be thought of as periodic polynomials encoding various dynamical and contact CS couplings of the 3d KK (or thermal) EFT arising from compactification on a circle \cite{Preskill:1990fr,DiPietro:2016ond,ArabiArdehali:2021nsx,Ardehali:2021irq}.} are given by
\begin{equation}
\begin{split}
Q_h(\mathbf{x}):=\frac{1}{12}\sum_{\chi}\sum_{\rho^{\chi}
}\kappa(\rho^{\chi}\cdot
\mathbf{x}+q^\chi\cdot\boldsymbol{\xi}),\label{eq:QhDef}
\end{split}
\end{equation}
\begin{equation}
\begin{split}
L_h(\mathbf{x}):= \frac{1}{2}\sum_{\chi}(1-r_\chi)\sum_{\rho^{\chi}
}\vartheta(\rho^{\chi}\cdot
\mathbf{x}+q^\chi\cdot\boldsymbol{\xi})-\sum_{\alpha_+}\vartheta(\alpha_+\cdot
\mathbf{x}).\label{eq:LhDef}
\end{split}
\end{equation}
Even though $\vartheta$ and $\kappa$ are piecewise quadratic and cubic respectively, anomaly cancellations make $L_h$ and $Q_h$ piecewise linear and quadratic (hence the letters). Moreover, it follows from their building blocks $\vartheta$ and $\kappa$ that $L_h$ is continuous while $Q_h$ has continuous first derivatives.

Note that the inner-patch contributions are neglected in \eqref{eq:simplifiedIntegral}, and $\epsilon$ is sent to zero (the integration domain is all of $\mathfrak{h}_\text{cl}$). This is because we are interested mainly in the leading parts of the asymptotics. As explained in \cite{ArabiArdehali:2015ybk,Ardehali:2021irq}, to obtain subleading asymptotics one has to take into account the inner patches. Typically the small-$\beta$ expansion of $\log\mathcal{I}^\gamma(q)$ is of the form:
\begin{equation}
    \log\mathcal{I}^\gamma(q)=\frac{\#}{\beta^2}+\frac{\#}{\beta}+\#\log\frac{1}{\beta}+\mathcal{O}(\beta^0).
\end{equation}
The inner-patch contributions become crucial at $\mathcal{O}(\beta^0)$ and higher. An example of the inner-patch analysis will be discussed in Appendix~\ref{app:I4toZ3}.

The integrand in \eqref{eq:simplifiedIntegral} is piecewise analytic, so can be treated via the saddle-point method in its domains of analyticity. The saddle-point method shows in these domains that the leading term of the exponent containing $i\,Q_h/\beta^2$ is the most important. So the saddles correspond to stationary points of $Q_h.$ In fact, since the first derivative of $Q_h$ is continuous, one can forget about non-analyticity and simply look for stationary points of $Q_h$ on all of $\mathfrak{h}_\text{cl}$. If the locus of such stationary points is extended, or consists of multiple points, then one has to find the subset of that locus where $L_h$ is minimized. Let us denote this subset by $\mathfrak{h}_\text{qu}.$ The saddle-point method then yields the asymptotic of the index (\emph{cf.} \cite{ArabiArdehali:2015ybk,Ardehali:2021irq}):
\begin{equation}
    \mathcal{I}^\gamma(q)\approx e^{-\frac{\pi^2}{3\beta}\mathrm{Tr}R} \left(\frac{2\pi}{\beta}\right)^{\mathrm{dim}\,\mathfrak{h}_{\text{qu}}}
e^{-\frac{4\pi^2}{\beta}L_h(\mathbf{x}^\ast)+i\frac{8\pi^3}{\beta^2}Q_h(\mathbf{x}^\ast)},\label{eq:saddleAsyGenLeading}
\end{equation}
where $\mathbf{x}^\ast$ is any point on $\mathfrak{h}_{\text{qu}}.$

The elementary fact that patch-wise application of the saddle-point method (together with the continuity argument for the derivative of $Q_h$) can determine the asymptotics of \eqref{eq:simplifiedIntegral} was missed in \cite{ArabiArdehali:2015ybk}, partly because the examples treated there were not rich enough to necessitate complex-analytic tools. The saddle-point analysis outlined above (augmented with a straightforward adaptation of the inner-patch results in \cite{Ardehali:2021irq}) resolves Problem~1 in \cite{ArabiArdehali:2015ybk}, which was to find the asymptotics of a general 4d index with a nontrivial $Q_h$ function.\footnote{Problem~1.1 in \cite{ArabiArdehali:2015ybk} will be addressed by the example in Section~\ref{subsec:FibBar}, which is the first case we have encountered where there is conflict between making $Q_h$ stationary and minimizing $L_h$, so that complex-analytic tools are required for the asymptotic analysis. Problems~2 and 3 there are still open; see Section~4.2 of \cite{ArabiArdehali:2023bpq} for more comments. Problem 4 appears straightforward if one imposes gauge-gravity-gravity and gauge-$R$-$R$ anomaly cancellation.} Note that while the asymptotic relation \eqref{eq:saddleAsyGenLeading} had appeared in a different guise in \cite{Ardehali:2021irq}, the real-analytic derivation there via minimization of $\mathrm{Re}(iQ_h/\beta^2)$ does not work when $\beta\in\mathbb{R}$, because then $\mathrm{Re}(iQ_h/\beta^2)$ would vanish; the complex-analytic, patch-wise saddle-point derivation here fills that gap.\\

\noindent\textbf{Dominant patches and their field content.} The asymptotic in \eqref{eq:saddleAsyGenLeading} can be improved to exponential accuracy by incorporating the contributions of all patches that intersect $\mathfrak{h}_\text{qu}.$ These will be the dominant patches. All other patches make exponentially smaller contributions to the asymptotic of the index.

To each patch, and in particular to each dominant patch, we can associate a \textbf{\small 3d $\mathcal{N}=2$ field content} as in \cite{Ardehali:2021irq}. Besides the $r_G$ photon multiplets present everywhere on $\mathfrak{h}_\text{cl},$ there will be light chiral multiplets for every instance of
\begin{equation}
    \rho^\chi\cdot\boldsymbol{x}+q^\chi\cdot\boldsymbol{\xi}\in\mathbb{Z}\,,
\end{equation}
inside the inner patch, and (pairs of) massless vector multiplets for every instance of
\begin{equation}
    \alpha_+\cdot\boldsymbol{x}\in\mathbb{Z}\,.
\end{equation}

\noindent\textbf{EFT couplings.} In the present work we focus on cases with $\mathrm{dim}\mathfrak{h}_\text{qu}=0$. In fact we will be considering examples where $\mathfrak{h}_\text{qu}$ consists of a single point $\boldsymbol{x}^\ast$ (modulo Weyl orbit redundancies). The small-$\beta$ asymptotics of the 4d index is then captured by the $S^3$ partition function of an EFT with field content as described above, and with various induced Chern-Simons couplings in the uv (\emph{i.e.} the scale set by the decomposition cut-offs). Denote the set of \emph{heavy} 4d multiplets (those that do not yield any light fields in the 3d EFT) at the saddle $\boldsymbol{x}^\ast$ by $H_\ast$. Denote the complement set, consisting of \emph{light} 4d multiplets, by $L_\ast$. The induced couplings can be obtained from the formulas \cite{Ardehali:2021irq}:
\begin{equation}
\begin{split}
    k^{\ast}_{ij}&=-\sum_{\chi}\sum_{\rho^\chi\in H_\ast}\overline{B}_1 \bigl(\rho^\chi \cdot \boldsymbol{x}^\ast  +q^\chi\cdot\boldsymbol{\xi} \bigr)\, \rho^{\chi}_i\, \rho^{\chi}_j,\\
    k^{\ast}_{j R}&=-\sum_{\chi}\sum_{\rho^\chi\in H_\ast}\overline{B}_1 \bigl(\rho^\chi \cdot \boldsymbol{x}^\ast  +q^\chi\cdot\boldsymbol{\xi} \bigr)\, \rho^{\chi}_j\,  (r_\chi-1)-\sum_{\alpha\in H_\ast}\overline{B}_1 \bigl(\alpha \cdot \boldsymbol{x}^\ast\bigr)\, \alpha_j\,  ,\\
    k^{\ast}_{R R}&=-\sum_{\chi}\sum_{\rho^\chi\in H_\ast}\overline{B}_1 \bigl(\rho^\chi \cdot \boldsymbol{x}^\ast  +q^\chi\cdot\boldsymbol{\xi} \bigr)\,   (r_\chi-1)^2,\\
    k^{\ast}_{\text{grav}}&=-2\sum_{\chi}\sum_{\rho^\chi\in H_\ast}\overline{B}_1 \bigl(\rho^\chi \cdot \boldsymbol{x}^\ast  +q^\chi\cdot\boldsymbol{\xi} \bigr).
    \end{split}\label{eq:EFT_CS_couplings}
\end{equation}
\begin{figure}[t]
\centering
\includegraphics[scale =0.5]{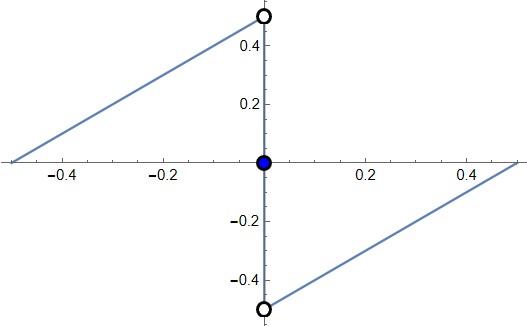}
\caption{The odd function $\overline{B}_1(x)$ versus $x$. It is periodic with period $1$, and discontinuous only across $\mathbb{Z}$, where it attains the average value of its left and right limits.}
\label{fig:Bbar1}
\end{figure}

The function $\overline B_1(x)$, displayed in Figure~\ref{fig:Bbar1}, is defined as\footnote{In computer implementations it may pay off to set  $\overline{B}_1=0$ in a tiny window (of width say $10^{-3}$) around $\mathbb{Z}$, banishing the discontinuities where they are unlikely to interfere with calculations.}
\begin{equation}
    \overline{B}_1(u):=\begin{cases}
    \{u\}-\frac{1}{2}\qquad\, \text{for $u\notin\mathbb{Z}$}\,,\\
    0 \qquad\qquad\text{for $u\in\mathbb{Z}$}\,.
    \end{cases}
\end{equation}
For the factor of $2$ arising for $k_{\text{grav}}$ and absent in $k_{ij},k_{jR},k_{RR}$, see \emph{e.g.} Appendix~A of \cite{Closset:2018ghr}. The corresponding CS terms are in our conventions
\begin{equation}
    \begin{split}
        &\frac{i}{4\pi}\int \mathrm{d}^3x\,\sqrt{g} \, \epsilon^{\mu\nu\rho}\mathcal{A}^i_\mu\,\partial_\nu\mathcal{A}^j_\rho\ ,\\
        &\frac{i}{4\pi}\int \mathrm{d}^3x\,\sqrt{g} \, \epsilon^{\mu\nu\rho}\mathcal{A}^j_\mu\,\partial_\nu\mathcal{A}^{(R)}_\rho\, ,\\
        &\frac{i}{4\pi}\int \mathrm{d}^3x\,\sqrt{g} \, \epsilon^{\mu\nu\rho}\mathcal{A}^{(R)}_\mu\,\partial_\nu\mathcal{A}^{(R)}_\rho\, ,\\
        &\frac{i}{192\pi}\int \mathrm{d}^3x\,\sqrt{g} \, \epsilon^{\mu\nu\rho}\ \mathrm{Tr}\bigl(\omega_\mu\,\partial_\nu\omega_\rho-\tfrac23 \omega_\mu\omega_\nu\omega_\rho\bigr)\, .
        \end{split}\label{eq:CS_actions}
\end{equation}
\vspace{.1cm}

From an EFT perspective, the formulas \eqref{eq:EFT_CS_couplings} arise from zeta-function regularization of the KK sums over the familiar contributions 
\begin{equation}
    \delta k_{xy}=\frac{1}{2}\,\mathrm{sign}(m)\,q_x\, q_y\,,
\end{equation}
to the one-loop-exact CS coupling generated by integrating out a fermion of real-mass $m$ with charges $q_{x,y}$ under the gauge fields $\mathcal{A}^{x,y}$ \cite{Niemi:1983rq,Redlich:1983kn,Coleman:1985zi,Closset:2012vp,Golkar:2012kb,DiPietro:2014bca}. For the contribution of a $\rho^\chi\in H_\ast$ to the gauge-gauge CS coupling for instance we have \cite{DiPietro:2016ond}
\begin{equation}
    \sum_{n\in\mathbb{Z}}\frac{1}{2}\,\mathrm{sign}\bigl(n+\rho^\chi \cdot \boldsymbol{x}^\ast  +q^\chi\cdot\boldsymbol{\xi} \bigr)\, \rho^{\chi}_i\, \rho^{\chi}_j\xrightarrow{\ \zeta\text{-reg.}\ }-\overline{B}_1 \bigl(\rho^\chi \cdot \boldsymbol{x}^\ast  +q^\chi\cdot\boldsymbol{\xi} \bigr)\, \rho^{\chi}_i\, \rho^{\chi}_j,
\end{equation}
where we have used $\sum_{n\in\mathbb{Z}}\mathrm{sign}(n+x)(n+x) \xrightarrow{\zeta\text{-reg.}\,} -2 \, \overline{B}_{1}(x)$.

The breakthrough realization of \cite{DiPietro:2014bca}, further corroborated in \cite{DiPietro:2016ond,ArabiArdehali:2021nsx,Cassani:2021fyv,Ardehali:2021irq}, was that by supersymmetrizing the CS actions \eqref{eq:CS_actions} (as well as those involving the KK photon) with the zeta-regularized couplings as in \eqref{eq:EFT_CS_couplings}, the 3d effective action is essentially fixed, at least as far as localization results are concerned.\footnote{This is modulo a lingering puzzle in the literature \cite{Cassani:2021fyv,Ardehali:2021irq} pertaining to geometries with nonzero angular twists $\Omega_{1,2}$. That puzzle seems irrelevant to the present work though, since we set $\Omega_{1,2}=0$.}\\

\noindent\textbf{Monopole operators.} Using the relations $\kappa(x)=2\overline{B}_3(x)$ and $\vartheta(x)=-\overline{B}_2(x)+\frac{1}{6}$
following from \eqref{eq:perBern}, together with $\overline{B}'_n(x)=n\,\overline{B}_{n-1}(x),$
we have $\kappa''(x)=12\overline{B}_1(x)$ and $\vartheta'(x)=-2\overline{B}_1(x)$. The formulas \eqref{eq:EFT_CS_couplings} then imply\footnote{Note that Eqs.~\eqref{eq:EFT_CS_couplings} together with the behavior of $\overline{B}_1$ as in Figure~\ref{fig:Bbar1}, imply that an inner patch interpolating between two outer patches has average values of $k_{ij}$ and $k_{jR}$.\label{fn:inner_average}}
\begin{equation}
    \boxed{\begin{split}
        \partial_i\partial_j Q_h\qquad&\xrightarrow{\text{\ \ \ \ patch-wise constant\ \ \ \ }} \qquad -k_{ij}\,,\\
        \partial_j L_h\qquad&\xrightarrow{\text{\ \ \ \ patch-wise constant\ \ \ \ }}\qquad -k_{jR}\,.
    \end{split}}\label{eq:kij&kjR_from_Q&L}
\end{equation}

We restrict to patches that intersect the stationary loci of $Q_h$ as discussed above. Since $Q_h$ is piecewise quadratic, if its second derivatives vanish on its stationary loci, it would be flat. Since a Coulomb branch in the $i$th direction of the 3d EFT is possible only if $k_{ij}=0$ for all $j,$ we conclude that the 3d EFT can only have viable Coulomb branch directions along flat directions of $Q_h.$ Alternatively, since \emph{on outer patches} the gauge charges of a 3d monopole $V_{\boldsymbol{m}}$ are given by (\emph{cf.}~Eq.~(3.18) in \cite{Intriligator:2013lca})
\begin{equation}
    c_i(\boldsymbol{m})=-\sum_j k^{}_{ij}\,{m}_j,
\end{equation}
the flat directions of $Q_h$ signal gauge-invariant monopoles in the 3d EFT.

There might still be obstructions to such Coulomb branch directions from contact terms on curved background due to $k_{jR}$ \cite{DiPietro:2016ond}. Alternatively, since \emph{on outer patches} the $R$-charge of a 3d monopole $V_{\boldsymbol{m}}$ is
\begin{equation}
    r(\boldsymbol{m})=-\sum_j k_{jR}^{}\, {m}_j,\label{eq:r(m)_out}
\end{equation}
the slope of $L_h$ along the flat directions of $Q_h$ determines whether there are gauge-invariant monopole operators of the right $R$-charge $(=2)$ to be dynamically generated in the superpotential, either \`{a}~la Affleck-Harvey-Witten \cite{Affleck:1982as}, or due to some KK- or multi-monopole generalization thereof \cite{Seiberg:1996nz,Lee:1997vp,Aharony:1997bx,Kraan:1998pm,Aharony:2013dha,Aharony:2013kma,Dorey:1999sj,Dorey:2001qj,Csaki:2017mik,Amariti:2019rhc,ArabiArdehali:2019zac}. Hence the unlifted Coulomb branch of the 3d EFT is expected to corresponds to $\mathfrak{h}_\text{qu}$, both on curved background where there are contact terms associated to non-flat $L_h$, and on $\mathbb{R}^3$ where the contact terms vanish but there can be monopole superpotentials diagnosed by the slope of $L_h$.

The general formulas valid \emph{on all patches} (see \emph{e.g.} \cite{Closset:2016arn,Borokhov:2002cg,Intriligator:2013lca})
\begin{equation}
    c_i(\boldsymbol{m})=-\sum_j k^{}_{ij}\,{m}_j-\frac{1}{2}\sum_{\chi}\sum_{\rho^\chi\in L_\ast}\rho^\chi_i|\rho^\chi(\boldsymbol{m})|,\label{eq:c(m)}
\end{equation}
\begin{equation}
    r(\boldsymbol{m})=-\sum_j k^{}_{jR}\,{m}_j-\frac{1}{2}\sum_{\chi}(r_{\chi}-1)\sum_{\rho^\chi\in L_\ast}|\rho^\chi(\boldsymbol{m})|-\frac{1}{2}\sum_{\alpha\in L_\ast} |\alpha(\boldsymbol{m})|,\label{eq:r(m)}
\end{equation}
together with \eqref{eq:EFT_CS_couplings} imply that monopole gauge and $R$ charges are continuous across patches. Therefore inner patches naturally inherit the dynamically generated monopole superpotentials of their neighboring outer patches. In addition, they might admit gauge-invariant superpotentials containing light matter fields of the patch (\emph{cf.} \cite{Aharony:1997bx,Aharony:1997gp,Intriligator:2013lca,Amariti:2015kha}). See \cite{ArabiArdehali:2019zac} for an example in a similar spirit to what we will encounter below.

The connection outlined above between Cardy limit of the 4d superconformal index and 3d monopoles was through CS terms in the KK effective action \cite{Intriligator:2013lca,DiPietro:2016ond}. The relation between $L_h$ and $R$-charges of monopoles can be understood also from an alternative perspective due to Shaghoulian \cite{Shaghoulian:2016gol} (see  \cite{Closset:2017bse} for a sharper BPS version). The idea is that \cite{Shaghoulian:2016gol}
\begin{equation*}
    S^1_{\beta\to0}\times S^3\approx S^1\times S^3/\mathbb{Z}_{p\to\infty}\,,
\end{equation*}
through a ``modular''
identification
\begin{equation}
\frac{\tilde{r}_{S^3}/p}{\tilde{r}_{S^1}}=\frac{r_{S^1}}{r_{S^3}}\,,\label{eq:modId}
\end{equation}
with the tilded parameters those of the latter space. This allows relating the Cardy limit of the index to the supersymmetric Casimir energy \cite{Assel:2015nca,Martelli:2015kuk} on $S^1\times S^3/\mathbb{Z}_{p\to\infty}\,$. There are discrete holonomy sectors associated to the torsion cycles of $S^3/\mathbb{Z}_{p},$ and in the limit $p\to\infty$ they can be thought of via Stokes' theorem as monopole sectors on $S^2\approx S^3/\mathbb{Z}_{p\to\infty}$. The supersymmetric Casimir energy of these flux sectors matches the $R$-charge of the corresponding monopole operator thanks to the BPS relation between energy and $R$-charge of 3d chiral operators in radial quantization. This ends up relating the ``high-temperature'' $L_h$ to the ``low-temperature'' $r(\boldsymbol{m})$ as in \cite{ArabiArdehali:2019zac,Dorey:2023jfw}. A similar understanding of $Q_h$ is lacking at present, but the results of \cite{Benini:2011nc} (see their Eq.~(25) in particular) appear to be a promising starting point.

\section{\texorpdfstring{$(A_1,A_2)$}{A1A2} and its Galois orbit}\label{sec:A1A2}

In this section, we analyze the Cardy limit of the index $\mathcal{I}^\gamma(q)$ of the $(A_1,A_2)$ Argyres-Douglas theory on its four inequivalent higher sheets, $\gamma=1,2,3,4$. This yields the EFTs arising from various $U(1)_r$-twisted circle reductions to 3d. These EFTs form the core result of this section. We also explain how they lead to 3d TQFTs and Galois orbits, using half-indices \cite{Gadde:2013wq,Dimofte:2017tpi} of natural boundary conditions in 3d EFTs as a tool for identifying the TQFTs.

On the second sheet (\emph{i.e.}, $\gamma=1$) we obtain the Gang-Yamazaki (GY) theory \cite{Gang:2018huc} as conjectured in \cite{Dedushenko:2023cvd}. The A-twisted (or H-twisted) GY theory supports the Lee-Yang VOA on its boundary \cite{Gang:2023rei}, as expected from the 4d/2d correspondence \cite{Dedushenko:2023cvd}. We refer to the A-twisted GY theory as the Lee-Yang TQFT. We also talk about the Lee-Yang modular tensor category (MTC), since the 3d TQFT is determined by the choice of MTC $\mathcal{C}$ and the central charge (compatible with the chiral central charge of $\mathcal{C}$ mod $8$).

On other sheets, using our EFTs, we obtain TQFTs closely related to the Galois conjugates of the Lee-Yang TQFT. Their modular $S$ and $T$ matrices, as evaluated through familiar Bethe root techniques \cite{Closset:2019hyt}, complete the Lee-Yang's Galois orbit \cite{Dedushenko:2018bpp,Harvey:2019qzs}, up to overall phases of $T$. These overall phases indicate an important subtlety, which we now explain. The Galois orbit of Lee-Yang MTC contains Fibonacci MTC, as well as the conjugates of these two. We sometimes denote the elements of the orbit as LY, Fib, $\overline{\text{LY}}$ and $\overline{\text{Fib}}$. It was originally conjectured in \cite{Dedushenko:2018bpp} that the second through fifth sheets of $(A_1, A_2)$ give precisely the Lee-Yang TQFT, Fibonacci TQFT, and their conjugates. Our findings corroborate this conjecture only up to multiplication by an invertible TQFT or spin-TQFT. Such invertible factors are responsible for the overall phases mentioned earlier. On the second and fifth sheets, Dirichlet half-indices yield LY and $\overline{\text{LY}}$, up to an invertible TQFT. On the third and fourth sheets, however, instead of the expected Fibonacci, we find fermionic theories: $\text{SM}(3,5)$ (an $\mathcal{N}=1$ supersymmetric minimal model that also coincides with the fermionized tricritical Ising model,) and its conjugate. This mismatch is resolved by realizing that SM$(3,5)$ coincides with $\overline{\text{Fib}}$ multiplied by an invertible spin-TQFT. 

Thus the modified proposal is that different sheet TQFTs, after possible multiplication by an invertible (spin-)TQFT, live on the same Galois orbit. To test this proposal, we also perform the half-index calculations yielding $(G_2)^{}_1 \cong \text{Fib}$ and fermionic $\mathrm{SM}(3,5)$ characters on different boundaries of the third and fourth-sheet TQFTs. The close connection between the spin-TQFT of $\mathrm{SM}(3,5)$ and the bosonic TQFT of $(F_4)^{}_1\cong \overline{\text{Fib}}$ is reflected in another close relation between the fermionic VOA $\mathrm{SM}(3,5)$ and the bosonic VOA $(G_2)^{}_1$, which will be discussed in Section~\ref{subsec:sense2dBCs}. Namely, they are commutants of each other inside the VOA of free fermions, similar to the case of ${\rm OSp}(1|2)_1$ and ${\rm LY}$ discussed in \cite{Ferrari:2023fez}.

We use the 4d $\mathcal{N}=1$ Maruyoshi-Song Lagrangian for the $(A_1, A_2)$ theory \cite{Maruyoshi:2016tqk}. The readers should consult this reference for details. Here we only explicitly quote the $\mathcal{N}=2$ index (see Eq.~(16) in \cite{Maruyoshi:2016tqk}):
\begin{equation}
    \begin{split}
     \mathcal{I}_t(p,q,t)&=(p;p)(q;q)\frac{\Gamma_e\big((\frac{pq}{t})^{\frac{6}{5}}\big)\Gamma_e\big((\frac{pq}{t})^{\frac{1}{5}}\big)}{\Gamma_e\big((\frac{pq}{t})^{2/5}\big)}\times\\
     &\ \oint\frac{\mathrm{d}z}{2\pi i z}\ \frac{\Gamma_e\big(z^{\pm1}(pq)^{\frac{2}{5}}t^{\frac{1}{10}}\big)\Gamma_e\big(z^{\pm1}(pq)^{-\frac{1}{5}}t^{\frac{7}{10}}\big)\Gamma_e\big(z^{\pm2}(\frac{pq}{t})^{\frac{1}{5}}\big)}{2\Gamma_e(z^{\pm2})}\,,
    \end{split}\label{eq:I_t for n=1}
\end{equation}
where $(\cdot\,;\cdot)$ and $\Gamma_e(\cdot)$ stand respectively for the $q$-Pochhammer symbol and the elliptic gamma function, and the integration contour is the unit circle.

\subsection{Second sheet: Lee-Yang}\label{subsec:LY}

Going to the 2nd sheet of the SUSY index by setting $\gamma=1$, the resulting twisted $\mathcal{N}=1$ index becomes\footnote{Note that a different ``2nd sheet'' index, namely $\mathcal{I}_t(p\, e^{2\pi i},q,e^{\frac{4\pi i}{3}}(pq)^{\frac{2}{3}})$, of the $(A_1,A_2)$ theory was studied in \cite{Kim:2019yrz}. That index corresponds to twisting the boundary condition around the circle with a 4d $\mathcal{N}=1$ $R$-charge rather than the $\mathcal{N}=2$ 
$U(1)_r$ charge of our interest here.}
\begin{equation}
    \begin{split}
     \mathcal{I}^{\gamma=1}(p,q)&=\mathcal{I}_t(p\, e^{2\pi i},q,(pq)^{2/3})\\
     &=(p;p)(q;q)\frac{\Gamma_e\big(e^{\frac{12\pi i}{5}}(pq)^{\frac{2}{5}}\big)\Gamma_e\big(e^{\frac{2\pi i}{5}}(pq)^{\frac{1}{15}}\big)}{\Gamma_e\big(e^{\frac{4\pi i}{5}}(pq)^{\frac{2}{15}}\big)}\times\\
     &\quad \oint\frac{\mathrm{d}z}{2\pi i z}\ \frac{\Gamma_e\big(z^{\pm1} e^{\frac{4\pi i}{5}} (pq)^{\frac{7}{15}}\big)\Gamma_e\big(z^{\pm1} e^{-\frac{2\pi i}{5}}(pq)^{\frac{4}{15}}\big)\Gamma_e\big(z^{\pm2}e^{\frac{2\pi i}{5}}(pq)^{\frac{1}{15}}\big)}{2\Gamma_e(z^{\pm2})}\,.   
    \end{split}\label{eq:I_4 for n=1}
\end{equation}
We further restrict to $p=q=e^{2\pi i\tau}$ for simplicity and denote the resulting index as ${\mathcal{I}}^{\gamma=1}(q)$. The more general index $\mathcal{I}^{\gamma=1}(p,q)$ can be studied similarly. 

\subsubsection{The holonomy saddle and the EFT data}

We now analyze the Cardy limit  $q=e^{-\beta}\to1$. For the leading asymptotics, \eqref{eq:simplifiedIntegral} gives:
\begin{equation}
    \mathcal{I}^{\gamma=1}(q)\approx\frac{2\pi}{\beta}\int_{-\frac{1}{2}}^{\frac{1}{2}}\mathrm{d}x\ e^{\frac{2\pi i}{90\tau}-\frac{4\pi^2}{\beta}L^{\gamma=1}_h(x)+i\frac{8\pi^3}{\beta^2}Q_h^{\gamma=1}(x)}\,,\label{eq:I_4QthLth}
\end{equation}
\begin{equation}
\begin{split}
Q_h^{\gamma=1}(x)&=\frac{1}{12}\bigg[\kappa\left(\frac{6}{5}\right)+\kappa\left(\frac{1}{5}\right)-\kappa\left(\frac{2}{5}\right)+\kappa\left(x+\frac{2}{5}\right)+\kappa\left(-x+\frac{2}{5}\right)\\
&\qquad +\kappa\left(x-\frac{1}{5}\right)+\kappa\left(-x-\frac{1}{5}\right)+\kappa\left(2x+\frac{1}{5}\right)+\kappa\left(-2x+\frac{1}{5}\right)
\bigg]\,,
\end{split}
\end{equation}
\begin{equation}
\begin{split}
L_h^{\gamma=1}(x)&=\left(1-\frac{4}{5}\right)\frac{\vartheta(\frac{6}{5})}{2}+\left(1-\frac{2}{15}\right)\frac{\vartheta(\frac{1}{5})}{2}-\left(1-\frac{4}{15}\right)\frac{\vartheta(\frac{2}{5})}{2}\\
&\quad+\left(1-\frac{14}{15}\right)\frac{\vartheta(x+\frac{2}{5})+\vartheta(-x+\frac{2}{5})}{2}+\left(1-\frac{8}{15}\right)\frac{\vartheta(x-\frac{1}{5})+\vartheta(-x-\frac{1}{5})}{2}\\
&\quad+\left(1-\frac{2}{15}\right)\frac{\vartheta(2x+\frac{1}{5})+\vartheta(-2x+\frac{1}{5})}{2}
-\vartheta(2x)\,,
\end{split}
\end{equation}
Recall that the functions $\vartheta(x)=\{x\}(1-\{x\})$ and $\kappa(x)=\{x\}(1-\{x\})(1-2\{x\})$, with $\{x\}:=x-\lfloor x\rfloor,$ are $1$-periodic.

\begin{figure}[t]
\centering
\includegraphics[scale =.7]{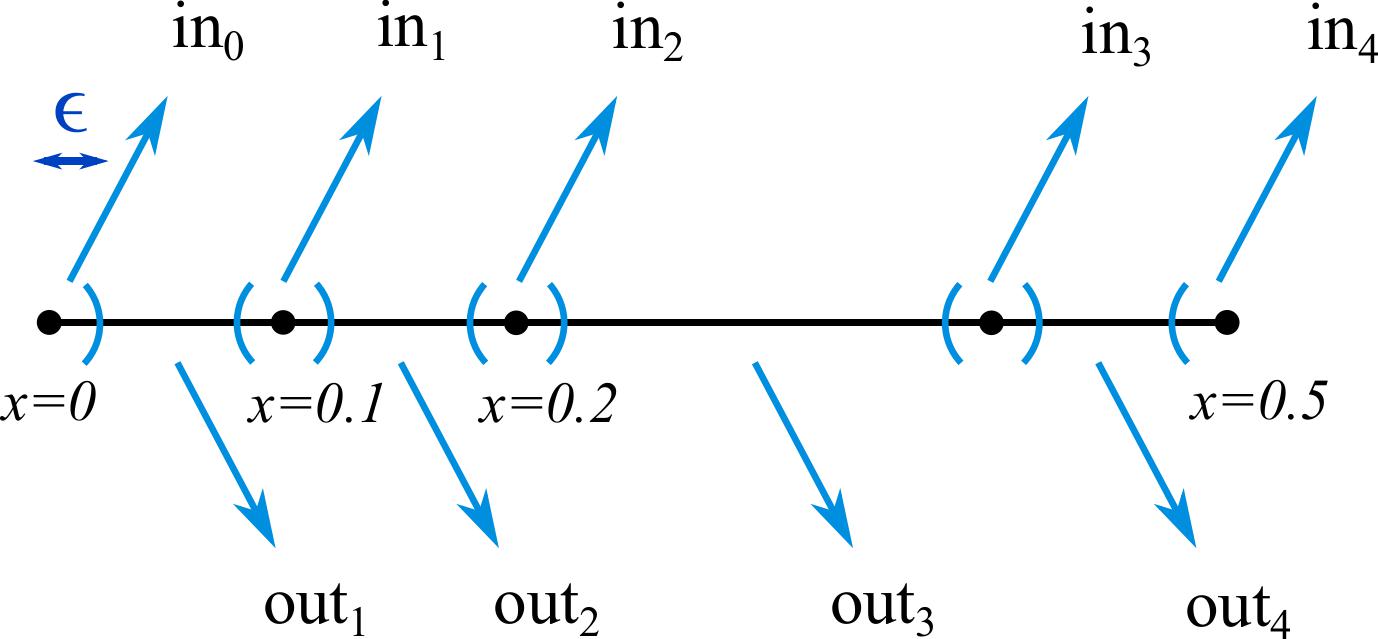}
\caption{Decomposition of the moduli-space of holonomies as in \cite{Ardehali:2021irq}, for the integral \eqref{eq:I_4 for n=1}. The domain $-1/2<x<0$ is neglected due to the $\mathbb Z_2$ Weyl redundancy. The constant $\epsilon$ is taken suitably small, for instance $\epsilon=0.01$}
\label{fig:patchesGamma=1}
\end{figure}

The result in \eqref{eq:I_4QthLth} is obtained using estimates inside the integrand that are uniformly accurate on the \emph{outer patch} (defined as the union of the disjoint outer patches). Those estimates lose uniform validity inside the \emph{inner patches}; see Figure~\ref{fig:patchesGamma=1}. Nevertheless, \eqref{eq:I_4QthLth} locates the holonomy saddle and captures the leading $e^{\frac{\#}{\beta^2}+\frac{\#}{\beta}}$ asymptotic of $\mathcal{I}^{\gamma}(q)$ correctly (\emph{cf.}~\cite{Ardehali:2021irq}). In Appendix~\ref{app:I4toZ3} we discuss how more accurate estimates inside the inner patches can yield the subleading $\mathcal{O}(\beta^0)$ asymptotic as well.

\begin{figure}[h]
\centering
\includegraphics[scale =0.5]{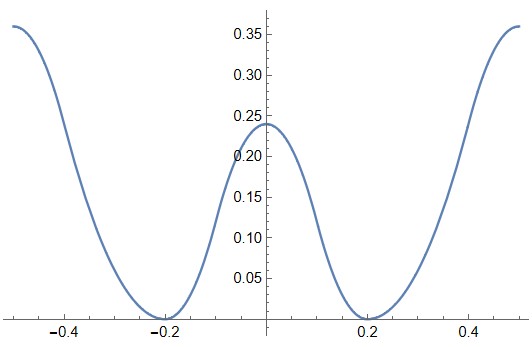}
\caption{The plot of $12Q^{\gamma=1}_h(x)$ versus $x$. The minima at $x=\pm2/10$ are exactly zero.}
\label{fig:QhtADn1}
\end{figure}

\begin{figure}[h]
\centering
\includegraphics[scale =0.5]{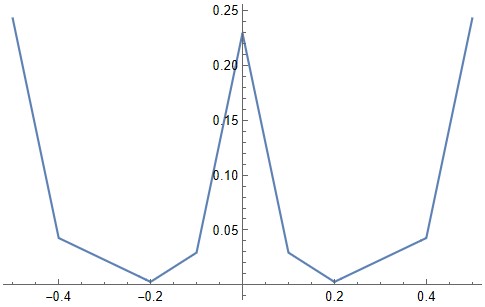}
\caption{The plot of $L^{\gamma=1}_h(x)$ versus $x$ for $(A_1,A_2)$. The minima at $x=\pm2/10$ are exactly zero.}
\label{fig:LhtADn1}
\end{figure}

The leading asymptotic of $\mathcal{I}^{\gamma=1}$ is dictated by the stationary point of $Q_h^{\gamma=1}$. In the best case scenario this would coincide with the locus of minima of $L_h^{\gamma=1}$. This best case scenario is realized here, at 
\begin{equation}
    x^\ast=\pm0.2\,,\label{eq:x*LY}
\end{equation}
as can be seen from the plots in Figures~\ref{fig:QhtADn1} and \ref{fig:LhtADn1}.

We can hence read the leading asymptotic from \eqref{eq:I_4QthLth} as\footnote{The remarkable-looking fact that $Q^{\gamma=1}_h
(.2) = 0$, which implies the index does not grow as $e^{\#/\beta^2},$ and also the fact that $L^{\gamma=1}_h
(.2) = 0$, will be explained in Section~\ref{subsec:sense4dBackground}.}
\begin{equation}
    \mathcal{I}^{\gamma=1}(q)\approx\, e^{\frac{2\pi i}{90\tau}-\frac{4\pi^2}{\beta}L^{\gamma=1}_h(x=.2)+i\frac{8\pi^3}{\beta^2}Q_h^{\gamma=1}(x=.2)}=e^{\frac{2\pi i}{90\tau}}.\label{eq:I_4n1SingAsy}
\end{equation}
The error is multiplicative $\mathcal{O}(\beta^0)$ (\emph{cf.} Eq.~(2.57) of \cite{Ardehali:2021irq}).\\

We conclude from \eqref{eq:x*LY} that \emph{the 3d effective theory lives on $\mathrm{in}_2$}. Its matter content can be read from \eqref{eq:I_4 for n=1} near $z=e^{2\pi i \frac{1}{5}}$: the only light multiplets are that of the photon as well as the chiral multiplet with gauge charge $+1$ and $R$-charge $8/15$. For the induced gauge-gauge, gauge-$R$, $R$-$R$, and gravitational CS couplings, the formulas in \eqref{eq:EFT_CS_couplings} give (see Appendix~\ref{app:effective_CS} for derivations):
\begin{equation}
    \begin{split}
        k_{gg}=-\frac{3}{2}\,,\qquad
    {k}_{gR}=\frac{1}{30}\,,\qquad
    {k}_{RR}=-\frac{31}{225}\,,\qquad
    k_{\text{grav}}=\frac{2}{5}\,.
    \end{split}\label{eq:CSlevels n=1 main}
\end{equation}

\subsubsection{SUSY enhancement \texorpdfstring{$\xrightarrow{\text{top. twist\ }}$}{ top. twist } non-unitary TQFT}\label{subsec:GY_susy_enhancement}

The 3d EFT found above is a 3d $\mathcal{N}=2$ $U(1)_{-3/2}$ gauge theory with a chiral multiplet $\Phi$ of gauge charge $+1$ and $R$-charge $8/15$. The effective Lagrangian at the uv scale\footnote{For inner-patches we distinguish between the ``uv'' cut-off $\Lambda\sim\frac{2\pi\epsilon}{\beta}$ of the 3d EFT \cite{Ardehali:2021irq}, and the ``UV'' scale $\propto \frac{1}{\beta}>\Lambda$ where the theory is four dimensional.} $\Lambda\sim \frac{2\pi\epsilon}{\beta}$ also contains a mixed gauge-$R$ CS coupling ${k}_{gR}=1/30$, as well as $k_{RR}=-31/225$ and $k_{\text{grav}}=2/5.$ This theory is the same as the one studied by Gang and Yamazaki in \cite{Gang:2018huc} (modulo the background CS levels which were irrelevant there). We thus refer to it henceforth as the GY theory.

It was discovered via $F$-maximization in \cite{Gang:2018huc} that the GY theory flows at low energies to a 3d $\mathcal{N}=4$ SCFT whose $\mathcal{N}=2$ superconformal $U(1)_R$ current is related to the uv $R$-current by mixing with the topological $U(1)_J.$ This mixing shifts the parameters of the IR SCFT with respect to those in the uv discussed above, such that in the IR we have  $r_\Phi=1/3$ for the 3d $\mathcal{N}=2$ $R$-charge of $\Phi,$ while $k_{gR}=0$. (Further mixing of the gauge and $U(1)_R$ symmetries, though physically inconsequential, allows for other ``$R$-charge schemes'' with different $k_{gR}$ and $r_\Phi.$ See Eqs.~\eqref{eq:c2} and \eqref{eq:kgR_mixings_gauge}.)

The resulting 3d $\mathcal{N}=4$ theory can then be subjected to the topological A-twist, also known as the H-twist, which employs the $\mathrm{SU}(2)_H$ $R$-symmetry. Thus it is convenient to use $U(1)_H\subset {\rm SU}(2)_H$ as the 3d $\mathcal{N}=2$ $R$-symmetry, yielding the following $\mathcal{N}=2$ uv data (see Appendix~\ref{app:Bethe_techniques} for the definitions of $k_{gg}^+$ and $k_{gR}^+$):
\begin{equation}
    \text{\textbf{LY}}:\qquad\qquad \mathbf{U(1)_{-3/2}}\ +\ \mathbf{\Phi_{+1}^{r=1}}\ \ \text{and}\ \ \mathbf{k_{gR}=0}\qquad(k_{gg}^+=-1,\ k_{gR}^+=0)\,.\label{eq:LY}
\end{equation}
We have indicated the gauge charge of the chiral multiplet in the subscript, and the $R$-charge in the superscript. Again, other $R$-charge schemes are possible via adding a multiple of the gauge charge to the $U(1)_R$ charge, but note that no such gauge mixing alone can relate the data in \eqref{eq:LY} to the saddle-point $\mathcal{N}=2$ data found above; a physical mixing with $U(1)_J$ is needed, to modify the saddle-point $U(1)_R$-charge and $k_{gR}$, to those of the A-twist, wherein the $U(1)_H$ is treated as the $\mathcal{N}=2$ $R$-symmetry.

We note in passing that the 3d A-model data in \eqref{eq:LY} was derived above via a multi-step procedure: 4d to 3d reduction, then $F$-maximization, then identification of the $U(1)_H$ $R$-symmetry appropriate for the A-twist. We will give a one-shot prescription for such derivations in Section~\ref{subsec:sense4dBackground}.

\subsection*{\texorpdfstring{$S$}{S} and \texorpdfstring{$T$}{T} matrices from handle-gluing and fibering operators}

The TQFT $S$ and $T$ matrices can be studied via localization and Bethe root techniques. Suppressing the contributions from $k_{RR}$ and $k_\text{grav},$ we first compute the effective twisted superpotential and the effective dilaton using \eqref{eq:WandOmega}:
\begin{equation}
    \begin{split}
        W(Z)&=-\pi i Z -\frac{1}{2} Z^2+\mathrm{Li}_2\big(e^{Z}\big)\,,\\
        \Omega(Z)&=0\,.
    \end{split}
\end{equation}
The Bethe equation $\exp\big(W'(Z_\alpha^\ast)\big)=1$ reads in terms of the charge $+1$ Wilson line\footnote{Instead of the holonomy saddle at $x=.2$, we could have picked the Weyl image at $x=-.2$, then the gauge charge of the chiral multiplet would become $-1$, and the Fibonacci fusion relation $ z^2=1+ z$ would arise for the charge $-1$ Wilson line.} variable $ z:=e^{Z}$ as
\begin{equation}
     z^2=1+ z\,.\label{eq:GYfusion}
\end{equation}
This is interpreted as a quantum relation either in the ring of BPS Wilson lines in 3d, or in the twisted chiral ring of the 2d theory obtained from reducing the 3d theory on a circle \cite{Kapustin:2013hpk}.

BPS surgery via fibering and handle-gluing operators \cite{Closset:2018ghr,Closset:2019hyt} yields as in \eqref{eq:HandF_vs_SandT} the following $S$ and $T$ matrices (up to an overall phase ambiguity for $T$ as in \cite{Gang:2023rei}):
\begin{equation}
\begin{split}
    S&=\left(
\begin{array}{cc}
 -\sqrt{\frac{2}{5-\sqrt{5}}} & \ \  \sqrt{\frac{2}{5+\sqrt{5}}} \\
 \sqrt{\frac{2}{5+\sqrt{5}}} & \ \ \sqrt{\frac{2}{5-\sqrt{5}}} \\
\end{array}
\right) = 
\frac{1}{\sqrt{2+\varphi }}
\begin{pmatrix}
   -\varphi &  1\\
   1 & \varphi
\end{pmatrix},\\
    T&=\left(
\begin{array}{cc}
 1 & \ \  0 \\
 0 & \ \ e^{2\pi i\,(-\frac{1}{5})} \\
\end{array}
\right),
    \end{split}\label{eq:S&T_GY}
\end{equation}
where $\varphi = \frac{1+\sqrt{5}}{2}$ is the golden ratio. Note that here the sign ambiguities in the entries of $S$ are removed by leveraging our knowledge that $i$) we are in the rational MTC context, where the $S$ matrix has one and only one positive row/column, and $ii$) we are in the topologically twisted SUSY enhanced setting yielding a non-unitary MTC, where the positive row is not the vacuum row. See Appendix~\ref{app:Bethe_techniques}.

The $S$ and $T$ matrix can be used to compute partition functions on three-manifolds. For example, the topological $S^3$ partition function can be either expressed as:
\begin{equation}
    Z^{}_{S^3}=S_{00}=\frac{-\varphi}{\sqrt{2+\varphi }}\,,
\end{equation}
 or as \eqref{eq:Z_surgery} (up to an overall phase).

The half-index of a $(0,2)$ boundary condition in a 3d $\mathcal{N}=2$ QFT \cite{Gadde:2013wq,Dimofte:2017tpi} gives the vacuum character of its boundary VOA. Dressing the half-index with Wilson lines gives access to the non-vacuum modules of the VOA as in \cite{Gang:2023rei}. In our case, the GY 3d $\mathcal{N}=2$ QFT possesses enhanced $\mathcal{N}=4$ SUSY in the IR, and is moreover subjected to the topological A-twist. Hence the boundary conditions we are interested in are $(0,4)$ boundary conditions subject to the Costello-Gaiotto deformation \cite{Costello:2018fnz} that makes them compatible with the A-twist. These are holomorphic boundary conditions in the TQFT, and their half-index also computes the vacuum character of the boundary VOA. How is this half-index related to that of $(0,2)$ boundary conditions in the 3d $\mathcal{N}=2$ theory? Firstly, the parameter $\epsilon$ of Costello-Gaiotto deformation breaks the $R$-symmetry of a $(0,4)$ boundary condition, yet it does not enter the half-index. Thus we may simply compute the half-index of $(0,4)$ boundary condition at $\epsilon=0$, but only with the fugacities for symmetries unbroken by the Costello-Gaiotto deformation present. Secondly, the $(0,4)$ boundary conditions look like $(0,2)$ from the 3d $\mathcal{N}=2$ perspective. Thus the TQFT half-index must be equal to the usual half-index of $(0,2)$ boundary conditions in the GY theory. We only have to make sure, first, that the $(0,2)$ boundary conditions enhance to $(0,4)$ in the IR, and second, that we only include fugacities for symmetries compatible with the Costello-Gaiotto deformation. For example, the topological symmetry $U(1)_J$ of the GY theory (identified with the anti-diagonal combination of $U(1)_H$ and $U(1)_C$ in the IR) is not one of them. As for the SUSY enhancement of $(0,2)$ boundary conditions to $(0,4)$, we do not study it in detail. We simply consider natural $(0,2)$ boundary conditions, and view various consistency checks, such as the anomaly matching and the expected answers for half-indices, as evidence that we chose the right boundary conditions.

The two characters of $M(2,5)$ have been reproduced via Dirichlet and A-twist on the left boundary in \cite{Gang:2023rei}, as well as via Neumann and B-twist on the right boundary in \cite{Ferrari:2023fez}. We have obtained the Lee-Yang characters in the A-twisted theory \eqref{eq:LY} via Dirichlet on the right boundary (due to our different parity conventions), with the computation paralleling the one in \cite{Gang:2023rei}. Quite interestingly, we were also able to reproduce them via the somewhat subtle Neumann boundary conditions on the right (enriched by the boundary chirals and Fermis), as explained in Section \ref{subsec:sense2dBCs}.

Finally, note that the induced gravitational CS level $k_\text{grav}=2/5$ in Eq.~\eqref{eq:CSlevels n=1 main} explains via inflow the boundary gravitational anomaly $48(c-a)=2/5$ discussed in \cite{Dedushenko:2023cvd}, resolving the anomaly mismatch puzzle raised in that work.


\subsection{Third sheet: conjugate Fibonacci}\label{subsec:FibBar}

Going to the 3rd sheet of the index by setting $\gamma=2$ we get:
\begin{equation}
    \begin{split}
     \mathcal{I}^{\gamma=2}(p,q)&=\mathcal{I}_t(p\, e^{4\pi i},q,t=(pq)^{2/3})\\
     &=(p;p)(q;q)\frac{\Gamma_e\big(e^{\frac{24\pi i}{5}}(pq)^{\frac{2}{5}}\big)\Gamma_e\big(e^{\frac{4\pi i}{5}}(pq)^{\frac{1}{15}}\big)}{\Gamma_e\big(e^{\frac{8\pi i}{5}}(pq)^{\frac{2}{15}}\big)}\times\\
     &\quad \oint\frac{\mathrm{d}z}{2\pi i z}\ \frac{\Gamma_e\big(z^{\pm1} e^{\frac{8\pi i}{5}} (pq)^{\frac{7}{15}}\big)\Gamma_e\big(z^{\pm1} e^{-\frac{4\pi i}{5}}(pq)^{\frac{4}{15}}\big)\Gamma_e\big(z^{\pm2}e^{\frac{4\pi i}{5}}(pq)^{\frac{1}{15}}\big)}{2\Gamma_e(z^{\pm2})}.   
    \end{split}\label{eq:Itilde_4 for n=1}
\end{equation}
Again for simplicity we restrict to $p=q$ and denote the resulting index as ${\mathcal{I}}^{\gamma=2}(q)$. 

\subsubsection{The dominant patch and the EFT data}

\begin{figure}[t]
\centering
\includegraphics[scale =0.5]{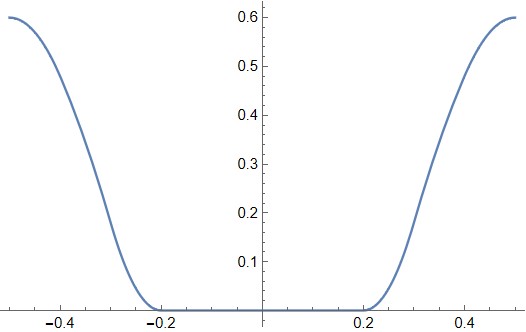}
\caption{The plot of $12Q^{\gamma=2}_h(x)$ versus $x$. The flat direction in the middle signals a gauge-invariant monopole.}
\label{fig:QhtTildeADn1}
\end{figure}

\begin{figure}[t]
\centering
\includegraphics[scale =0.5]{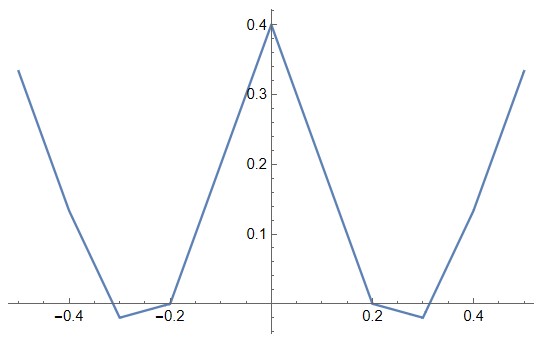}
\caption{The plot of $L^{\gamma=2}_h(x)$ versus $x$. At $x=\pm2/10$, it is exactly zero. As explained below \eqref{eq:r(m)_out}, the slope of $L^{\gamma=2}_h$ being $\pm2$ along the flat direction of $Q^{\gamma=2}_h$ signals Affleck-Harvey-Witten type superpotentials.}
\label{fig:LhtTildeADn1}
\end{figure}

Asymptotics of the index \eqref{eq:Itilde_4 for n=1} is obtained using the estimates in \cite{ArabiArdehali:2015ybk} as
\begin{equation}
    {\mathcal{I}}^{\gamma=2}(q)\approx\frac{2\pi}{\beta}\int_{-\frac{1}{2}}^{\frac{1}{2}}\mathrm{d}x\ e^{\frac{2\pi i}{90\tau}-\frac{4\pi^2}{\beta}L^{\gamma=2}_h(x)+i\frac{8\pi^3}{\beta^2}Q_h^{\gamma=2}(x)},\label{eq:I_4QtTildehLtTildeh}
\end{equation}
\begin{equation}
\begin{split}
Q_h^{\gamma=2}(x)&=\frac{1}{12}\bigg[\kappa\left(\frac{12}{5}\right)+\kappa\left(\frac{2}{5}\right)-\kappa\left(\frac{4}{5}\right)+\kappa\left(x+\frac{4}{5}\right)+\kappa\left(-x+\frac{4}{5}\right)\\
&\qquad +\kappa\left(x-\frac{2}{5}\right)+\kappa\left(-x-\frac{2}{5}\right)+\kappa\left(2x+\frac{2}{5}\right)+\kappa\left(-2x+\frac{2}{5}\right)
\bigg].
\end{split}
\end{equation}
\begin{equation}
\begin{split}
L_h^{\gamma=2}(x)&=\left(1-\frac{4}{5}\right)\frac{\vartheta(\frac{12}{5})}{2}+\left(1-\frac{2}{15}\right)\frac{\vartheta(\frac{2}{5})}{2}-\left(1-\frac{4}{15}\right)\frac{\vartheta(\frac{4}{5})}{2}\\
&\quad+\left(1-\frac{14}{15}\right)\frac{\vartheta(x+\frac{4}{5})+\vartheta(-x+\frac{4}{5})}{2}+\left(1-\frac{8}{15}\right)\frac{\vartheta(x-\frac{2}{5})+\vartheta(-x-\frac{2}{5})}{2}\\
&\quad+\left(1-\frac{2}{15}\right)\frac{\vartheta(2x+\frac{2}{5})+\vartheta(-2x+\frac{2}{5})}{2}
-\vartheta(2x),
\end{split}
\end{equation}
These functions are plotted in Figures~\ref{fig:QhtTildeADn1} and \ref{fig:LhtTildeADn1}.

Where $L_h^{\gamma=2}$ is minimized, $Q_h^{\gamma=2}$ is not stationary. So we have to asymptotically analyze the integral in \eqref{eq:I_4QtTildehLtTildeh} via complex-analytic tools. We appeal to the saddle-point method, which in the present context is equivalent to the steepest-descent analysis.

\begin{figure}[t]
\centering
\includegraphics[scale =0.7]{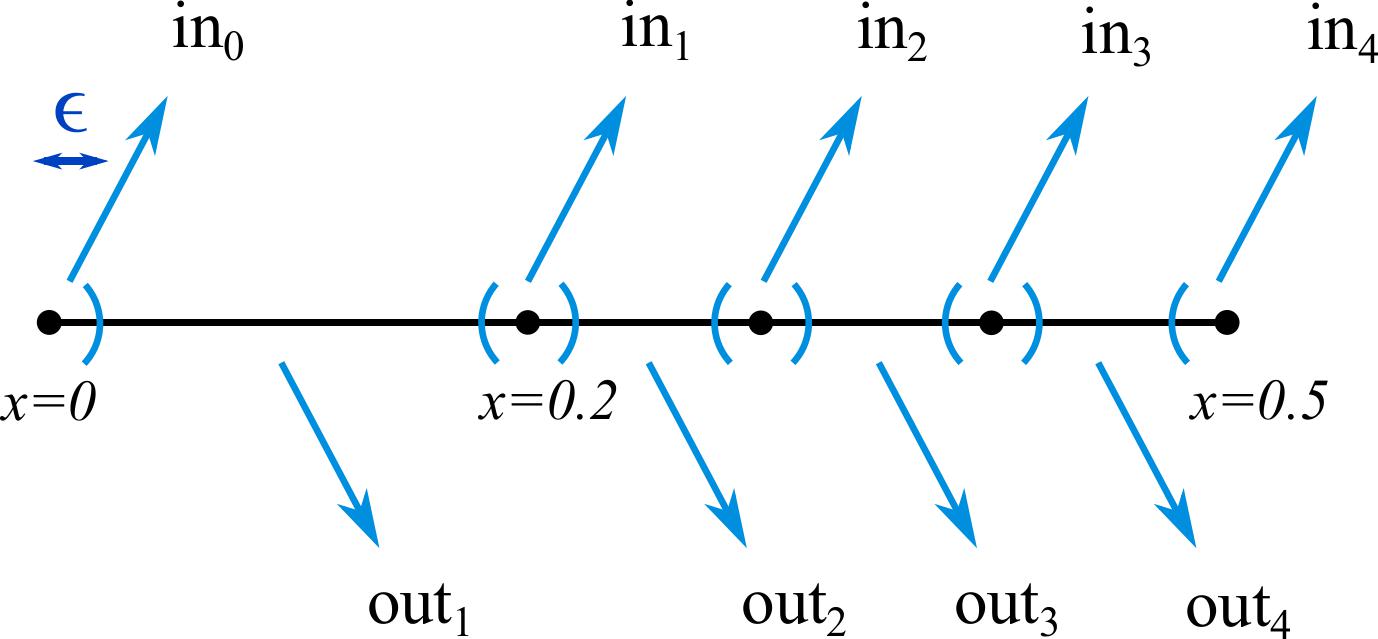}
\caption{Decomposition of the moduli-space of holonomies as in \cite{Ardehali:2021irq}, for the integral \eqref{eq:Itilde_4 for n=1}. The domain $-1/2<x<0$ is neglected due to the $\mathbb Z_2$ Weyl redundancy. The constant $\epsilon$ is taken suitably small, for instance $\epsilon=0.01$}
\label{fig:patchesGamma=2}
\end{figure}

The integrand in \eqref{eq:I_4QtTildehLtTildeh} is piecewise analytic, so we have to decompose the integration domain to sub-intervals (or patches) where the integrand is analytic. See Figure~\ref{fig:patchesGamma=2}. We spell out the details of the computation only for the patch $\frac{2}{10}+\epsilon<x<\frac{3}{10}-\epsilon$, denoted $\mathrm{out}_2$ in Figure~\ref{fig:patchesGamma=2}. The contribution of the other patches will be briefly outlined.

For $.2<x<.3$ we have:
\begin{equation}
    \begin{split}
        Q_h^{\gamma=2}(x)&=\frac{3}{50}(1-5x)^2,\\
    L_h^{\gamma=2}(x)&=\frac{1}{25}(1-5x).\end{split}
\end{equation}
The saddle of the integrand in \eqref{eq:I_4QtTildehLtTildeh} is found to be at
\begin{equation}
    x^\ast_2=\frac{1}{5}+\frac{\tau}{15},\label{eq:x1FibBar}
\end{equation}
where $\tau=i\beta/2\pi$. The corresponding growth of the index is\footnote{Similarly to the 2nd sheet case, an explanation for the fact that $Q^{\gamma=2}_h
(.2) = 0$, which implies the index does not grow as $e^{\#/\beta^2},$ and the fact that $L^{\gamma=2}_h
(.2) = 0$ will be given in Section~\ref{subsec:sense4dBackground}.}
\begin{equation}
    {\mathcal{I}}^{\gamma=2}(q)\longrightarrow e^{\frac{2\pi i}{90\tau}+\mathcal{O}(\beta^0)}.\label{eq:CardyAsy}
\end{equation}

As for the other patches, it turns out that $\mathrm{out}_1$ gives a contribution similar to \eqref{eq:CardyAsy} coming from a saddle on its right end 
\begin{equation}
    x^\ast_1=\frac{1}{5},\label{eq:x0FibBar}
\end{equation}
while $\mathrm{out}_3$ and $\mathrm{out}_4$ give contributions that are exponentially smaller (\emph{i.e.} suppressed as $e^{-\#^{>0}/\beta}$ compared to \eqref{eq:CardyAsy}).\\

The conclusion is that \emph{the 3d effective theory lives on $\mathrm{in}_1$}.
The EFT matter content can be read from \eqref{eq:Itilde_4 for n=1} near $z=e^{2\pi i \frac{1}{5}}$. The only light multiplets are: (1) the photon multiplet; (2) the chiral multiplet with gauge charge $+1$ and $R$-charge $14/15$; (3) the chiral multiplet with gauge charge $-2$ and $R$-charge $2/15$. For the various induced CS couplings we get (see Appendix~\ref{app:effective_CS} for the derivations):
\begin{equation}
    \begin{split}
        k_{gg}=-\frac{3}{2},\qquad
    {k}_{gR}=\frac{11}{10},\qquad k_{RR}=-\frac{13}{450},\qquad k_{\text{grav}}=-\frac{1}{5}.
    \end{split}
\end{equation}

\subsubsection{Mass gap \ \texorpdfstring{$\Longrightarrow$}{ } \ unitary TQFT}

Let us summarize the 3d EFT found in the previous section. It is a 3d $\mathcal{N}=2$ $U(1)_{-3/2}$ gauge theory with two chiral multiplets $\Phi^{r=14/15}_{+1}$ and $\Phi^{r=2/15}_{-2}$. The effective Lagrangian at the uv scale $\Lambda\sim \frac{2\pi\epsilon}{\beta}$ also contains a mixed gauge-$R$ CS coupling ${k}_{gR}=11/10$, as well as $k_{RR}=-13/450$ and $k_{\text{grav}}=-1/5.$ For reasons alluded to in the introduction and spelled out below, we will refer to this theory as the $\overline{\text{Fib}}$ theory.

To determine  the low-energy dynamics, we first note that the theory has monopole operators whose charges can be found from \eqref{eq:c(m)} and \eqref{eq:r(m)}:
\begin{equation}
    \begin{split}
        &V_+:\qquad\text{gauge charge $3$ and $R$-charge }\ -\frac{1}{5}\,,\\
        &V_-:\qquad\text{gauge invariant with $R$-charge }\ 2\,.
    \end{split}
\end{equation}

From the charge assignments above we see that there are two possible gauge-invariant terms of $R$-charge $2$ in the uv:
\begin{equation}
    \Phi^2_{+1} \Phi^{}_{-2} \,,\qquad\, V_-\ .
\end{equation}
Importantly, these are also invariant under the 4d $\mathcal{N}=1$ flavor symmetry $U(1)_f$, which is part of the 4d $\mathcal{N}=2$ $R$-symmetry: invariance of $\Phi^2_{+1} \Phi^{}_{-2}$ follows from the flavor charges $1/10$ and $-1/5$ as seen in \eqref{eq:I_t for n=1}, while invariance of $V_-$ follows from the well-known formula \eqref{eq:f(m)} for the flavor charges of BPS monopoles, as explained in \eqref{eq:f(-1)=0}.
Naturalness therefore implies a superpotential
\begin{equation}
    \mathcal{W}^{}_{\overline{\text{Fib}}}\,\sim\,\Phi^2_{+1} \Phi^{}_{-2} \,+\, V_-\ ,\label{eq:FibBarW}
\end{equation}
with $V_-$ arising via the Affleck-Harvey-Witten type mechanism \cite{Affleck:1982as} in the UV $\mathrm{SU}(2)$ gauge theory on a circle \cite{Seiberg:1996nz,Aharony:1997bx,Aharony:2013dha}. 
The first term in the superpotential prevents a $U(1)$ flavor symmetry in the matter sector, as well as a Higgs branch in the 3d theory, while the second breaks the topological $U(1)_J$ and lifts the Coulomb branch. We thus end up with a theory lacking a moduli-space of vacua, as expected from the $R$-twisted reduction \cite{Dedushenko:2018bpp}.

With the $U(1)_J$ broken in the 3d EFT due to the monopole superpotential, one may wonder what happens to the 4d $\mathcal{N}=1$ flavor symmetry $U(1)_f$. The twisted reduction is not expected to break 4d global symmetries (\emph{cf.}~\cite{Aharony:2013dha}). Since our 3d EFT has no global $U(1)$ symmetry to accommodate the $U(1)_f$, the only remaining possibility is that $U(1)_f$ acts trivially in the dynamical sector of the EFT. With an analysis of the $S^3$ partition function, we argue in Appendix~\ref{app:flavor} that this possibility is indeed realized.

Computing the 3d superconformal index \cite{Bhattacharya:2008zy},
\begin{equation}
    \tilde{I}^{}(q):=\mathrm{Tr}^{}_{S^2}(-1)^R q^{R/2+j_3},\label{eq:3dIndexDef}
\end{equation}
of the theory via the well-known formula \eqref{eq:2ndSheetGenConv} for gauge-theory indices,
we find:
\begin{equation}
    \tilde{I}^{}_{\overline{\text{Fib}}}(q)=1\,.
\end{equation}
This strongly indicates that the 3d EFT is gapped and has a topological vacuum.\\

\subsection*{\texorpdfstring{$S$}{S} and \texorpdfstring{$T$}{T} matrices from handle-gluing and fibering operators}

In order to leverage Bethe root techniques, we mix gauge and $U(1)_R$ to make the $R$-charges of the chiral multiplets integer. For the calculation in the present subsection, we choose the mixing scheme
$$R\text{-charge}\to R\text{-charge}+\frac{1}{15}\times\text{gauge charge},$$
so that now $\Phi_{+1}$ has $R$-charge $1$, and $\Phi_{-2}$ has $R$-charge $0.$ The mixing implies also (see Eq.~\eqref{eq:kgR_mixings_gauge}):
\begin{equation}
    \begin{split}
        k_{gR}&\to k_{gR}+\frac{1}{15}\times k_{gg}=1\,.
    \end{split}
\end{equation}
We thus have:
\begin{equation}
    \overline{\text{\textbf{Fib}}}:\qquad\qquad \mathbf{U(1)_{-3/2}}\ +\ \mathbf{\Phi_{+1}^{r=1}}+\, \mathbf{\Phi_{-2}^{r=0}}\ \ \text{and}\ \ \mathbf{k_{gR}=1}\qquad(k_{gg}^+=1,\ k_{gR}^+=2)\,,\label{eq:fibBarData}
\end{equation}
with the superpotential as in \eqref{eq:FibBarW}.

The twisted superpotential and the effective dilaton are found from the general 3d $\mathcal{N}=2$ gauge-theory formulae in \eqref{eq:WandOmega} to be:
\begin{equation}
    \begin{split}
        W(Z)&=\pi i Z + \frac{1}{2}Z^2+\mathrm{Li}_2\big(e^{Z}\big)+\mathrm{Li}_2\big(e^{-2Z}\big)\,,\\
        \Omega(Z)&=2Z+\log\big(1-e^{-2Z}\big)\,.
    \end{split}
\end{equation}
The Bethe equation $\exp\big(W'(Z_\alpha^\ast)\big)=1$ reads in terms of the charge $1$ Wilson line variable $ z:=e^{Z}$ as
\begin{equation}
     z^2=1+ z\,.\label{eq:FibBarFusion}
\end{equation}
The 3d $\mathcal{N}=2$ formula \eqref{eq:HandF_vs_SandT} gives the $S$ and $T^2$ matrices (up to an overall phase for $T^2$) as:
\begin{equation}
\begin{split}
    S&=\left(
\begin{array}{cc}
 \sqrt{\frac{2}{\sqrt{5}+5}} & \ \  \sqrt{\frac{2}{5-\sqrt{5}}} \\
 \sqrt{\frac{2}{5-\sqrt{5}}} & \ \ -\sqrt{\frac{2}{\sqrt{5}+5}} \\
\end{array}
\right) = 
\frac{1}{\sqrt{2+\varphi }}
\begin{pmatrix}
   1 &  \varphi\\
   \varphi & -1
\end{pmatrix},\\
    T^2&=\left(
\begin{array}{cc}
 1 & \ \  0 \\
 0 & \ \ e^{2\pi i(\frac{6}{5})} \\
\end{array}
\right).
    \end{split}\label{eq:S&T_Fibbar}
\end{equation}
The sign ambiguities in the entries of $S$ are removed as explained below \eqref{eq:S&T_GY}, though this time leveraging our knowledge that we are in a gapped unitary setting, where the positive row/column corresponds to the vacuum. Also, we are considering $T^2$ instead of $T$ since as explained below we conjecture that the bulk TQFT is a spin-TQFT.

As a check, the 3d $\mathcal{N}=2$ formula \eqref{eq:Z_surgery} for the $S^3$ partition function gives:
\begin{equation}
    |Z^{}_{S^3}|=\frac{1}{\sqrt{2+\varphi}}\,,
\end{equation}
matching $S_{00}$, as it should. These indeed correspond, up to the overall phase of $T^2$, to the $\overline{\text{Fib}}$ (or Rep$(F_4)^{}_1$) modular data \cite{Harvey:2018rdc}.

\subsection*{Boundary VOA from the half-index}

We now present Dirichlet half-index calculations yielding characters of
\begin{equation}
    \mathrm{SM}(3,5)\times \mathrm{SO}(1)_1\,,
\end{equation}
where $\mathrm{SM}(3,5)$ is an $\mathcal{N}=1$ supersymmetric minimal model \cite{Friedan:1984rv}, also known as the fermionized tricritical Ising model $M(4,5)/\mathbb Z_2^f$, and $\mathrm{SO}(1)_1$ is a free Majorana fermion.

By the RCFT/TQFT correspondence, $\mathrm{SM}(3,5)$ yields a spin-TQFT, which is equivalent to $(F_4)^{}_1 = \overline{\rm Fib}$, up to an invertible spin-TQFT factor, and $\mathrm{SO}(1)_1$ is itself an invertible spin-TQFT. See Section~\ref{subsec:sense2dBCs} for more on the relation between the TQFTs as well as the VOAs.

The calculation is done with the $\mathcal{N}=(0,2)$ Dirichlet boundary conditions on the gauge multiplet and on $\Phi_{+1}$, and with modified Dirichlet (or D$_c$) on $\Phi_{-2}$,\footnote{One may be tempted to impose D$_c$ on $\Phi_{+1}$ and D on $\Phi_{-2}$. The scalar potential $|\phi^2_{+1}|^2+|\phi_{+1}\phi_{-2}|^2$ following from $\Phi^2_{+1}\Phi_{-2}\in\mathcal{W}_{\,\overline{\text{Fib}}}$ implies that D$_c$ on $\Phi_{+1}$ breaks supersymmetry. We have checked that the half-index with such boundary condition vanishes, consistent with supersymmetry breaking.} on the right boundary. The boundary gauge anomaly is:
\begin{equation}
    \underbrace{\frac{3}{2}\,\mathbf{f}^2}_{-k_{gg}}-\underbrace{2\,\mathbf{f}\cdot\mathbf{r}}_{-k_{gR}}+\underbrace{\frac{1}{2}\,\mathbf{f}^2}_{\Phi_{+1}}+\underbrace{\frac{1}{2}\,(-2\,\mathbf{f}-\mathbf{r})^2}_{\Phi_{-2}}=4\,\mathbf{f}^2\,.
\end{equation}
So we can use the Dirichlet half-index formula \eqref{eq:gen-half-index-D} with $k_{gg}^+\to4$ and $k_{gR}^+\to0:$
\begin{equation}
    I\!\!I^{R}_{\mathcal{D},D,D}(q)=\frac{1}{(q;q)}\sum_{m\in\mathbb{Z}}q^{2m^2} z^{4m}\, \big(- z^{-1}q^{1/2-m};q\big)\,\big( z^{2}q^{1+2m};q\big).\label{eq:Fibo3}
\end{equation}
Sending the gauge fugacity $ z\to1$ due to the D$_c$ condition on $\Phi_{-2}$ breaking the boundary global $U(1)$ descending from the bulk gauge symmetry, we get: 
\begin{equation}
    I\!\!I^{R}_{\mathcal{D},D,D_c}(q)=\frac{1}{(q;q)}\sum_{m\in\mathbb{Z}}q^{2m^2}\, \big(-q^{1/2-m};q\big)\,\big(q^{1+2m};q\big)=:\chi_0(q).\label{eq:Fibo4}
\end{equation}
This is a fermionic character, and we would like to determine the corresponding VOA.

The non-vacuum character can be obtained by inserting a Wilson line of gauge charge $1$:
\begin{equation}
    \chi_1(q):=\frac{1}{(q;q)}\sum_{m\in\mathbb{Z}}q^{2m^2-m}\, \big(-q^{1/2-m};q\big)\,\big(q^{1+2m};q\big).\label{eq:FiboChi1}
\end{equation}
Introduction of $q^{-m}$ inside the summand, instead of $q^m$ as prescribed in \cite{Dimofte:2017tpi}, is because we are considering the right boundary.

Requiring modular covariance of the two characters
\begin{equation}
    \begin{pmatrix}
        \tilde q^{-c/24}\chi_0(\tilde q)\\
        \tilde q^{-c/24+h}\chi_1(\tilde q)
    \end{pmatrix}=\left(
\begin{array}{cc}
 \sqrt{\frac{2}{\sqrt{5}+5}} & \ \  \sqrt{\frac{2}{5-\sqrt{5}}} \\
 \sqrt{\frac{2}{5-\sqrt{5}}} & \ \ -\sqrt{\frac{2}{\sqrt{5}+5}} \\
\end{array}
\right) \cdot \begin{pmatrix}
         q^{-c/24}\chi_0( q)\\
         q^{-c/24+h}\chi_1( q)
    \end{pmatrix},
\end{equation}
where $\tilde q:=e^{-2\pi i/\tau}$, we find that
\begin{equation}
    c=\frac{6}{5},\label{eq:spinTQFTc}
\end{equation}
and $h=1/10$. 

Guided by these data, we recognize that the two characters are actually the NS characters of the following fermionic RCFT:
$$
\text{the FRCFT} = \mathrm{SM}(3,5)\otimes(\text{free fermion RCFT}).$$
More precisely, we have
\begin{equation}
    \begin{split}
        q^{-\frac{c}{24}}\chi_0(q)&=\big(\chi^{}_{(s,t)=(1,1)}(q)\text{ of }\mathrm{SM}(3,5)\big)\times \chi^{}_F(q),\\
        q^{-\frac{c}{24}+h}\chi_1(q)&=\big(\chi^{}_{(s,t)=(1,3)}(q)\text{ of }\mathrm{SM}(3,5)\big)\times \chi^{}_F(q),
    \end{split}
\end{equation}
where $c=6/5= c(3,5)+1/2$ and $h=1/10=h_{(1,3)}^{(3,5)}.$ The characters can be obtained as a special case of the general $\mathrm{SM}(p,p')$ NS character formula \cite{feigin1982invariant,Goddard:1986ee,meurman1986highest}:
\begin{equation}
    \chi^{}_{(s,t)}=q^{h_{(s,t)}^{p,p'}-\frac{c(p,p')}{24}}\frac{(-q^{\frac{1}{2}};q)}{(q;q)}\sum_{l\in\mathbb Z}\left(q^{\frac{l(lpp'+sp'-tp)}{2}}-q^{\frac{(lp+s)(lp'+t)}{2}}\right),
\end{equation}
with
\begin{equation}
    c(p,p')=\frac{3}{2}\left(1-\frac{2(p'-p)^2}{pp'}\right),\qquad h^{(p,p')}_{(s,t)}=\frac{(p's-pt)^2-(p-p')^2}{8pp'}.
\end{equation}
The conformal primaries $O_{(s,t)}$ are labeled by two integers $1\le s\le p$ and $1\le t\le p'$, with an equivalence relation $O_{(s,t)}=O_{(p-s,p'-t)}$. The NS primaries are $O_{(s,t)}$ with $s-t\in2\mathbb Z$. See \cite{Baek:2024tuo} for other recent examples of supersymmetric minimal models arising from 3d TQFTs.

By ``free fermion RCFT'' we mean the free Majorana fermion theory with $c=1/2$, whose NS character reads
\begin{equation}
    \chi^{}_F=q^{-\frac{1}{48}}\prod_{n=0}^\infty(1+q^{n+1/2}).
\end{equation}

\subsubsection*{Bosonic VOA on the opposite boundary}

The left enriched Neumann boundary conditions support ${\rm Fib}\cong(G_2)^{}_1$, as we now demonstrate.

The $\mathcal{N}=(0,2)$ Neumann boundary conditons on all the multiplets induce the boundary gauge anomaly:
\begin{equation}
    \underbrace{-\frac{3}{2}\,\mathbf{f}^2}_{k_{gg}}+\underbrace{2\,\mathbf{f}\cdot\mathbf{r}}_{k_{gR}}-\underbrace{\frac{1}{2}\,\mathbf{f}^2}_{\Phi_{+1}}-\underbrace{\frac{1}{2}\,(-2\,\mathbf{f}-\mathbf{r})^2}_{\Phi_{-2}}=-4\,\mathbf{f}^2\,.
\end{equation}
This can be cancelled by adding four boundary fermi multiplets of gauge charge $-1$ (or $1$) and $R$-charge $0$. The Neumann half-index formula in Appendix~\ref{app:half-index} then gives:
\begin{equation}
    I\!\!I^L_{\mathcal{N},N,N}=(q,q) \oint \frac{\mathrm{d}z}{2 \pi i z} \frac{\theta_0(-q^{1/2}z;q)^4}{(-q^{1/2}z;q)(z^{-2};q)} = 1 + 14 q + 42 q^2 + 140q^3 + \dots,
\end{equation}
matching the $(G_2)^{}_1$ vacuum character. See \emph{e.g.}~Table~1 in \cite{Mukhi:2022bte}.  In evaluating the integral we have excluded the pole at $z=1$ via an $\varepsilon$-prescription $z\to z\,q^{\,\varepsilon<0}$. Alternatively, one can add some multiple of gauge charge to the $R$-charge of the bulk chiral multiplets to bring them inside the safe interval $0<r_\chi<2$.

The non-vacuum character can be obtained by inserting a Wilson line of gauge charge $+1$:
\begin{equation}
    (q,q) \oint z\,\frac{ \mathrm{d}z}{2 \pi i z} \frac{\theta_0(-q^{1/2}z;q)^4}{(-q^{1/2}z;q)(z^{-2};q)} = q^{\frac{1}{2}}\big(7 + 34 q + 119 q^2 + 322q^3 + \dots\big)\,.
\end{equation}
This matches the non-vacuum character of $(G_2)^{}_1$, up to the overall $q^{1/2}$ factor. Explaining this factor takes two steps: The Wilson line, in the presence of gauge CS level $k_{gg}^+=1$, supports a magnetic flux $1$, which then, through the mixed CS level $k^+_{gR}=2$ in \eqref{eq:fibBarData}, generates an $R$-symmetry Wilson line. The latter contributes since the half-index background includes the $R$-symmetry holonomy $q^{1/2}$ around the $S^1$. Thus the line that actually realizes the non-vacuum module is a gauge Wilson line combined with the $R$-symmetry Wilson line canceling $q^{1/2}$ (as in \cite{Ferrari:2023fez}).

\subsubsection*{Discussion}
We found that the (left) enriched Neumann boundary supports $(G_2)_1$ VOA, while the (right) Dirichlet boundary supports $\mathrm{SM}(3,5)\otimes \mathrm{SO}(1)_1$. Since the former is ${\rm Fib}$ and the latter is $\overline{\rm Fib}\otimes (\text{invertible spin-TQFT})$, this is quite natural at the level of MTC, which does not see the invertible factor. If we want to precisely identify the bulk TQFT, the invertible factor matters. Which of the two (if any) captures the bulk TQFT?

We conjecture that
\begin{equation}
\label{conj_spinTQFT}
    \boxed{\text{bulk TQFT} \cong \mathrm{SM}(3,5)\otimes \mathrm{SO}(1)_1}
\end{equation}
There are a few reasons to believe in this conjecture. First note, at the most naive level, that a uv 3d $\mathcal{N}=2$ theory depends on the spin structure, so the spin-TQFT $\mathrm{SM}(3,5)\otimes \mathrm{SO}(1)_1$ is, generically, more natural than the bosonic TQFT $(G_2)_1$. More seriously, the structures we get are consistent with such a conjecture. In the known examples, the Dirichlet boundary captures the bulk TQFT. For example, consider 3d $\mathcal{N}=2$ $G_k$ super Yang-Mills with $k>h^\vee$, which is gapped. The $\mathcal{N}=(0,2)$ boundary is known to support the affine VOA $L_{k-h^\vee}(\mathfrak{g})$ \cite{Costello:2020ndc}, capturing the IR TQFT given by the $G_{k-h^\vee}$ bosonic Chern-Simons theory. At the same time, the Neumann boundary with $k-h^\vee$ fundamental Fermi multiplets (say in the $G=SU(N)$ case) would support some fermionic VOA $V$. By putting the theory on the interval, with the $L_{k-h^\vee}(\mathfrak{g})$-carrying Dirichlet b.c. on the one side and the $V$-carrying enriched Neumann on the opposite side, we, on the one hand, obviously get free fermion VOA $F={\rm Ferm}^{\otimes \#}$ in the IR. On the other hand, $L_{k-h^\vee}(\mathfrak{g})$ and $V$ embed into $F$ as commutants of each other, $V = F / L_{k-h^\vee}(\mathfrak{g})$ and $L_{k-h^\vee}(\mathfrak{g})=F/V$. One can in fact view $V = F / L_{k-h^\vee}(\mathfrak{g})$ as a gauging prescription, defining how the bulk TQFT (corresponding to $L_{k-h^\vee}(\mathfrak{g})$) gauges the boundary Fermi multiplets $F$, resulting in the boundary VOA $V$. We find similar structures in our case, after putting our theory on the interval with $\mathrm{SM}(3,5)\otimes \mathrm{SO}(1)_1$ on the right and $(G_2)_1$ on the left. Due to the boundary Fermi multiplets,  $\mathrm{SM}(3,5)\otimes \mathrm{SO}(1)_1$ and $(G_2)_1$ are commutants of each other in ${\rm Ferm}^{\otimes 4}$, which will be discussed more in Section \ref{subsec:sense2dBCs}. 
We could try to consider different boundary conditions on the chiral multiplets, but we would like to ensure that the boundary $U(1)$ symmetry is broken, since our 3d theory has no flavor symmetries. As we explained, using D$_c$ for $\Phi_{+1}$ is not an option as it breaks SUSY, so we had to break $U(1)$ by imposing D$_c$ on $\Phi_{-2}$. We impose D on $\Phi_{+1}$, and replacing it by Neumann would not be a good idea, as the superpotential would evaluate to $\Phi_{+1}^2 c$ at such a boundary, which breaks SUSY without introduction of additional boundary Fermi multiplets and superpotentials \cite{Dimofte:2017tpi}. The latter would, however, add extra boundary degrees of freedom not captured by the bulk. Overall, the $(\mathcal{D}, {\rm D}, {\rm D}_c)$ boundary conditions we use seem like the best choice. Clearly, it would be interesting to study the 3d IR physics of our theory in more detail and verify the conjecture \eqref{conj_spinTQFT} more convincingly.

\subsection{Fourth sheet: Fibonacci}\label{subsec:Fib}

Without repeating all the derivation steps, we emphasize that the fourth sheet theory is quite similar to the one on the third sheet, but with the following modifications:
\begin{equation}
    {\text{\textbf{Fib}}}:\qquad\qquad \mathbf{U(1)_{3/2}}\ +\ \mathbf{\Phi_{-1}^{r=1}}+\, \mathbf{\Phi_{+2}^{r=0}}\ \ \text{and}\ \ \mathbf{k_{gR}=1}\qquad(k_{gg}^+=4,\ k_{gR}^+=0)\,.\label{eq:FibData}
\end{equation}
The superpotential is
\begin{equation}
    \mathcal{W}^{}_{{\text{Fib}}}\,\sim\,\Phi^2_{-1} \Phi^{}_{+2} \,+\, V_-\ .
\end{equation}

The effective twisted superpotential and the effective dilaton are found from \eqref{eq:WandOmega}:
\begin{equation}
    \begin{split}
        W(Z)&=4\pi i Z + 2Z^2+\mathrm{Li}_2\big(e^{-Z}\big)+\mathrm{Li}_2\big(e^{2Z}\big)\,,\\
        \Omega(Z)&=\log\big(1-e^{2Z}\big)\,.
    \end{split}
\end{equation}
The Bethe equation $\exp\big(W'(Z_\alpha^\ast)\big)=1$ reads in terms of the charge $1$ Wilson line variable $ z:=e^{Z}$ as
\begin{equation}
     z^2=1+ z\,.\label{eq:FibFusion}
\end{equation}
Equation \eqref{eq:HandF_vs_SandT} gives the $S$ and $T^2$ matrices (up to an overall phase for $T^2$) as:
\begin{equation}
\begin{split}
    S&=\left(
\begin{array}{cc}
 \sqrt{\frac{2}{\sqrt{5}+5}} & \ \  \sqrt{\frac{2}{5-\sqrt{5}}} \\
 \sqrt{\frac{2}{5-\sqrt{5}}} & \ \ -\sqrt{\frac{2}{\sqrt{5}+5}} \\
\end{array}
\right) = 
\frac{1}{\sqrt{2+\varphi }}
\begin{pmatrix}
   1 &  \varphi\\
   \varphi & -1
\end{pmatrix},\\
    T^2&=\left(
\begin{array}{cc}
 1 & \ \  0 \\
 0 & \ \ e^{2\pi i(\frac{4}{5})} \\
\end{array}
\right),
    \end{split}
\end{equation}
with the sign ambiguities in the entries of $S$ fixed as explained below \eqref{eq:S&T_Fibbar}. Again we are considering $T^2$ instead
of $T$ since we conjecture that the bulk TQFT is a spin-TQFT. Eq.~\eqref{eq:Z_surgery} gives for the topological $S^3$ partition function:
\begin{equation}
    |Z^{}_{S^3}|=\frac{1}{\sqrt{2+\varphi}}\,,
\end{equation}
matching $S_{00}$ as it should. These indeed correspond, up to the overall phase of $T^2$, to the ${\text{Fib}}$ (or Rep$(G_2)^{}_1$) modular data \cite{Harvey:2018rdc}.

\subsection*{Boundary VOA from the half-index}

We first reproduce Fib (or $(G_2)^{}_1$) characters on the {right boundary}. The anomaly with $(\mathcal{N},N,N)$ boundary conditions on the right boundary is:
\begin{equation}
    -k_{gg}^+\,\mathbf{f}^2-2k_{gR}^+\,\mathbf{f}\cdot\mathbf{r}=-4\mathbf{f}^2\,.
\end{equation}
This can be cancelled by adding four boundary fermi multiplets of gauge charge $-1$ (or $1$) and $R$-charge $0$. The half-index can be found as in Appendix~\ref{app:half-index} to be:
\begin{equation}
    I\!\!I^R_{\mathcal{N},N,N}=(q,q) \oint \frac{\mathrm{d}z}{2 \pi i z} \frac{\theta_0(-q^{1/2}z;q)^4}{(-q^{1/2}z;q)(z^{-2};q)} = 1 + 14 q + 42 q^2 + 140q^3 + \dots,
\end{equation}
matching the $(G_2)^{}_1$ vacuum character.

The non-vacuum character can be obtained by considering a Wilson line of gauge charge $-1$ (inserting $z$ instead of $z^{-1}$ as prescribed in \cite{Dimofte:2017tpi} since we are considering the right boundary):
\begin{equation}
    (q,q) \oint z\,\frac{ \mathrm{d}z}{2 \pi i z} \frac{\theta_0(-q^{1/2}z;q)^4}{(-q^{1/2}z;q)(z^{-2};q)} = q^{\frac{1}{2}}\big(7 + 34 q + 119 q^2 + 322q^3 + \dots\big)\,.
\end{equation}
This matches the non-vacuum character of $(G_2)^{}_1$, again up to the overall $q^{1/2}$ factor corresponding to the $R$-symmetry Wilson loop induced by the Chern-Simons couplings.

\subsubsection*{Fermionic VOA on the opposite boundary}

On the left boundary, using the general formula \eqref{eq:gen-half-index-D} for the 3d half-index with Dirichlet boundary conditions on all fields we get:
\begin{equation}
    I\!\!I^{L}_{\mathcal{D},D,D}(q)=\frac{1}{(q;q)}\sum_{m\in\mathbb{Z}}q^{2m^2} z^{4m}\, \big(- z^{-1}q^{1/2-m};q\big)\,\big( z^{2}q^{1+2m};q\big).\label{eq:Fibo3}
\end{equation}
Sending $ z\to1$ due to the D$_c$ condition on $\Phi_{-2}$ breaking the boundary global $U(1)$ descending from the bulk gauge symmetry, we get: 
\begin{equation}
    I\!\!I^{L}_{\mathcal{D},D,D_c}(q)=\frac{1}{(q;q)}\sum_{m\in\mathbb{Z}}q^{2m^2}\, \big(-q^{1/2-m};q\big)\,\big(q^{1+2m};q\big)=:\chi_0(q).\label{eq:Fibo4}
\end{equation}
This is the vacuum character of $\mathrm{SM}(3,5)\times\mathrm{Majorana}$.

The non-vacuum character can be obtained by inserting a Wilson line of gauge charge $-1$:
\begin{equation}
    \chi_1(q):=\frac{1}{(q;q)}\sum_{m\in\mathbb{Z}}q^{2m^2-m}\, \big(-q^{1/2-m};q\big)\,\big(q^{1+2m};q\big),\label{eq:FiboChi1}
\end{equation}
giving the non-vacuum character of $\mathrm{SM}(3,5)\otimes\mathrm{Majorana}.$

\subsubsection*{Discussion}

We found the same VOAs as on the third sheet, but on the opposite boundaries. Via the same reasoning, we now conjecture that the bulk TQFT is captured by the conjugate of $\mathrm{SM}(3,5)\otimes\mathrm{Majorana}.$

\subsection{Fifth sheet: conjugate Lee-Yang}\label{subsec:5thSheet}

Again, without repeating the derivation steps, we emphasize that the fifth sheet theory is very similar to the one on the second sheet, but with the following modifications:
\begin{equation}
    \overline{\text{\textbf{LY}}}:\qquad\qquad \mathbf{U(1)_{3/2}}\ +\ \mathbf{\Phi_{-1}^{r=1}}\ \ \text{and}\ \ \mathbf{k_{gR}=0}\qquad(k_{gg}^+=2,\ k_{gR}^+=0)\,.
\end{equation}
This is the uv data of the fifth-sheet theory appropriate for the A-twist. We have checked that the same TQFT (up to an invertible factor) arises from the B-twist of the 2nd-sheet theory. As in Gang-Yamazaki, there is no 3d superpotential.

The twisted superpotential and the effective dilaton are found from \eqref{eq:WandOmega} to be:
\begin{equation}
    \begin{split}
        W(Z)&=2\pi i Z +Z^2+\mathrm{Li}_2\big(e^{-Z}\big)\,,\\
        \Omega(Z)&=0\,.
    \end{split}
\end{equation}
The Bethe equation $\exp\big(W'(Z_\alpha^\ast)\big)=1$ reads in terms of the charge $1$ Wilson line variable $ z:=e^{Z}$ as
\begin{equation}
     z^2=1+ z\,.\label{eq:LYbarFusion}
\end{equation}
Equation \eqref{eq:HandF_vs_SandT}  gives the $S$ and $T$ matrices (up to an overall phase for $T$) as:
\begin{equation}
\begin{split}
    S&=\left(
\begin{array}{cc}
 -\sqrt{\frac{2}{5-\sqrt{5}}} & \ \  \sqrt{\frac{2}{5+\sqrt{5}}} \\
 \sqrt{\frac{2}{5+\sqrt{5}}} & \ \ \sqrt{\frac{2}{5-\sqrt{5}}} \\
\end{array}
\right) = 
\frac{1}{\sqrt{2+\varphi }}
\begin{pmatrix}
   -\varphi &  1\\
   1 & \varphi
\end{pmatrix},\\
    T&=\left(
\begin{array}{cc}
 1 & \ \  0 \\
 0 & \ \ e^{2\pi i(-\frac{4}{5})} \\
\end{array}
\right),
    \end{split}
\end{equation}
where the sign ambiguities in the entries of $S$ are fixed as explained below \eqref{eq:S&T_GY}. 
Eq.~\eqref{eq:Z_surgery} gives the topological $S^3$ partition function:
\begin{equation}
    |Z^{}_{S^3}|=\frac{\varphi}{\sqrt{2+\varphi}}\,,
\end{equation}
which matches $S_{00}$, as it should. These indeed correspond, up to the overall phase of $T$, to the $\overline{\text{LY}}$ modular data \cite{Harvey:2018rdc}.

We note in passing that the Fibonacci-type fusion rule $\phi\times\phi\sim I+\phi$ has another MTC solution corresponding to the $\mathrm{Rep}(E_{7\frac{1}{2}})^{}_1$ intermediate vertex operator algebra \cite{kawasetsu2014intermediate,Harvey:2019qzs}. It would be interesting to realize the corresponding characters via suitable boundary conditions in one of the TQFTs encountered in this section.

\section{\texorpdfstring{$(A_1,A_{2n})$}{A1A4} with \texorpdfstring{$n\geqslant2$}{nge2}}\label{sec:A1A4}

For the $(A_1,A_{2n})$ theory, we again use the $\mathcal{N}=1$ Lagrangian of \cite{Maruyoshi:2016aim}, quoting only the $\mathcal{N}=2$ index (setting $z_j=e^{2\pi i x_j}$):
\begin{equation}
\begin{split}
\mathcal{I}_{t}^{(A_1, A_{2n})}(p,q,t) &=\big((p;p)(q;q)\big)^n\, \left[ \prod_{i=1}^n \frac{ \Gamma_e \left( (\frac{pq}{t})^{\alpha_i} \right)}{ \Gamma_e \left( (\frac{pq}{t})^{\beta_i} \right)}\right] \Gamma_e\left( (\frac{pq}{t})^{\frac{1}{2n+3}} \right)^n  \\
  &\int \frac{\mathrm{d}^n x}{2^n n!}
  \left[\prod_{i=1 }^n \Gamma_e \left(z_i^{\pm1} (\frac{pq}{t})^{\frac{n+1}{2n+3}} t^{\frac{1}{2}} \right) \Gamma_e \left(z_i^{\pm1} (\frac{pq}{t})^{\frac{-n}{2n+3}} t^{\frac{1}{2}} \right)\right]\\
  &\qquad     \left[\prod_{i=1}^n \frac{\Gamma_e \left({z_i}^{\pm2} (\frac{pq}{t})^{\frac{1}{2n+3}} \right)}{\Gamma_e({z_i}^{\pm2})}\right]\left[\prod_{1\le i<j\le n} \frac{\Gamma_e \left(z_i^{\pm1}z_j^{\pm1} (\frac{pq}{t})^{\frac{1}{2n+3}} \right)}{\Gamma_e(z_i^{\pm1}z_j^{\pm1})}\right],
  \end{split}\label{eq:A1A2nIndex}
\end{equation}
with the integral over $-\frac{1}{2}<x_j<\frac{1}{2}$, while $\alpha_i:=\frac{2(n+i+1)}{2n+3}$ and $\beta_i:=\frac{2i}{2n+3}.$

From the exponents in the arguments of the gamma functions in \eqref{eq:A1A2nIndex}, or from the lowest common denominator of $r$-charges being $2n+3,$ we see that there are $2n+3$ inequivalent sheets.
For $(A_1,A_4)$ we get seven sheets. Discarding the trivial sheet corresponding to $\gamma=0$ (which is well-understood \cite{Benvenuti:2018bav}), we end up with six sheets, or three up to conjugation. Below we will study the three sheets corresponding to $\gamma=1,2,3$. The conjugate sheets arising for $\gamma=6,5,4$ can be studied similarly.

Our main focus will in fact be on the second sheet, $\gamma=1$, where we will make contact with the $\mathcal{T}_2$ theory of Gang-Kim-Stubbs \cite{Gang:2023rei}. The third and fourth sheets of $(A_1,A_4)$ will be discussed briefly.

\subsection{Second sheet: SUSY enhancement with AHW superpotentials}

\begin{figure}[h]
\centering
\includegraphics[scale =0.5]{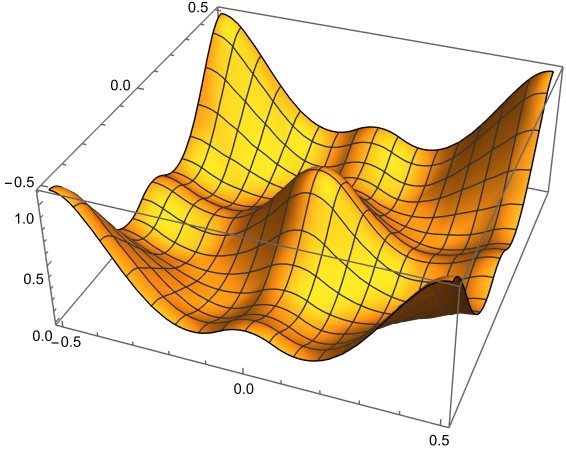}
\caption{The plot of $12Q^{\gamma=1}_h(x_1,x_2)$ versus $(x_1,x_2)$ for $(A_1,A_4)$.}
\label{fig:A1A4Q2nd}
\end{figure}

\begin{figure}[h]
\centering
\includegraphics[scale =0.5]{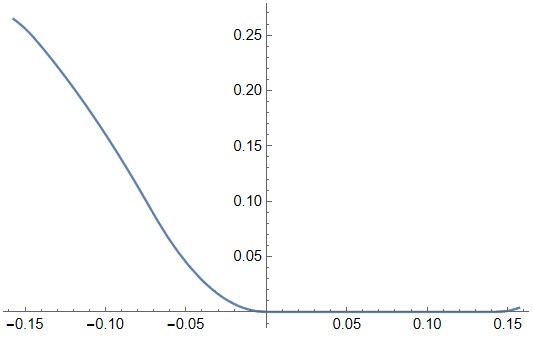}
\caption{The plot of $12Q^{\gamma=1}_h(\frac{1}{7}+x,\frac{2}{7}-x)$ versus $x$ for $(A_1,A_4)$. The minimum is exactly zero, and the flat direction to the right (\emph{i.e.} increasing $x$) signals a gauge-invariant monopole.}
\label{fig:A1A4Q2ndFlat}
\end{figure}

There are two gauge holonomies $x_1,x_2$ for $(A_1,A_4)$. The associated function $Q_h^{\gamma=1}$ is depicted in Figure~\ref{fig:A1A4Q2nd}. Although somewhat invisible to the naked eye, it has a flat direction around each minimum, as can be seen more clearly in Figure~\ref{fig:A1A4Q2ndFlat}. The flat direction signals a gauge-invariant monopole $V^\text{out}_{1,-1}$ in the corresponding outer patch, where the subscript indicates the magnetic charges of the monopole.

\begin{figure}[h]
\centering
\includegraphics[scale =0.5]{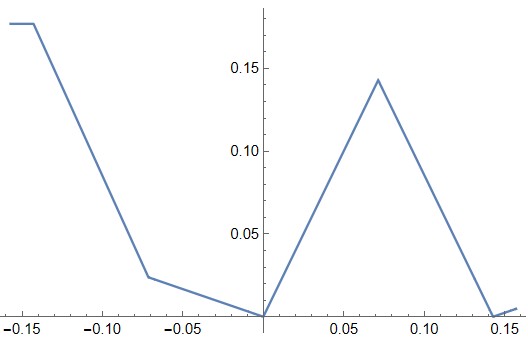}
\caption{The plot of $L^{\gamma=1}_h(\frac{1}{7}+x,\frac{2}{7}-x)$ versus $x$ for $(A_1,A_4)$. The minimum at $x=0$ is zero, and the slope to the right (\emph{i.e.} increasing $x$) being $2$ signals a monopole superpotential.}
\label{fig:A1A4L2ndFlat}
\end{figure}

The associated function $L_h^{\gamma=1}$ evaluated along the flat direction of $Q_h^{\gamma=1}$ is depicted in Figure~\ref{fig:A1A4L2ndFlat}. It shows that there is a holonomy saddle at $(x_1,x_2)=(1/7,2/7)$, and determines the dominant inner patch.\footnote{There are Weyl images of the saddle, one of them visible at $x=1/7$ in Figure~\ref{fig:A1A4L2ndFlat}. Since it is enough to consider one member of the Weyl orbit, we discard the rest.} The slope of $L_h^{\gamma=1}$ being $2$ along the flat direction of $Q_h^{\gamma=1}$ implies that the gauge-invariant monopole $V^\text{out}_{1,-1}$ has $R$-charge $2.$

Having found the holonomy saddle at $(x_1,x_2)=(1/7,2/7)$, the inner-patch EFT field content can be read easily from the index. We have a 3d $\mathcal{N}=2$ $U(1)\times U(1)$ gauge theory with two light chiral multiplets $\Phi_{1,-1}$ and $\Phi_{0,1}$. The gauge charges are indicated as subscripts, and the $R$-charges are $2/21$ and $10/21$, respectively.

The EFT couplings can be computed using the formulas in Section~\ref{sec:Cardy}. The matrix of gauge-gauge CS couplings comes out
\begin{equation}
    k_{ij}=-\begin{pmatrix}
        \frac{3}{2}&\frac{1}{2}\\
        \frac{1}{2}&1
    \end{pmatrix},\label{eq:kij_T2}
\end{equation}
while the mixed gauge-$R$ CS couplings read $k_{1R}=-11/42$ and $k_{2R}=4/7.$

An instructive consistency check of the inner-patch EFT data is to match the quantum numbers it yields for $V^{\text{in}}_{1,-1}$ with those of $V^{\text{out}}_{1,-1}$, which is the gauge-invariant monopole of $R$-charge $2$ identified from the plots above. The interested reader can compute the gauge and $R$ charges of $V^{\text{in}}_{1,-1}$ via the EFT data using \eqref{eq:c(m)} and \eqref{eq:r(m)}. 

We have checked via the formulas \eqref{eq:k_jf} and \eqref{eq:f(m)} that $V:=V^{\text{in}}_{1,-1}$ is also invariant under the 4d $\mathcal{N}=1$ flavor symmetry. Since $V$ is gauge-invariant, has $R$-charge $2,$ and is compatible with the flavor symmetry, it should be included in the superpotential:
\begin{equation}
    \mathcal{W}\,\sim\, V\,.
\end{equation}
This concludes our determination of the 3d EFT.\\

Let us now compare with the $\mathcal{T}_2$ theory of Gang-Kim-Stubbs~\cite{Gang:2023rei}. We only need to change the gauge variables and instead of the first $U(1)$ factor of the gauge group, work with the difference of the two factors $\sigma_1-\sigma_2=:y_1\,.$ Comparison of the CS quadratic forms (appearing, for instance, in the $S^3$ partition function):
\begin{equation}
    (\sigma_1,\sigma_2)\cdot k_{ij} \cdot(\sigma_1,\sigma_2)^T\longleftrightarrow (y_1,y_2)\cdot K_{ij} \cdot(y_1,y_2)^T\!,
\end{equation}
by setting $\sigma_1=y_1+y_2$ and $\sigma_2=y_2 $, shows that
\begin{equation}
    K_{ij}=-\begin{pmatrix}
        3/2&2\\
        2&7/2
    \end{pmatrix}.
\end{equation}
To find the fluxes of the monopole $V$ in the new gauge coordinates, we use the fact that the fluxes are valued in the co-character lattice to write:
\begin{equation}
    m^\text{new}_i=\sum_j\frac{\partial y_i}{\partial\sigma_j}\, m_j^\text{old}\,.\label{eq:monopoleTransformation}
\end{equation}
This gives
\begin{equation}
    \begin{pmatrix}
        m_1\\
        m_2
    \end{pmatrix}=\begin{pmatrix}
        1&-1\\
        0&1
    \end{pmatrix}\begin{pmatrix}
        1\\
        -1
    \end{pmatrix}=\begin{pmatrix}
        2\\
        -1
    \end{pmatrix}.
\end{equation}
We thus end up with the description of $\mathcal{T}_2$ in \cite{Gang:2023rei}, up to a convention-dependent parity transformation yielding an overall sign for the matrix of CS levels. Additionally, we have here a microscopic mechanism for dynamical generation of the monopole superpotential in the compactified Maruyoshi-Song theory
\`{a}~la~Affleck-Harvey-Witten \cite{Affleck:1982as,Seiberg:1996nz}. It should be possible to corroborate this result via the Nye-Singer index theorem \cite{Nye:2000eg,Poppitz:2008hr} as in \cite{ArabiArdehali:2019zac}, but we do not attempt that here.

As explained in Section~3.1 of \cite{ArabiArdehali:2024vli}, similar mechanisms are at work on the 2nd sheet of $(A_1,A_{2n})$ for $n>2.$ The 3d EFT yields the $\mathcal{T}_n$ theory of \cite{Gang:2023rei} for arbitrary $n\in\mathbb{N}$, possibly up to a convention-dependent parity transformation.

\subsection*{TQFT $S$ and $T$ matrices}

The uv data of the 2nd-sheet theory appropriate for the A-twist is given by
\begin{equation}
    \text{\textbf{triLY}}:\qquad\quad \mathbf{U(1)\times U(1)}\ +\ \mathbf{\Phi_{1,-1}^{r=0}}\ +\ \mathbf{\Phi_{0,1}^{r=1}}\ \ \text{with}\ \ \mathbf{k_{1R}=-\frac{1}{2}},\  \mathbf{k_{2R}=\frac{1}{2}}\,,\label{eq:tLY}
\end{equation}
with the matrix of gauge-gauge CS couplings as in \eqref{eq:kij_T2}. By triLY we mean \emph{tricritical Lee-Yang}, due to the connection between $M(2,7)$ and the tricritical Yang-Lee edge singularity \cite{Lencses:2022ira}.

A quick calculation shows:
\begin{equation}
    k^+_{ij}=\begin{pmatrix}
        -1&-1\\
        -1&0
    \end{pmatrix},\label{eq:k+ij_T2}
\end{equation}
while $k^+_{1R}=-1$ and $k^+_{2R}=1.$ Therefore,
\begin{equation}
    \begin{split}
        W(Z)&=-\frac{1}{2}Z_1^2-Z_1 Z_2-\pi i Z_1+\mathrm{Li}_2\big(e^{Z_1-Z_2}\big)+\mathrm{Li}_2\big(e^{Z_2}\big)\,,\\
        \Omega(Z)&=-Z_1+Z_2+\log\big(1-e^{Z_1-Z_2}\big)\,.
    \end{split}
\end{equation}

We now identify the Wilson lines realizing simple objects $L_\alpha$ in the triLY MTC. They play dual role: when piercing the boundary, they give simple modules of the boundary VOA, and when inserted parallel to the boundary, they realize Verlinde lines \cite{Verlinde:1988sn,Witten:1988hf,Elitzur:1989nr,Moore:1989yh,Moore:1988qv}. Wrapping $L_\alpha$ on any loop in $S^3$ leads to the vev \cite{Witten:1988hf}:
\begin{equation}
    \langle L_\alpha\rangle^{}_{S^3}=\frac{S_{0\alpha}}{S_{00}}\,,\label{eq:LineSurgery}
\end{equation}
which in terms of the handle-gluing operator $\mathcal{H}$ and the Bethe roots $Z^\ast_\alpha$ reads (\emph{cf.~}\cite{Gang:2023rei}):
\begin{equation}
    \langle L_\alpha\rangle^{}_{S^3}=L_\alpha(Z^\ast_0)=\pm\frac{\mathcal{H}(Z^\ast_\alpha)^{-1/2}}{\mathcal{H}(Z^\ast_0)^{-1/2}}\,.\label{eq:identifyLines}
\end{equation}
These equations allow us to identify the appropriate Wilson lines as ${z}_1=e^{Z_1}$ and $\tilde{z}_2=e^{Z_1+Z_2}$, in addition to the trivial line $L_0=1$. In terms of these, the Bethe equations read\footnote{From the Bethe equations, instead of \eqref{eq:triLYfusion2}, we actually find $ \tilde z_2^2=1+ z_1  \tilde z_2$, but the resulting systems are equivalent under the condition that $z_1\neq0.$ We have included \eqref{eq:triLYfusion2} instead, because it is a fusion rule. See \emph{e.g.}~\cite{Closset:2023vos} for a computational commutative algebraic approach to such systems of polynomial Bethe equations.}
\begin{subequations}
\begin{eqnarray}
     z_1^2&=1+ \tilde z_2\,, \label{eq:triLYfusion1}\\
     z_1 \tilde{z}_2 &= {z}_1+\tilde{z}_2\,.\label{eq:triLYfusion2}
\end{eqnarray}\label{eq:triLYfusion}
\end{subequations}
Next, using \eqref{eq:WL_surgery}:
\begin{equation}
    S_{\alpha\beta}=L_\alpha(Z^\ast_\beta)S_{0\beta}\,,\label{eq:Smatrix_formula}
\end{equation}
together with \eqref{eq:HandF_vs_SandT} and \eqref{eq:Z_surgery}, we find\footnote{The order of the non-vacuum objects is chosen so that the fusion rules \eqref{eq:triLYfusion} match the Verlinde formula based on the expected $S$ matrix. The sign ambiguities in the entries of $S$ are fixed as follows. Since we are in a non-unitary rational setting, either the 2nd or the 3rd row can be strictly positive. This turns out to leave only two possibilities compatible with $S^2$ being a permutation matrix, of which one yields negative fusion coefficients via the Verlinde formula, so we end up with $S$ as given.} the $S$ and $T$ matrix to be (up to an overall phase ambiguity in $T$):
\begin{equation}
\begin{split}
    S&= 
\frac{2\sin\big(\frac{\pi}{7}\big)}{\sqrt{7}}
\begin{pmatrix}
   d &1-d^2&  1\\
   1-d^2 & -1&d\\
   1&d&d^2-1
\end{pmatrix},\\
    T&=\left(
\begin{array}{ccc}
 1 &   \ \ 0 &0 \\
 0 & \ \ \ \ e^{2\pi i(-\frac{2}{7})}&0 \\
 0 &\ \ 0& \!\! e^{2\pi i(-\frac{3}{7})}
\end{array}
\right),
\end{split}\label{eq:triLY_SandT}
\end{equation}
with $d\!=\!2\cos\big(\frac{\pi}{7}\big)$. The $S^3$ partition function computed via \eqref{eq:Z_surgery} gives:
\begin{equation}
    |Z^{}_{S^3}|=\frac{2\sin\big(\frac{2\pi}{7}\big)}{\sqrt{7}},\label{eq:triLY_Z}
\end{equation}
matching $S_{00}$ as it should. These indeed correspond to the Virasoro minimal model $M(2,7)$~\cite{DiFrancesco:1997nk}, up to the said ambiguities.

\subsection{Higher sheets of $(A_1,A_4)$}

We next discuss the 3rd and 4th sheets of $(A_1,A_4).$ We will skip the other three sheets since their EFT data can be reached via simple conjugations from the sheets discussed.

\subsubsection{Third sheet: duality up to an overall phase and the B-twist}

Here we get a saddle at $(x_1,x_2)=(3/7,-1/7).$ The EFT is a 3d $\mathcal{N}=2$ $U(1)\times U(1)$ gauge theory with
\begin{equation}
    k=\begin{pmatrix}
        1&\frac{1}{2}\\
        \frac{1}{2}&-1
    \end{pmatrix}.\label{eq:kij_Ttilde2}
\end{equation}
The matter content is
\begin{center}
\begin{tabular}{|c|c|c|c|c|}
  \hline
  Chiral  & $\Phi_{-1,0}$ &$\Phi_{0,-1}$& $\Phi_{-1,-1}$ &$\Phi_{0,2}$ \\
  \hline
  $R$-charge  &$\frac{10}{21}$ &$\frac{20}{21}$&$\frac{2}{21}$ & $\frac{2}{21}$\\
  \hline
\end{tabular}
\end{center}
with the gauge charges indicated as subscripts. We also have $k_{1R}=-\frac{9}{7}$ and $k_{2R}=-\frac{13}{21}$. The monopole superpotential is
\begin{equation}
    \mathcal{W}\sim V_{1,0}+V_{0,1}\, .\label{eq:A1A4_3rd_Wv}
\end{equation}
As the $R$-charges indicate, there is also a natural possibility of adding
\begin{equation}
    \mathcal{W}\, \overset{}{\ni}\, \Phi_{0,-1}^2\Phi_{0,2}\,.\label{eq:A1A4_3rd_Wphi}
\end{equation}
Both \eqref{eq:A1A4_3rd_Wv} and \eqref{eq:A1A4_3rd_Wphi} are invariant under the $U(1)_f$ descending from 4d $\mathcal{N}=1$ flavor symmetry (that is part of the $\mathcal{N}=2$ $R$-symmetry). $F$-maximization with respect to this $U(1)_f$ gives the $R$-charges $\frac{3}{14}$, $\frac{13}{14}$, $\frac{1}{7}$, $\frac{1}{7}$ for $\Phi_{-1,0}$, $\Phi_{0,-1}$, $\Phi_{-1,-1}$, $\Phi_{0,2}$, respectively, in a gauge-$R$ mixing scheme where $k_{1R}=-\frac{33}{28}$ and $k_{2R}=-\frac{19}{28}.$

Comparison of the superconformal indices suggests that this theory, which we denote by $\widetilde{\mathcal{T}}_2$, is dual to the $\mathcal{T}_2$ theory \cite{Gang:2023rei} that we obtained on the 2nd sheet, except for a difference in background CS couplings.  In particular, $\widetilde{\mathcal{T}}_2$ is an $\mathcal{N}=4$ SCFT. The differing background CS couplings manifest themselves in overall phases of the $S$ and $T$ matrices, and any partition function obtained after the A-twist.

This \emph{duality up to overall phase} between the second and third sheets of $(A_1,A_4)$ may appear disappointing at first, if one were hoping to find an entirely new TQFT. We will argue below, however, that they are in fact \emph{mirror} dual to each other, which readers might find more appealing. So for the second sheet theory $\mathcal{T}_2$, we find its mirror $\widetilde{\mathcal{T}}_2$ on the third sheet and its conjugate on the seventh sheet. Note that in the Lee-Yang case, the mirror and the conjugate theories coincided and were identified as $\overline{\rm LY}$ on the fifth sheet of $(A_1,A_2)$. The appearance of mirror dual on the Galois orbit of TQFT seems intriguing.

\subsection*{TQFT $S$ and $T$ matrices}

The uv data of the 3rd-sheet theory appropriate for the A-twist is given by\footnote{Verifying that this is a TQFT data via superconformal index calculations in Mathematica can be simplified with a minor gauge-$R$ mixing so that poles are removed from the unit circle contours.}
\begin{equation}
    \widetilde{\text{\textbf{triLY}}}:\  \mathbf{U(1)\!\times\! U(1)}\, +\, \mathbf{\Phi_{-1,-1}^{r=0}} +\, \mathbf{\Phi_{0,2}^{r=0}} + \,\mathbf{\Phi_{-1,0}^{r=1}} +\, \mathbf{\Phi_{0,-1}^{r=1}}\,, \ \ \ \mathbf{k_{1R}\!=-\frac{3}{2}},\,  \mathbf{k_{2R}\!=-\frac{1}{2}}\,,\label{eq:tLYtilde}
\end{equation}
with the matrix of gauge-gauge CS couplings as in \eqref{eq:kij_Ttilde2}. It is easily seen from \eqref{eq:k+} that
\begin{equation}
    k^+_{ij}=\begin{pmatrix}
        2&1\\
        1&2
    \end{pmatrix},\label{eq:k+ij_T2}
\end{equation}
while $k^+_{1R}=-1$ and $k^+_{2R}=-1.$ Therefore
\begin{equation}
    \begin{split}
        W(Z)&=Z^2_1+Z^2_2+Z_1Z_2+2\pi i(Z_1+Z_2)\\
        &\hspace{2.3cm}+\mathrm{Li}_2\big(e^{-Z_1-Z_2}\big)+\mathrm{Li}_2\big(e^{2Z_2}\big)+\mathrm{Li}_2\big(e^{-Z_1}\big)+\mathrm{Li}_2\big(e^{-Z_2}\big)\,,\\
        \Omega(Z)&=-Z_1-Z_2+\log\big(1-e^{-Z_1-Z_2}\big)+\log\big(1-e^{2Z_2}\big)\,.
    \end{split}
\end{equation}

Identifying the Wilson lines corresponding to the simple modules via \eqref{eq:identifyLines},
we find\footnote{The negative sign here is introduced to enforce positive coefficients in the fusion rule \eqref{eq:triLYtildeFusion2}.} $\tilde{z}_1=-e^{Z_2}$ and $\tilde{z}_2=e^{-Z_1}$, in terms of which the Bethe equations read:
\begin{subequations}
\begin{eqnarray}
     \tilde z_1^2&=1+\tilde z_2\,, \label{eq:triLYtildeFusion1}\\
     \tilde z_1 \tilde{z}_2 &= \tilde{z}_1+\tilde{z}_2\,.\label{eq:triLYtildeFusion2}
\end{eqnarray}\label{eq:triLYtildeFusion}
\end{subequations}

Next, using \eqref{eq:Smatrix_formula} and \eqref{eq:HandF_vs_SandT}, and fixing the sign ambiguities in the entries of $S$ as in the triLY case, we find (up to the overall phase ambiguity in $T$):
\begin{equation}
\begin{split}
    S&= 
\frac{2\sin\big(\frac{\pi}{7}\big)}{\sqrt{7}}
\begin{pmatrix}
   d^2-1 &  1&-d\\
   1&d & d^2-1\\
   -d&d^2-1&-1&
\end{pmatrix},\\
    T&=\left(
\begin{array}{ccc}
 1 &   \ \ 0 &0 \\
 0 & \ \ \ \ e^{2\pi i(\frac{3}{7})}&0 \\
 0 &\ \ 0& \!\! e^{2\pi i(\frac{1}{7})}
\end{array}
\right).
\end{split}\label{eq:triLYtilde_SandT}
\end{equation}
\vspace{.1cm}

The $T$ matrix coincides with the square of the $T$ matrix on the 2nd sheet \eqref{eq:triLY_SandT}, confirming the general expectation that the $T$ matrix on the $(\gamma+1)$st sheet is given by the $\gamma$th power of the 2nd-sheet $T$ matrix.

From the $S$ matrix note that for $|S_{00}|$ we get $\frac{2\sin\frac{\pi}{7}}{\sqrt{7}}(d^2-1)$, which is different from that of triLY in \eqref{eq:triLY_Z}. This implies that triLY and $\widetilde{\text{triLY}}$ have different $S^3$ partition functions, and are hence truly distinct TQFTs. Since they arise from the A-twist of dual 3d $\mathcal{N}=4$ SCFTs $\mathcal{T}_2$ and $\widetilde{\mathcal{T}}_2$, we suggest that $\widetilde{\text{triLY}}$ arises from B-twisting $\mathcal{T}_2$. In other words, we conjecture that $\mathcal{T}_2$ and $\widetilde{\mathcal{T}}_2$ are mirror dual, which we confirm by checking that their flavored indices coincide, up to the inversion of the flavor fugacity corresponding to the flip $U(1)_H \longleftrightarrow U(1)_C$. Since the A and B twists are truly distinct (and not just conjugate) for all $\mathcal{T}_n$ with $n>1$, we conjecture that more generally, B-twisted $\mathcal{T}_n$ will be on the Galois orbit of the A-twisted $\mathcal{T}_n$.

\subsubsection{Fourth sheet: non-abelian TQFT and fractional monopoles}

Here we get a saddle at $(x_1,x_2)=(2/7,2/7).$ The EFT is a 3d $\mathcal{N}\!=\!2$ $\mathrm{SU}(2)\!\times\! U(1)$ gauge theory. Before recovering the $\mathrm{SU}(2),$ we have a $U(1)\times U(1)$ theory with
\begin{equation}
    k=-\begin{pmatrix}
        2&\frac{1}{2}\\
        \frac{1}{2}&2
    \end{pmatrix}.
\end{equation}
The matter content is:
\begin{center}
\begin{tabular}{|c|c|c|c|c|c|}
  \hline
  Chiral & $\Phi_{1,1}$ & $\Phi_{2,0}$ & $\Phi_{0,2}$&$\Phi_{-1,0}$&$\Phi_{0,-1}$ \\
  \hline
  $R$-charge &$\frac{2}{21}$ &$\frac{2}{21}$ &$\frac{2}{21}$ &$\frac{20}{21}$&$\frac{20}{21}$ \\
  \hline
\end{tabular}
\end{center}
as well as the light W-bosons with charges $(1,-1)$ and $(-1,1)$, which will be responsible for the gauge symmetry enhancement. We also have $k_{1R}=k_{2R}=-\frac{34}{21}$. The monopole superpotential is
\begin{equation}
    \mathcal{W}\sim V_{1,0}+V_{0,1}\, .\label{eq:4thSheetA1A4W}
\end{equation}
It is also natural to add:
\begin{equation}
    \mathcal{W}\, \overset{}{\supset}\, \mathcal{W}_\Phi=\Phi_{-1,0}\Phi_{0,-1}\Phi_{1,1}+ \Phi_{-1,0}^2\Phi_{2,0}+ \Phi_{0,-1}^2\Phi_{0,2}\,.
\end{equation}
For going to gauge coordinates where the $\mathrm{SU}(2)$ is manifest, consider:
\begin{equation*}
    \begin{split}
    Z^{}_{S^3}&=\int\mathrm{d}\sigma_1\mathrm{d}\sigma_2\, e^{-2\pi i\, k_{ij}\frac{\sigma_i\sigma_j}{2}+2\pi\, k^{}_{jR}\sigma_j}\\
    &\quad\frac{\Gamma_h\big(\frac{2}{21}i+\sigma_1+\sigma_2\big)\Gamma_h\big(\frac{2}{21}i+2\sigma_1\big)\Gamma_h\big(\frac{2}{21}i+2\sigma_2\big)\Gamma_h\big(\frac{20}{21}i-\sigma_1\big)\Gamma_h\big(\frac{20}{21}i-\sigma_2\big)}{\Gamma_h\big(\pm(\sigma_1-\sigma_2)\big)}.
    \end{split}
\end{equation*}
We change variables to $y:=y_1=(\sigma_1-\sigma_2)/2$ and $x:=y_2=(\sigma_1+\sigma_2)/2$:
\begin{equation*}
    \begin{split}
    Z^{}_{S^3}&=\int\frac{\mathrm{d}y}{2}\mathrm{d}x\, e^{2\pi i\, (3\frac{y^2}{2}+5\frac{x^2}{2})-2\pi\, \frac{68}{21}x}\\
    &\quad\frac{\Gamma_h\big(\frac{2}{21}i+2x\big)\Gamma_h\big(\frac{2}{21}i+2x+2y\big)\Gamma_h\big(\frac{2}{21}i+2x-2y\big)\Gamma_h\big(\frac{20}{21}i-x-y\big)\Gamma_h\big(\frac{20}{21}i-x+y\big)}{\Gamma_h\big(\pm2y\big)}.
    \end{split}
\end{equation*}
A further shift of $x$ by $-i/21$ makes the $R$-charges $0,1$, while $k_{xR}=-3$ and $k_{yR}=0.$ We thus get:
\begin{equation}
      {\text{\textbf{TQFT}}^{\gamma=3}_{A_1A_4}}:\qquad\mathbf{\frac{{SU}(2)_{-3}\times U(1)_{-5}}{\mathbb{Z}_2}}\ +\ \mathbf{\Phi_{\Box\!\Box,2}^{r=0}}\, +\, \mathbf{\Phi_{\Box,-1}^{r=1}}\ \ \text{with}\ \  \mathbf{k_{xR}=-3}\,,\label{eq:TQFT_A1A4_gamma=3}
\end{equation}
with the $\mathbb{Z}_2$ identification to be explained momentarily. The superpotential becomes:
\begin{equation}
    \mathcal{W}\,\sim\, \sqrt{Y_- V_+}+\sqrt{Y_+ V_+}+\mathcal{W}_\Phi\ \xrightarrow[\text{Weyl redundancy}]{\text{removing the $\mathbb{Z}_2$\ }}\,\sqrt{Y\, V_+}+\mathcal{W}_\Phi\,,\label{eq:sqRW}
\end{equation}
where $V_+$ is the $U(1)$ monopole with flux $m_x=1$, and $Y$ is the $\mathrm{SU}(2)$ monopole with GNO charge $m_y=1$. To obtain the monopole superpotential in \eqref{eq:sqRW} from \eqref{eq:4thSheetA1A4W}, we have used \eqref{eq:monopoleTransformation}:
\begin{equation}
    \begin{pmatrix}
        m_y\\
        m_x
    \end{pmatrix}=\begin{pmatrix}
        \frac{1}{2}&-\frac{1}{2}\\
        \frac{1}{2}&\,\ \frac{1}{2}
    \end{pmatrix}\begin{pmatrix}
        m_1\\
        m_2
    \end{pmatrix}.
\end{equation}

We have checked that the superconformal index of this theory is trivial:
\begin{equation}
    \tilde{I}(q)=1\,,
\end{equation}
indicating that it is gapped and flows to a TQFT. Note that in the computation of the index, we have to sum over $m_y,m_x\in\frac{1}{2}\mathbb{Z}$ subject to $m_y+m_x\in\mathbb{Z}$, as dictated by 
$m_1,m_2\in\mathbb{Z}$ from the UV completion \cite{Preskill:1984gd}. This restriction is reflected in the $\mathbb{Z}_2$ identification in \eqref{eq:TQFT_A1A4_gamma=3}.

It would be interesting to perform half-index calculations for this theory and see whether the three-component vector-valued modular forms (vvmfs) from \cite{hikami2005quantum,Hampapura:2016mmz} mentioned in Section~5.3 of \cite{Harvey:2018rdc} arise. 

Our preliminary calculations do not reproduce the expected fusion rules and modular data from the Bethe root techniques in this case. We leave clarification of the relation between the 4th-sheet TQFT and triLY$, \widetilde{\text{triLY}}$ to future work.

\section{Discussion and open questions}\label{sec:openQs}

Building on \cite{Dedushenko:2023cvd}, we have further developed the
\begin{equation}
    \text{4d } \underset{}{\xrightarrow{S^1\text{ reduction}\,}} \text{ 3d } \underset{\text{}}{\xrightarrow{\text{ boundary}\ }} \text{ 2d}\label{eq:4d3d2d}
\end{equation}
picture of the SCFT/VOA correspondence \cite{Beem:2013sza}. We studied here the $U(1)_r$-twisted circle reductions that leave only finitely many points of the Coulomb branch unlifted in 3d \cite{Fredrickson:2017yka}, focusing specifically on theories without Higgs branches. Then, either via the topological A-twist (when we have SUSY enhancement to 3d $\mathcal{N}=4$), or via flowing to gapped phases, we obtained 3d TQFTs without local operators. The former TQFTs are non-unitary, and the latter are unitary, but in either case, they are controlled by some modular tensor categories \cite{Moore:1988qv}. On their holomorphic boundaries, our TQFTs support VOAs whose characters are accessible via line-decorated half-indices. The minimal $U(1)_r$ twist with $\gamma=1$ yields the VOAs of \cite{Beem:2013sza}, while other choices yield other VOAs related to those of \cite{Beem:2013sza} via Galois/Hecke-type transformations \cite{Dedushenko:2018bpp,Harvey:2018rdc,Harvey:2019qzs,Lee:2022yic}.

At each step in \eqref{eq:4d3d2d}, there are various choices to be made that we did not spell out in the main text. We address some of them below.

\subsection{Topological twist and Bethe roots technique}

We use the Bethe roots technique \cite{Nekrasov:2009uh,Closset:2019hyt} as formulated in \cite{Closset:2019hyt} to compute the TQFT $S$ and $T$ matrices. There is, however, a technical subtlety that we skipped. This technique was developed for the partial topological, or quasi-topological, twist in 3d $\mathcal{N}=2$ theories, sometimes also called 3d $\mathcal{N}=2$ A-twist. In this paper, on the other hand, we never work with this twist. We are either interested in the fully topological twist, or we consider gapped theories that are topological in the IR on their own, without any twist. Then are our results reliable? We believe that when studying partition functions on three-manifolds that are total spaces of circle fibrations, this distinction is irrelevant. Applying the 3d $\mathcal{N}=2$ A-twist to a gapped theory is almost vacuous, and will at most result in the overall phase of $T$, which we ignore anyways. The distinction between the topological and quasi-topological twist is slightly more subtle. By deforming the metric on the total space of circle fibration, we can make sure that the topological and quasi-topological backgrounds agree almost everywhere, except the location of fibering operators. This implies that the computation of handle-gluing operators is reliable, and our $S$-matrix is fully correct. At the same time the fibering operators are likely to receive some additional phases in the fully topological background, capturing the overall phase of $T$. It would be useful to clarify this issue.

\subsection{Sensitivity to 2d boundary conditions}\label{subsec:sense2dBCs}

In the main text of the paper, we only studied the simplest $\mathcal{N}=(0,2)$ boundary conditions, with either Dirichlet or Neumann on all fields, with the boundary Fermi multiplets canceling anomalies when necessary. The
hope was that such boundary conditions could be used to probe the possible VOAs and the bulk TQFT. However, this does not exhaust the possible boundary conditions. Furthermore, the cigar reduction in \cite{Dedushenko:2023cvd} implied that there exist preferred, or canonical, boundary conditions $H_\varepsilon$ for the second sheet theory, guaranteed to carry the VOA of the 4d SCFT. It was also argued that the half-index of $H_\varepsilon$ --- or the TQFT partition function on solid torus with the $H_\varepsilon$ boundary --- computes the Schur index.

In the context of Lagrangian theories, such preferred boundary conditions $H_\varepsilon$ were identified in \cite{Dedushenko:2023cvd} as the $\mathcal{N}=(0,4)$ Neumann, deformed to be compatible with the topological 3d A-twist. In the notation $H_\varepsilon$, $\varepsilon$ stands for this deformation, referred to as the Costello-Gaiotto deformation \cite{Costello:2018fnz,Costello:2018swh}.  Before discussing the possible modifications in the non-Lagrangian context of our main interest here, let us explain how the $(0,4)$ Neumann boundary conditions reproduce the SCFT Schur index in Lagrangian cases.

First, our $r$-twisting is trivial in Lagrangian theories due to the $U(1)_r$ charge quantization of $\mathcal{N}=2$ multiplets, meaning that there exists only one sheet, corresponding to the ordinary supersymmetric circle reduction. It always gives a (not necessarily dominant \cite{ArabiArdehali:2023bpq}) holonomy saddle at the origin, yielding a 3d $\mathcal{N}=4$ theory with the same field content as that obtained from the naive dimensional reduction. 
The $(0,4)$ Neumann boundary conditions amount to $(0,2)$ Neumann on all multiplets, except for the adjoint chirals in the 3d $\mathcal{N}=4$ vectors that should have $(0,2)$ Dirichlet. For compatibility with the A-twist, we use $U(1)_H$ as the $\mathcal{N}=2$ $R$-symmetry, and compute the half-index using formulas from \cite{Dimofte:2017tpi}. The result is:
\begin{equation}
\mathcal{I}(q)=\frac{(q;q)^{2r_G}/|W|}{\theta_0(q^{1/2};q)^{n_{\rho_0}/2}}\int_{\mathfrak{h}_{\text{cl}}} \mathrm{d}^{r_G}x\ \frac{\prod_{\alpha}\theta_0(z^{\alpha};q)}{\prod_{\rho_+^\chi}\theta_0(q^{1/2}z^{\rho^\chi_+};q)},\label{eq:indexTheta0}
\end{equation}
matching the Schur index of the Lagrangian 4d $\mathcal{N}=2$ SCFT. The weights $\rho_+^\chi$ above go over all the positive weights of the gauge group representation of the chiral multiplets inside the (half-) hypermultiplets, and $n^{}_{\rho_0}$ is the number of zero weights in the chiral multiplets inside (half-) hypers. Note that we are not including among $\rho_+^\chi$ the weights of the chiral multiplets inside 4d $\mathcal{N}=2$ vector multiplets.
From the 3d $\mathcal{N}=2$ perspective, half the numerator contribution $\theta_0(z^\alpha; q)=(z^\alpha; q)(q z^{-\alpha};q)$ in \eqref{eq:indexTheta0} comes from the 3d $\mathcal{N}=2$ vector, while the other half is from its chiral partner.

Can we similarly determine the preferred boundary conditions $H_\varepsilon$ in the non-Lagrangian examples of our main interest? The answer is almost certainly yes, though we leave this question to future work, only explaining the main idea here. Starting with the 4d $\mathcal{N}=1$ Lagrangians of Maruyoshi-Song \cite{Maruyoshi:2016aim,Maruyoshi:2016tqk}, we are supposed to dimensionally reduce them on the cigar with the topological $U(1)_r$ twist, following the procedure in \cite{Dedushenko:2023cvd}. The corresponding background and Lagrangians were described in \cite{Longhi:2019hdh}. In the 3d limit, as in this paper, the dominant contribution will come from certain gauge field configurations. Namely, there will be a nonzero gauge flux through the cigar, breaking the gauge group down to its maximal torus, and screening the $U(1)_r$ flux in such a way that some 4d chiral multiplets will possess zero modes on the cigar, resulting in the 3d chiral multiplets. Our $\Phi_{+1}$ in the Gang-Yamazaki theory is one such multiplet, and the boundary condition on $\Phi_{+1}$ engineered by the tip of cigar clearly is the $(0,2)$ Neumann. Similarly, the surviving $U(1)$ gauge multiplet also obeys the $(0,2)$ Neumann. Such boundary conditions, as we know, are anomalous. Since the starting 4d theory is anomaly-free, there is only one resolution: The tip of the cigar must support additional localized normalizable modes that in the 3d limit become boundary modes. Indeed, such a possibility can be inferred from the analysis in \cite{Longhi:2019hdh}.

In the GY theory, imposing $(0,2)$ Neumann boundary conditions on both the gauge and chiral multiplets on the right boundary results in the boundary gauge anomaly 
\begin{equation}
    \frac32 \mathbf{f}^2 - 2\,\mathbf{f}\cdot\mathbf{f}_x -\frac12 \mathbf{f}^2 = \mathbf{f}^2 - 2\,\mathbf{f}\cdot\mathbf{f}_x\,,\label{eq:Misha_AP}
\end{equation}
where $\mathbf{f}$ is the gauge field strength and $\mathbf{f}_x$ is for $U(1)_J$. We cancel the gauge anomaly by adding a boundary $(0,2)$ chiral multiplet of gauge$\,\times\, U(1)_J\times U(1)_R$ charges $(1,-1,1)$, which carries anomaly $-(\mathbf{f}-\mathbf{f}_x)^2$. We also include an extra boundary Fermi multiplet of charges $(0, 1, 1)$ to ensure that the $x\to 1$ limit is regular (where $x$ is the $U(1)_J$ fugacity).

Now let us compute the half-index. Since we have a 2d chiral multiplet, we need to have a way of dealing with the infinite number of poles in the integral.
It is known how to deal with this problem in the context of the 2d elliptic genera, and the JK prescription appears as the result of carefully dealing with the zero modes \cite{Benini:2013xpa,Benini:2016qnm}.
We use the two-step procedure of \cite{Dimofte:2017tpi} to compute the 3d half-index as in Appendix~\ref{app:half-index}. First, compute the half-index with Dirichlet condition on the gauge field, which in the present case implies the same boundary anomaly as in \eqref{eq:Misha_AP}. Then dress it with the 2d contributions from the anomaly-canceling matter, and finally gauge the boundary $U(1)$ symmetry descending from the bulk gauge field via a 2d gauge multiplet. The result is (after a change $z\to z (-q^{-1/2})$ of the gauge variable)
\begin{equation}
    \begin{split}
      I\!\!I_{\text{Neu}}(q)&=\lim_{x\to1}\ \theta_0(x^{-1};q) (q;q)^2 \oint_{\underset{z=x}{\text{JK res}}} \frac{dz}{2\pi i z} \frac{1}{\theta_0(zx^{-1};q)} \frac{1}{(q;q)} \sum_{m\in\mathbb{Z}} \frac{q^{m^2/2} z^m (-q^{-1/2}x^{-1})^m}{(z q^m;q)}\\
       &=\lim_{x\to1}\  -\,\frac{(x^{-1};q)(xq;q)}{(q;q) }\sum_{m\in\mathbb{Z}} q^{m^2/2} (-q^{-1/2})^m \,\frac{1 }{(x q^m;q)}\\
&\overset{}{=}         \sum_{m\in\mathbb{Z}_{\ge0}}\frac{q^{m^2+m}}{(q;q)_m},
    \end{split}
\end{equation}
which matches the vacuum character $ \chi_0^{M(2,5)}(q)$. In going from the second line to the third we used: 
\begin{equation}
    \lim_{x\to1}\frac{(x^{-1};q)}{(x\,q^m;q)} =\begin{cases}
        \frac{(-1)^{m+1} q ^{m(m+1)/2}}{(q,q)_m} \quad&\text{for}\quad m\leq0,\\
        \qquad\quad 0 \quad&\text{for}\quad m>0.
    \end{cases}
\end{equation}

To get the non-vacuum character, we insert a charge $-1$ Wilson line. This amounts to an insertion of $q^m$ in the summand of the previous computation, changing it to:
\begin{equation}
    \begin{split}
        \lim_{x\to1}\  -\,\frac{(x^{-1};q)(xq;q)}{(q;q) }\sum_{m\in\mathbb{Z}} q^{\frac{m^2}{2}+m} (-q^{-1/2})^m \,\frac{1 }{(x q^m;q)}=        \sum_{m\in\mathbb{Z}_{\ge0}}\frac{q^{m^2}}{(q;q)_m},
    \end{split}
\end{equation}
which matches the non-vacuum character $\chi_1^{M(2,5)}(q)$. 

Since we found the correct characters, we conjecture that the boundary chiral and Fermi multiplets that we included by hand must appear naturally as normalizable edge modes in the reduction of $\mathcal{N}=1$ Maruyoshi-Song Lagrangian on the cigar.

We emphasize that these characters can be obtained from the $(\mathcal{D}, {\rm D}_c)$ boundary conditions in the GY theory \cite{Gang:2023rei}. This suggests that such boundary conditions are dual to the enriched Neumann that we just considered. It would be quite interesting to study these issues further. It is especially interesting to systematically derive the $H_\varepsilon$ type boundary conditions by the topological cigar reduction of the Maruyoshi-Song $\mathcal{N}=1$ Lagrangians. As said earlier, we conjecture that the above Neumann boundary conditions with the boundary chiral and Fermi should arise in such a way.


Note that the described procedure gives the preferred boundary conditions $H_\varepsilon$ for the second sheet theory. In fact, it also works for its conjugate, or ``last'' sheet --- the fifth sheet in the $(A_1,A_2)$ case. Indeed, the second sheet has the holonomy $e^{2\pi i/N}$ originating from the topological twist along the cigar, and the last sheet has $e^{2\pi i (N-1)/N}=e^{-2\pi i/N}$, clearly originating from the anti-topological twist along the cigar.\footnote{These are the 2d ${\rm B}$ and $\overline{\rm B}$ twists along the cigar \cite{Dedushenko:2023cvd}, which are switched by the charge conjugation.} The intermediate higher-sheet theories do not seem to posses such preferred boundary conditions like $H_\varepsilon$ descending from 4d.

Thus, for the third-sheet $\overline{\text{Fib}}$ (resp. fourth-sheet ${\text{Fib}}$) theory, we studied the Dirichlet as well as Neumann half-indices simply for the reasons of naturalness. We also had a prior expectation, based on our experience with the second sheet, as well as other examples, that the Dirichlet boundary conditions (including ${\rm D}_c$ on some chirals) are likely to capture the bulk TQFT. This reasoning was explained around the conjecture \eqref{conj_spinTQFT}. While with Neumann conditions we found the expected \cite{Dedushenko:2018bpp} $(G_2)^{}_1$ characters on the left (resp.~right) boundary, with Dirichlet conditions we obtained characters of the fermionized tricritical Ising model times a free Majorana fermion on the opposite boundary. This motivated us to conjecture that the bulk TQFT is actually a spin-TQFT $\mathrm{SM}(3,5)\otimes \mathrm{SO}(1)_1$ as in \eqref{conj_spinTQFT} on the third sheet (or its conjugate on the fourth sheet).

The conjecture is corroborated by the following considerations. First, assuming that the third-sheet TQFT is $\mathrm{SM}(3,5)\otimes \mathrm{SO}(1)_1$ with $c_\text{bulk}=6/5$ as in \eqref{eq:spinTQFTc}, we see that appearance of the $(G_2)^{}_1$ VOA on its boundary is compatible with our addition of the four boundary Fermi multiplets:\footnote{The negative sign is because we are considering the Weyl anomaly induced on the opposite boundary. Compare with Section~4.4 in \cite{Ferrari:2023fez}, in particular their Eq.~(4.49).}
\begin{equation}
    -c_\text{bulk}+c_\text{boundary} = -\frac{6}{5} + 4 = \frac{14}{5}=c^{}_{(G_2)^{}_1}.
    \label{eq:confemdG2SM}
\end{equation}
The four added fermions on the opposite boundary yield $U(4)_1$, which is the only chiral fermionic CFT of central charge $4$ \cite{Creutzig:2017fuk,BoyleSmith:2023xkd,Rayhaun:2023pgc,Hohn:2023auw}. The corresponding bulk spin-TQFT is invertible. These considerations point to the possibility of realizing $(G_2)^{}_1$ as:
\begin{equation}
    (G_2)_1 = (U(4)_1)/(\mathrm{SM}(3,5)\otimes \mathrm{SO}(1)_1) .\label{eq:G2vsSM35}
\end{equation}
The bosonic counterpart $(G_2)^{}_1 = (E_8)^{}_1/(F_4)^{}_1$ is of course standard.

We can consider the 3d system on an interval with the enriched Neumann boundary conditions on one boundary and Dirichlet on the other.\footnote{For intervals with both boundaries being Dirichlet or Neumann see \emph{e.g.} \cite{Dedushenko:2022fmc,Sugiyama:2020uqh}.}
As those boundaries are mutually exclusive, after the interval reduction, the only surviving degrees of freedom are the boundary fermions.
Indeed, the product of VOAs $(G_2)_1 \otimes \mathrm{SM}(3,5) \otimes \mathrm{SO}(1)_1$ conformally embeds into $U(4)_1$, as \eqref{eq:confemdG2SM} suggests.
The $G_2$ VOA at level 1 has two modules, and we denote the second, non-vacuum, module by $L_{\hat{\omega}_2}(\widehat{G_2})$.
It is straightforward to check at the level of characters the following decomposition of the four-fermion vacuum module:\footnote{We take the liberty of denoting the affine VOA as $G_k$, and using the same symbol for both the VOA and its representation category. At the same time, $U(4)_1$ is understood as a spin-TQFT, whose corresponding VOA of free fermions is an extension of the affine VOA of $U(4)$ at level one. We hope this frivolous approach to notations will not cause confusion.}
\begin{equation}
\begin{split}
   U(4)_1 = \big((G_2)_1 \otimes V(1,1)^{(3,5)} \oplus L_{\hat{\omega}_2}(\widehat{G_2}) \otimes V(1,3)^{(3,5)}\big) \otimes Ff^{\mathrm{SO}(1)},
\end{split}
\end{equation}
where $Ff^{\mathrm{SO}(1)}$ denotes the Majorana fermion VOA.
To the best of our knowledge, this relation has not appeared in the literature before.
While we have checked this relation at the level of characters, we believe it indeed holds at the level of VOAs.
This relation is, of course, consistent with the expectation that Wilson lines form bimodules of the boundary algebras, and extend them in the 2d limit (see \cite{Dimofte:2017tpi,Alekseev:2022gnr}).

This result also implies that the representation categories of those vertex algebras are
braided-reverse equivalent: $(G_2)_{-1} \simeq \mathrm{SM}(3,5)$, up to an invertible factor. More precisely, they are spin-TQFTs and depend on the choice of spin structure:
 \begin{equation}
     (G_2)^{}_{-1}\otimes U(4)_1 \simeq \mathrm{SM}(3,5) \otimes \mathrm{SO}(1)_1\,.
 \end{equation}
Since $(G_2)^{}_{1}= (E_8)^{}_{1}/(F_4)^{}_{1}$ and $U(4)_1= \mathrm{SO}(8)_1$, we can alternatively write:
\begin{equation}
    \underbrace{(F_4)^{}_1}_{c\,=\,\frac{26}{5}}\otimes\underbrace{\frac{\mathrm{SO}(8)_1}{(E_8)^{}_1}}_{c\,=\,-4}=\underbrace{\mathrm{SM}(3,5)\otimes \mathrm{SO}(1)_1}_{c\,=\,\frac{6}{5}}\,,
\end{equation}
at the level of spin-TQFTs.

This is reminiscent of level-rank dualities, which have been recently discussed in closely related contexts in \cite{Ferrari:2023fez,Creutzig:2024ljv}.

\subsection{Sensitivity to 3d superpotentials}\label{subsec:sense3dSuperpotentials}

We have discussed two kinds of 3d superpotentials in this work: matter superpotentials $\mathcal{W}_\Phi$ and monopole superpotentials $\mathcal{W}^{}_V$. We have not considered dressed monopole superpotentials containing both monopoles and matter fields, because in our settings they would either have wrong $R$-charge or nonzero spin. Whether spin-singlets can be formed from higher powers of such terms with the right $R$-charge is an intriguing possibility that we leave for future studies.

Already our derivations of $\mathcal{W}_\Phi$ and $\mathcal{W}^{}_V$ may be questioned since they relied largely on naturalness. As for $\mathcal{W}_\Phi$, it can actually be easily checked that they can be obtained from the corresponding superpotentials in 4d \cite{Maruyoshi:2016tqk,Maruyoshi:2016aim,Benvenuti:2018bav}. For example, the superpotential $\mathcal{W}^{\text{Fib}}_\Phi=\Phi_{-1}^2\Phi_{+2}$ of the Fibonacci theory of Section~\ref{subsec:Fib}, arises from the term denoted $\mathrm{tr}(p\phi p)$ $\big(\!=p_{-1}^2\phi_2+p_1^2\phi_{-2}-2\,p_1p_{-1}\phi_0\,\big)$ in Eq.~(7.3) in \cite{Benvenuti:2018bav}. Nevertheless, we have also numerically investigated relevance of $\mathcal{W}^{\text{Fib}}_\Phi$ to the IR phase of the theory on $\mathbb{R}^3$ as follows. Dropping it, a flavor $U(1)_s$ arises in the 3d theory, under which $\Phi_{-1}$ has charge $2$ and $\Phi_{2}$ has charge $-1$ (in a scheme where $k_{gs}=0$). Numerically $F$-maximizing with respect to $U(1)_s$, we found a superconformal fixed point without extended SUSY. In other words, the fixed point obtained upon dropping $\mathcal{W}^{\text{Fib}}_\Phi$ would neither be gapped to yield a unitary TQFT, nor have extended SUSY to yield a non-unitary TQFT after twisting.

As for $\mathcal{W}^{}_V,$ the fact that the corresponding superpotentials should be generated on the outer patches essentially follows from the Affleck-Harvey-Witten mechanism \cite{Affleck:1982as}, but why the inner patches always inherit the monopole superpotentials of their neighboring outer patches deserves further scrutiny. They are certainly needed (in all cases discussed above, except for Gang-Yamazaki) if the 3d $\mathcal{N}=2$ Coulomb branch of the reduced Maruyoshi-Song theory is to be completely lifted on $\mathbb{R}^3$. For example in the $\overline{\text{Fib}}$ theory, as can be seen in Figure~\ref{fig:QhtTildeADn1}, only the 3d $\mathcal{N}=2$ Coulomb branch to the right of the $x^\ast=.2$ saddle is lifted by the CS coupling; for the part to the left of the saddle to be lifted on $\mathbb{R}^3$, the monopole superpotential $V_-$ is necessary. On curved backgrounds where the contact terms of \cite{DiPietro:2016ond} associated with $L_h$ as in Figure~\ref{fig:LhtTildeADn1} are active, they would of course suffice for lifting the 3d $\mathcal{N}=2$ Coulomb branch, and the superpotential $V_-$ would not be needed for that purpose in the theory.

Proper superpotentials can be essential for SUSY enhancement to $\mathcal{N}=4$, or for  a mass gap in the 3d theory.
Most significantly, the superpotentials prevent extra symmetries from emerging in the 3d EFT.
Such symmetries would widen the possibilities of $F$-maximization, potentially leading to new IR phases for the 3d EFT on $\mathbb{R}^3$ (different from the SUSY enhanced or gapped phases found above). If they had mixed boundary anomalies with the gauge symmetry, they might also necessitate a different set of anomaly canceling boundary multiplets when considering Neumann conditions on the gauge fields. For the purpose of locating the TQFT point on the moduli-space of $R$-mixings, however, the widened set of possibilities would actually be only a minor inconvenience. In fact even that minor inconvenience can be completely bypassed using a better index as the starting point in 4d, as explained below.

\subsection{Sensitivity to 4d background}\label{subsec:sense4dBackground}

In this work we have mainly focused on the index
\begin{equation}
    \mathcal{I}^{\gamma}(q)=\mathcal{I}_t(q\,e^{2\pi i\gamma},q,q^{4/3}),
\end{equation}
corresponding to the $\mathcal{N}=1$ background of \cite{Assel:2014paa}, albeit with the $U(1)_r$-twisted boundary conditions around the circle. From the point of view of the SCFT/VOA correspondence, there are several other 4d backgrounds that constitute more natural starting points for the twisted reduction. We now discuss some of those and their respective advantages compared to that of $\mathcal{I}^\gamma(q)$.\\

\noindent\textbf{Cigar$_\varepsilon$ $\times$ Riemann surface and the 4d A-model.} The present work is to a large extent a follow-up to \cite{Dedushenko:2023cvd} and \cite{Dedushenko:2018bpp}. The former used an $\Omega$-deformed cigar $\times$ Riemann surface background, reducing on the angular direction of the cigar, while the latter used the 4d A-model, of which the $T^2\times \Sigma$ topologically twisted index \cite{Benini:2015noa,Closset:2017bse} is a prominent example. For the 4d/3d/2d
picture that we have painted here, the cigar $\times$ Riemann surface background appears to be the most appropriate. Using that background as our starting point though, would require evaluation of the contact terms in \cite{DiPietro:2016ond} on the corresponding rigid supergravity background, which has not yet been done. The $T^2\times \Sigma$ index on the other hand, has been examined from an EFT perspective in \cite{DiPietro:2016ond}, but only partially, and in particular its large gauge flux sectors on $\Sigma$ remain to be understood. Our focus on $\mathcal{I}^\gamma(q)$ was because its 3d EFT is quite well-understood, and once the 3d EFT is obtained, general local QFT considerations imply that it can be put on any 3d background, including those arising from circle reductions of cigar $\times$ Riemann surface and $T^2\times \Sigma\,$. The 3d EFTs obtained via reduction on different circles of different backgrounds can of course differ by various flavor-$R$ mixings, which can indeed be significant for our intended applications. This aspect of the problem, however, can be put into sharp focus by considering other limits of the $\mathcal{N}=2$ index as we discuss next.

\vspace{.5cm}
\noindent\textbf{The Schur and Coulomb backgrounds.} 
A natural alternative to $\mathcal{I}^\gamma(q)$ from the viewpoint of the 4d/2d correspondence is the $R$-twisted Schur index:
\begin{equation}
    \mathcal{I}_\text{Schur}^{\gamma}(q):=\mathcal{I}_t(q\,e^{2\pi i\gamma},q,q).\label{eq:SchurGamma}
\end{equation}
Thinking of the Schur index as a partition function on $S^3\times S^1$, taking the Cardy limit shrinks the $S^1$, which is one of the directions where the VOA lives (\emph{cf.}~\cite{Pan:2019bor,Dedushenko:2019yiw}). So we would be deviating from the picture~\eqref{eq:4d3d2d} (\emph{cf.}~\cite{Dedushenko:2019mzv}), but only momentarily, since we expect the resulting EFT can then be placed on other backgrounds, and in particular on the background of the 3d half-index.

Studying the Cardy limit of $\mathcal{I}_\text{Schur}^{\gamma}(q)$ turns out to be quite illuminating. First, it is straightforward to check that $\mathcal{I}_\text{Schur}^{\gamma}(q)$ yields exactly the same $Q_h^\gamma$ function as $\mathcal{I}^{\gamma}(q)$. It also yields analogs of the $L_h^\gamma$ functions that differ from the ones we obtained above only by an overall factor of $3/2$ (similarly to what was noted for Schur indices in \cite{ArabiArdehali:2015ybk}). Consequently, all our results could have been obtained equally well starting from $\mathcal{I}_\text{Schur}^{\gamma}(q)$ instead of $\mathcal{I}^{\gamma}(q)$.

Although it is not obvious from the way we have defined $\mathcal{I}_\text{Schur}^{\gamma}(q)$ in \eqref{eq:SchurGamma}, explicit calculation for the theories we have considered shows that $\mathcal{I}_\text{Schur}^{\gamma}(q)$ exactly coincides (after an insignificant shift of all gauge holonomies by $1/2$) with the usual Schur index on its higher sheets: $\mathcal{I}^{\gamma=0}_\text{Schur}(q\, e^{2\pi i\gamma}).$

This close connection between $\mathcal{I}_\text{Schur}(q\, e^{2\pi i\gamma})$ and $\mathcal{I}^{\gamma}(q)$ provides an explanation for what appeared to be remarkable accidents in Sections~\ref{sec:A1A2} and \ref{sec:A1A4}: that $Q_h^\gamma$ and $L_h^\gamma$ vanished on the holonomy saddles. This follows from the fact that the theories we considered have single-valued Schur indices (\emph{i.e.} $\mathcal{I}_\text{Schur}(q\, e^{2\pi i\gamma})=\mathcal{I}_\text{Schur}(q)$ for all $\gamma\in\mathbb{Z}$), unlike what their UV Maruyoshi-Song appearances might suggest. As a result, $\mathcal{I}^\gamma_\text{Schur}(q)$ should have the same $q\to1^-$ asymptotic for all $\gamma.$ Since for $\gamma=0$ they are known to exhibit the Di~Pietro-Komargodski-type asymptotic \cite{Buican:2015hsa}, and since the saddle values of $Q_h^\gamma$ and $L_h^\gamma$ quantify deviations from that asymptotic formula \cite{ArabiArdehali:2015ybk}, it follows that $Q_h^\gamma=L_h^\gamma=0$ on the holonomy saddles for all $\gamma\in\mathbb{Z}.$

Another advantage of working with $\mathcal{I}^\gamma_\text{Schur}(q)$ is that its Cardy limit lands us directly on the $F$-maximized point of the 3d EFT! In particular, Cardy limit of the 2nd sheet index $\mathcal{I}^{\gamma=1}_\text{Schur}(q)$ gives directly the $R$-charges and mixed gauge-$R$ CS couplings of the 3d $\mathcal{N}=4$ SCFT,\footnote{This is known \emph{not to be the case} on the 1st sheet where $\gamma=0$ \cite{Buican:2015hsa,Dedushenko:2019mnd}. See Section~2 of \cite{ArabiArdehali:2024vli} for more on this point.}
relieving us from the burden of $F$-maximization that we would need to do if we reduced the $\mathcal{N}=1$ index as in Section~\ref{subsec:GY_susy_enhancement} for example. This observation is quite helpful in deriving new families of 3d SUSY enhancing theories from 4d, as will be demonstrated in \cite{ArabiArdehali:2024vli}.

Cardy limit of the $R$-twisted \emph{regularized Coulomb index}:
\begin{equation}
    \boxed{\mathcal{I}_\text{Coul}^{\gamma}(q):=\lim_{t\to q^2}\mathcal{I}_t(q\,e^{2\pi i\gamma},q,t),}\label{eq:CoulGamma}
\end{equation}
does even better: it lands us directly on the A-twisted TQFT. This is in fact how most of the TQFT data in the main text (such as \eqref{eq:tLYtilde}) were obtained. So it is actually the Cardy limit of this index that bridges the 4d/2d correspondence most practically.

The limit in \eqref{eq:CoulGamma} needs a few clarifying remarks. First, convergence requires that it be more precisely $q^2/t\to1^-.$ This effectively assigns 4d $\mathcal{N}=1$ $R$-charge $r=1$ to the fundamental chiral multiplets (analogous to the chirals in hypers) as can be seen from \eqref{eq:A1A2nIndex}, and $r\to0^+$ to all other matter multiplets (which are analogous to adjoint chirals in the 4d $\mathcal{N}=2$ vector multiplets). This seems to be related to the $Z_{\text{top.}}(S^3\times S^1)$ of \cite{Dedushenko:2018bpp}. But it differs slightly from the usual Coulomb index \cite{Gadde:2011uv}, wherein $q^2/t$ is set to a constant $u$, and then $q$ is sent to zero.

Recall that the Schur index is associated to the 4d \emph{Higgs} branch \cite{Beem:2013sza,Beem:2017ooy}.  We have shown how its $q$- (or ``low-temperature'') expansion can be obtained via the half-index calculation in the A-twist TQFT arising from the Cardy (or ``high-temperature'') limit of the $R$-twisted \emph{Coulomb} index. Analogous crossed-channel relations were noticed in the context of the 4d A-model in \cite{ArabiArdehali:2019orz}. This raises the possibility that the 4d A-model perspective can shed light, via a combination of modularity and mirror symmetry, on the intriguing Coulomb/Higgs relations discovered in \cite{Cordova:2015nma,Shan:2023xtw}. We leave clarification of this for the future.

\begin{acknowledgments}
We are grateful to Anirudh~Deb for collaboration during the early stages of this work, and to C.~Closset, S.~Chen, D.~Delmastro, T.~Gannon, N.~Garner, H.~Kim, Z.~Komargodski, J.~Kulp, N.~Mekareeya, D.~Pei, N.~Rajappa, L.~Rastelli, B.~Rayhaun, S.~Razamat, M.~Sacchi, S.~Shao, and Y.~Zheng for helpful correspondences and conversations related to this project. AA owes his involvement in this project to encouraging remarks from L.~Rastelli, and is indebted to Z.~Komargodski for clarifying questions. The work of AA was supported in part by the NSF grant PHY-2210533 and the Simons Foundation grants 397411 (Simons Collaboration on the Nonperturbative Bootstrap) and 681267 (Simons Investigator Award). 	The work of DG  is supported in part by the National Research Foundation of Korea grant  NRF-2022R1C1C1011979. DG also acknowledges support by the National Research Foundation of Korea (NRF) Grant No. RS-2024-00405629. ML was supported by
the National Science Foundation under Award PHY 2210533.
\end{acknowledgments}

\appendix

\section{Reducing the index to the $S^3$ partition function}\label{app:I4toZ3}

In this appendix we illustrate how the 2nd sheet index $\mathcal{I}^{\gamma=1}$ of the $(A_1,A_2)$ theory can be reduced in the Cardy-like limit to the supersymmetric $S^3$ partition function of the Gang-Yamazaki theory.

\subsection*{Subleading asymptotics of the index}

The $\mathbb{Z}_2$ Weyl redundancy implies we can focus on $0\le x\le\frac{1}{2}$. Therefore we only consider the saddle at $x=0.2$ and multiply its contribution by two to account for its $\mathbb{Z}_2$ image at $x=-0.2$.

To find the contribution from a small neighborhood of $x=0.2$, we need to apply inside the integrand of \eqref{eq:I_4 for n=1} asymptotic estimates  that are valid uniformly near $x=0.2.$ As direct examination shows, to all the elliptic gamma functions in \eqref{eq:I_4 for n=1}, except $\Gamma_e\big(z\, e^{-\frac{2\pi i}{5}}(pq)^{\frac{4}{15}}\big)$ to which we will return shortly, we can apply the following estimate (see Eq.~(2.12) of \cite{Ardehali:2021irq}):
\begin{equation}
\begin{split}
    \Gamma_e\big((pq)^{\frac{r}{2}}e^{2\pi i u}\big)&= \exp \biggl(-2\pi i \, \biggl(\,\frac{1}{\sigma\tau}\, \frac{\overline{B}_3(u)}{3!}
    + \frac{1}{\sigma\tau}\big(\frac{\sigma+\tau}{2}\big)\, (r -1) \frac{\overline{B}_2(u)}{2!}\\
    &\quad\qquad\qquad\qquad+\frac{3(r-1)^2 (\sigma+\tau)^2-(\sigma^2+\tau^2)}{24\sigma\tau} \, \overline{B}_1(u)+\mathcal{O}(\beta)\biggr) \biggr) ,\label{eq:outerEst}
    \end{split}
\end{equation}
valid for any $r\in\mathbb{R}$, and point-wise for $u\in\mathbb R\setminus\mathbb Z\,$. The functions $\overline{B}_{1,2,3}$ above are the \emph{periodic Bernoulli polynomials}, explicitly given by
\begin{equation}
\begin{split}
    \overline{B}_3(u)&:=B_3(\{u\})=\frac{1}{2}\{u\}(1-\{u\})(1-2\{u\}),\\
    \overline{B}_2(u)&:=B_2(\{u\})=-\{u\}(1-\{u\})+\frac{1}{6},\\
    \overline{B}_1(u)&:=\begin{cases}
    B_1(\{u\})=\{u\}-\frac{1}{2}\qquad\, \text{for $u\notin\mathbb{Z}$},\\
    0 \qquad\qquad\qquad\qquad\qquad\text{for $u\in\mathbb{Z}$}.
    \end{cases}
    \end{split}\label{eq:perBern}
\end{equation}
It is a simple exercise to show the compatibility of \eqref{eq:outerEst} with \eqref{eq:GammaAsy}.

The remaining elliptic gamma function in \eqref{eq:I_4 for n=1}, namely $\Gamma_e\big(z\, e^{-\frac{2\pi i}{5}}(pq)^{\frac{4}{15}}\big)$, corresponds to a \emph{light multiplet} in the dimensionally reduced theory. That is because the real mass of the 3d multiplet would be $\propto x-0.2$, which is small near $x=0.2.$ In mathematical terms, application of \eqref{eq:outerEst} to this elliptic gamma function is not justified because the uniform validity of \eqref{eq:outerEst} breaks down at $u=0$, corresponding to $u=x-0.2=0$ when applied to $\Gamma_e\big(z\, e^{-\frac{2\pi i}{5}}(pq)^{\frac{4}{15}}\big)$. We have to use instead (see Eq.~(2.31) of \cite{Ardehali:2021irq})
\begin{equation}
\begin{split}
    \Gamma_e\big((pq)^{\frac{r}{2}}e^{2\pi i u}\big)&= \exp \biggl(-2\pi i \, \biggl(\,\frac{1}{\sigma\tau}\, \frac{\overline{K}_3(u)}{3!}
    + \frac{1}{\sigma\tau}\big(\frac{\sigma+\tau}{2}\big)\, (r -1) \frac{\overline{K}_2(u)}{2!}\\
    &\quad\qquad\qquad\qquad+\frac{3(r-1)^2 (\sigma+\tau)^2-(\sigma^2+\tau^2)}{24\sigma\tau} \, \overline{K}_1(u)+\mathcal{O}(\beta)\biggr) \biggr)\\
    &\quad\times\Gamma_h\big(\frac{2\pi}{\beta}  u^{}_\mathbb{Z}+\big(\frac{\omega_1+\omega_2}{2}\big)\, r\, ;\omega_1,\omega_2\big),\label{eq:innerEst}
    \end{split}
\end{equation}
valid for any $r\in\mathbb{R}$, and point-wise for any $u\in\mathbb{R}$. The functions $\overline{K}_j$, which we call \emph{modified periodic Bernoulli polynomials}, are defined as
\begin{equation}
    \overline{K}_j(u):=\overline{B}_j(u)+\frac{j}{2}\mathrm{sign}(u^{}_{\mathbb{Z}}) (u^{}_\mathbb{Z} )^{j-1}.\label{eq:KbarDef}
\end{equation}
Here $u^{}_\mathbb{Z} :=u-\mathrm{nint}(u)$, with nint$(\cdot)$ the \emph{nearest integer} function. The parameters $\omega_{1,2}$ above are defined as $\omega_1:=2\pi\sigma/\beta,$ $\omega_2:=2\pi\tau/\beta.$  For simplicity we assume below that $\Omega_{1,2}=0$, so that $\omega_1=ib,$ $\omega_2=ib^{-1}.$

Applying the above estimates, we get (see Eq.~(2.47) of \cite{Ardehali:2021irq}):
\begin{equation}
    \mathcal{I}^{\gamma=1}(p,q)\approx e^{-V^\text{out}(x^\ast)}\, Z^{\text{in}}_{3d}(b), \label{eq:fullRedI_4}
\end{equation}
where $-V^\text{out}(x^\ast)=\frac{2\pi i}{90}\frac{\sigma+\tau}{2\tau\sigma},$ up to a constant shift (related to the induced $k_{RR}$ and $k_\text{grav}$) that we do not consider here, and (see Eq.~(2.48) of \cite{Ardehali:2021irq}):
\begin{equation}
    \begin{split}
    Z^{\text{in}}_{3d}(b)&=\int_{-\infty}^{\infty}\mathrm{d}\tilde{x}\, \, e^{-2\pi i\, k_{gg}\, \frac{\tilde{x}^2}{2}\, -2\pi i\omega\, \tilde{k}^{}_{gR}\, \tilde{x}}\ \Gamma_h\big(\frac{8}{15}\omega+\tilde{x}\big), \label{eq:3dZinInfty} \end{split}
\end{equation}
where $k_{gg}=-3/2,\ \tilde{k}_{gR}=1/30,$ and $\omega:=(\omega_1+\omega_2)/2=i(b+b^{-1})/2.$ The new variable $\tilde{x}$ above is defined via
\begin{equation}
    \tilde x:=\frac{2\pi}{\beta}(x-x^\ast),
\end{equation}
where $x^\ast=0.2$

\subsection*{$R$-current mixing and $S^3$ partition function comparisons}\label{app:Rmixing}

We would like to compare our result $Z^{\text{in}}_{3d}(b)$ above with the $S_b^3$ partition function of the GY theory at the superconformal point $k_{gR}=0$ and $r_\chi=1/3$. With a real-mass $m_J$ turned on for the topological $U(1)_J$ symmetry, the said partition function is \cite{Kapustin:2009kz,Imamura:2011wg,Hama:2011ea,Aharony:2013dha} (see Appendix~\ref{app:S3_ptn}):
\begin{equation}
    \label{eq:Z_S3forU(1)_3/2}
    Z(b,m^{}_J)=\int_{-\infty}^\infty \mathrm{d}\sigma\ e^{-2\pi i\, k_{gg}\frac{\sigma^2}{2}}\ e^{2\pi i\,  m^{}_J\,\sigma} \  \Gamma_h(r_\chi\omega+\sigma),
\end{equation}
where $k_{gg}=-3/2$.

We first perform a change of variables in \eqref{eq:Z_S3forU(1)_3/2} from $\sigma$ to $\tilde{x}$ via
\begin{equation}
    \sigma=\tilde{x}-\omega(r_\chi-r_0),
\end{equation}
where we have denoted the $R$-charge $8/15$ appearing in \eqref{eq:3dZinInfty} by $r_0$.
The necessity of this transformation for a successful comparison of $Z^{\text{in}}_{3d}(b)$ and $Z(b,m^{}_J)$ as in Eq.~\eqref{eq:Z_S3forU(1)_3/2} signals that the $R$-charges used in the two expressions differ by mixing with the $U(1)$ gauge (besides a mixing with the $U(1)_J$ that will be discussed momentarily). In terms of the new variable, we get:
\begin{equation}
    Z(b,m^{}_J)=\int_{-\infty}^\infty \mathrm{d}\tilde{x}\ e^{-2\pi i\, k_{gg}\frac{\tilde{x}^2}{2}}\ e^{2\pi i\,  (m^{}_J-k_{gg}\omega(r_0-r_\chi))\,\tilde{x}} \  \Gamma_h(r_0\omega+\tilde{x}),
\end{equation}
up to an overall phase (related to the background fields) that we do not consider here. It is now clear that we have a match with $\eqref{eq:3dZinInfty}$ if we take:
\begin{equation}
    m^{}_J=-\omega\tilde{k}_{gR}+ k_{gg}(r_0-r_\chi)\omega.
\end{equation}

We now explain why this sort of relation signals the fact that the $R$-charges used in \eqref{eq:Z_S3forU(1)_3/2} and \eqref{eq:3dZinInfty} differ by a mixing with the topological $U(1)_J$. See \cite{Buican:2015hsa,Dedushenko:2019mnd} for earlier related discussions in the context of reduction on the first sheet ($\gamma=0$).

Consider a 3d $\mathcal{N}=2$ $U(1)_{k_{gg}}$ gauge theory with a chiral multiplet of gauge charge $g_\chi$ and a mixed gauge-$R$ CS coupling $\tilde{k}_{gR}$. Assume the $U(1)_R$ mixes with the topological $U(1)_J$ and the gauge $U(1)$ as follows
\begin{equation}
   R_{\text{new}}=R+c_1\cdot J+c_2\cdot g,  
\end{equation}
where $J,g$ stand for the topological $U(1)_J$ and gauge $U(1)$ charges, respectively.

The 3d $\mathcal{N}=2$ chiral multiplet is not charged under $U(1)_J$. Therefore $c_2$ is fixed by the shift in its $R$-charge as
\begin{equation}
    c_2=\frac{r_\chi-r_0}{g_{\chi}}.\label{eq:c2}
\end{equation}
Note that we have taken its $R_\text{new}$ to be $r_\chi$, while its $R$ is $r_0,$ and its $g$ is $g_\chi.$

To fix $c_1$, we find the shift in the $R$-charge of the $J=1$ monopole. The old $R$-charge of the $J=1$ monopole is $-\tilde{k}_{gR}-\frac{|g_\chi|}{2}(r_0-1).$ Assume the new $R$-current has a mixed CS level $k_{gR}$ with the gauge $U(1)$. The new $R$-charge of the $J=1$ monopole is then $-k_{gR}-\frac{|g_\chi|}{2}(r_\chi-1).$ Since the gauge charge of the $J=1$ monopole is $g_m=-k_{gg}-\frac{1}{2}g_\chi|g_\chi|,$ we get an equation
\begin{equation}
   \underbrace{-k_{gR}-\frac{|g_\chi|}{2}(r_\chi-1)}_{R_\text{new}}=\underbrace{-\tilde{k}_{gR}-\frac{|g_\chi|}{2}(r_0-1)}_{R}+\,c_1\cdot1+c_2\cdot(-k_{gg}-\frac{1}{2}g_\chi|g_\chi|).
\end{equation}
Plugging in $c_2$ from \eqref{eq:c2}, we obtain:
\begin{equation}
    c_1=-\frac{k_{gg}}{g_\chi}(r_0-r_\chi)-k_{gR}+\tilde{k}_{gR}.\label{eq:deltaKgR&c1}
\end{equation}
If $c_1$ were zero, this would be a special case of the formula \cite{Closset:2012vp}:
\begin{equation}
    k^\text{new}_{gR}={k}^\text{old}_{gR}+k_{gg}\,c_2.\label{eq:kgR_mixings_gauge}
\end{equation}
We see from \eqref{eq:deltaKgR&c1} that the nonzero mixing with $U(1)_J$ via $c_1$ has the effect of inducing an additional $\Delta k_{gR}:$
\begin{equation}
    k^\text{new}_{gR}={k}^\text{old}_{gR}+k_{gg}\,c_2-c_1.\label{eq:kgR_mixings}
\end{equation}

A comparison of the partition functions:
\begin{equation}
    Z_{3d}(b)=\int_{-\infty}^\infty \mathrm{d}\tilde{x}\ e^{-2\pi i\, k_{gg}\frac{\tilde{x}^2}{2}}\ e^{-2\pi i\,  \omega\, \tilde{k}^{}_{gR}\tilde{x}} \  \Gamma_h(r_0\omega+g_\chi\tilde{x}),
\end{equation}
and
\begin{equation}
    Z_\text{new}(b,m^{}_J)=\int_{-\infty}^\infty \mathrm{d}\sigma\ e^{-2\pi i\, k_{gg}\frac{\sigma^2}{2}}\ e^{2\pi i\,  (m^{}_J-\omega\, k^{}_{gR})\,\sigma} \  \Gamma_h(r_\chi\omega+g_\chi\sigma),
\end{equation}
now establishes equivalence (possibly up to an overall constant related to the background fields) upon identifying:
\begin{equation}
    \sigma=\tilde{x}-c_2\omega,\qquad m^{}_J=-c_1\omega.\label{eq:compareParameters}
\end{equation}
The former relation implements an (``unphysical'') change of gauge-$R$ mixing scheme, while the latter compensates for the (``physical'') difference $k_{gR}-\tilde{k}_{gR}$ due to the $U(1)_J$-$R$ mixing.

Note that the change of integration contour that $\sigma=\tilde{x}-c_2\omega$ yields can be undone via contour deformation assuming that $r_0,r_\chi\in(0,2).$ This follows from the fact that for generic $b\in\mathbb{R}_{>0}$, the function
$\Gamma_h(x)$ has simple zeros at
$x=ib\mathbb{Z}^{\ge1}+ib^{-1}\mathbb{Z}^{\ge1}$ and simple
poles at $x=ib\mathbb{Z}^{\le0}+ib^{-1}\mathbb{Z}^{\le0}$.


\subsection{Turning on flavor fugacities/real-masses}\label{app:flavor}

Let us set $t=(pq)^{2/3}\xi$ in \eqref{eq:I_4 for n=1}. The fugacity $\xi$ corresponds to the part of the Cartan of the 4d $\mathcal{N}=2$ $\mathrm{SU}(2)_R\times U(1)_r$ $R$-symmetry that is flavor from an $\mathcal{N}=1$ perspective. We denote this flavor by $U(1)_f$.

Introducing $m_f$ via
\begin{equation}
    \xi=e^{i\beta m^{}_f},
\end{equation}
and performing the reduction similarly to how it was done above, we get:
\begin{equation}
    \begin{split}
    Z^{\text{in}}_{3d}(b,m^{}_f)&=\int_{-\infty}^{\infty}\mathrm{d}\tilde{x}\, \, e^{-2\pi i\, k_{gg}\, \frac{\tilde{x}^2}{2}\, -2\pi i(\omega\tilde{k}^{}_{gR}+m^{}_f\tilde{k}^{}_{gf}) \tilde{x}}\ \Gamma_h\big(\frac{8}{15}\omega+\tilde{x}+\frac{7}{10}m^{}_f\big), \label{eq:3dZinInftyXi} \end{split}
\end{equation}
instead of \eqref{eq:3dZinInfty}. The effective mixed gauge-flavor CS coupling $\tilde{k}^{}_{gf}$ can be obtained similarly to \eqref{eq:EFT_CS_couplings} from:
\begin{equation}
    k^{\ast}_{j f}=-\sum_{\chi}\sum_{\rho^\chi\in H_\ast}\overline{B}_1 \bigl(\rho^\chi \cdot \boldsymbol{x}^\ast  +q^\chi\cdot\boldsymbol{\xi} \bigr)\, \rho^{\chi}_j\,  q^{}_f\,.\label{eq:k_jf}
\end{equation}
In the present case this gives $\tilde{k}^{}_{gf}=-1/20.$

We then shift 
\begin{equation}
    \tilde x\to\tilde x-q^{}_f\,m^{}_f\,,
\end{equation}
with $q^{}_f=7/10,$ which amounts to adding a multiple of the gauge charge to the flavor charge (in effect, going to a different gauge-flavor mixing scheme). This allows us to rewrite the above integral as
\begin{equation}
    \begin{split}
    Z^{\text{in}}_{3d}(b,m^{}_f)&=\int_{-\infty}^{\infty}\mathrm{d}\tilde{x}\, \, e^{-2\pi i\, k_{gg}\, \frac{\tilde{x}^2}{2}\, +2\pi i(-\omega\tilde{k}^{}_{gR}+\zeta^{}_f) \tilde{x}}\ \Gamma_h\big(\frac{8}{15}\omega+\tilde{x}\big), \label{eq:3dZinInftyXi2} \end{split}
\end{equation}
where
\begin{equation}
    \zeta^{}_f:=-\big(\tilde{k}^{}_{gf}-k_{gg}\,q{}_f\big)m^{}_f\,.
\end{equation}

Since $\zeta^{}_f$ can be considered as the real mass associated with the $U(1)_J$, we see that the four-dimensional $U(1)_f$ descends effectively to the $U(1)_J$. (One can think of $\zeta^{}_f$ as an effective three-dimensional FI parameter as well.)

Moreover, the dependence of the \emph{dynamical} part of $Z^{\text{in}}_{3d}(b,m^{}_f)$ on the real mass $m^{}_f$ descending from 4d is entirely through  $\zeta^{}_f$, with a proportionality factor $k^{}_{gf}:=\tilde{k}^{}_{gf}-k_{gg}\,q{}_f=1$.

Our emphasis on the word dynamical is because there are background CS actions involving $k_{fR}$ and $k_{ff}$ that we have suppressed above for simplicity. Including them gives extra dependence on $m_f$ through the multiplicative factors:
\begin{equation}
    e^{-2\pi i \omega\tilde{k}^{}_{fR}m^{}_f-2\pi i \tilde{k}^{}_{ff}\frac{m^{2}_f}{2}}\xrightarrow{\tilde{x}\,\to\,\tilde{x}-q^{}_f\,m^{}_f}e^{-2\pi i \omega\tilde{k}'_{fR}m^{}_f-2\pi i \tilde{k}'_{ff}\frac{m^{2}_f}{2}},
\end{equation}
where $\tilde{k}_{fR}=17/300$ and $\tilde{k}_{ff}=73/200$, while $\tilde{k}'_{fR}=\tilde{k}_{fR}-q^{}_f\tilde{k}_{gR}=1/30$ and $\tilde{k}'_{ff}=\tilde{k}_{ff}-2q^{}_f\tilde{k}_{gf}+q^2_f{k}_{gg}=21/50$.

\subsection*{When the 4d flavor symmetry disappears in 3d}

Now we consider the case where the 3d EFT is gapped, and the 4d $U(1)_f$ disappears in 3d. More precisely, it acts trivially in the dynamical sector of the EFT below the uv scale $\propto \epsilon/\beta$.

We skip the details of the reduction, but it should be clear that for the third sheet of $(A_1,A_2)$, the analog of \eqref{eq:3dZinInftyXi} becomes:
\begin{equation}
    \begin{split}
    Z^{\text{in}}_{3d}(b,m^{}_f)=\int_{-\infty}^{\infty}\mathrm{d}\tilde{x}\, \, &e^{-2\pi i\, k_{gg}\, \frac{\tilde{x}^2}{2}\, -2\pi i(\omega\tilde{k}^{}_{gR}+m^{}_f\tilde{k}^{}_{gf}) \tilde{x}}\\
    &\times\Gamma_h\big(\frac{14}{15}\omega+\tilde{x}+\frac{1}{10}m^{}_f\big)\,\Gamma_h\big(\frac{2}{15}\omega-2\tilde{x}-\frac{1}{5}m^{}_f\big). \label{eq:3dZinInftyXiFibBar} \end{split}
\end{equation}
The flavor charges $1/10,\,-1/5$ can be found by examining \eqref{eq:I_t for n=1}. The mixed gauge-flavor CS coupling is found in this case to be $\tilde{k}^{}_{gf}=-3/20.$ Combined with the flavor analog of \eqref{eq:r(m)}:
\begin{equation}
    f(\boldsymbol{m})=-\sum_j k^{}_{jf}\,{m}_j-\frac{1}{2}\sum_{\chi}f_{\chi}\sum_{\rho^\chi\in L_\ast}|\rho^\chi(\boldsymbol{m})|,\label{eq:f(m)}
\end{equation}
the value $\tilde{k}^{}_{gf}=-3/20$ implies that the monopole $V_-$ has zero flavor charge:
\begin{equation}
    f(-1)=\frac{3}{20}(-1)-\frac{1}{2}\frac{1}{10}|1|-\frac{1}{2}(-\frac{1}{5})|-2|=0.\label{eq:f(-1)=0} 
\end{equation}

A quick examination of \eqref{eq:3dZinInftyXiFibBar} shows that a shift $\tilde x\to\tilde x-q^{}_f\,m^{}_f$, with $q^{}_f=1/10$, removes $m_f$ from the arguments of the hyperbolic gamma functions. Therefore, as in the second-sheet case discussed above, the dependence of $Z^{\text{in}}_{3d}(b,m^{}_f)$ on the real mass $m^{}_f$ descending from 4d is entirely through the effective FI parameter $\zeta^{}_f$. This time, however, regardless of $m^{}_f,$ the FI parameter $\zeta^{}_f$ vanishes! The reason is that the proportionality constant becomes
\begin{equation}
    k_{gf}=\tilde{k}^{}_{gf}-k_{gg}\,q{}_f=-\frac{3}{20}+\frac{3}{2}\cdot\frac{1}{10}=0.
\end{equation}

In other words, despite the initial appearances in \eqref{eq:3dZinInftyXiFibBar}, the dynamical part of the $S_b^3$ partition function of the third-sheet EFT is completely independent of $m^{}_f$. This suggests that the 4d $U(1)_f$ acts trivially in the dynamical sector of the 3d EFT.\\

We emphasize, though, that there are background CS terms in the 3d EFT that do depend on $m^{}_f$, but have been suppressed for simplicity. These are seen if we do not suppress the contributions from flavor-$R$ and flavor-flavor CS actions:
\begin{equation}
    \begin{split}
    Z^{\text{in}}_{3d}(b,m^{}_f)=\int_{-\infty}^{\infty}\mathrm{d}\tilde{x}\, \, &e^{-2\pi i\, k_{gg}\, \frac{\tilde{x}^2}{2}\, -2\pi i(\omega\tilde{k}^{}_{gR}+m^{}_f\tilde{k}^{}_{gf}) \tilde{x}{\color{olive}\,-2\pi i \omega\tilde{k}^{}_{fR}m^{}_f-2\pi i \tilde{k}^{}_{ff}\frac{m^{2}_f}{2}}}\\
    &\times\Gamma_h\big(\frac{14}{15}\omega+\tilde{x}+\frac{1}{10}m^{}_f\big)\,\Gamma_h\big(\frac{2}{15}\omega-2\tilde{x}-\frac{1}{5}m^{}_f\big)\\
    =\int_{-\infty}^{\infty}\mathrm{d}\tilde{x}\, \, &e^{-2\pi i\, k_{gg}\, \frac{\tilde{x}^2}{2}\, -2\pi i\omega\tilde{k}^{}_{gR} \tilde{x}{\color{olive}\,-2\pi i \omega\tilde{k}'_{fR}m^{}_f-2\pi i \tilde{k}'_{ff}\frac{m^{2}_f}{2}}}\\
    &\times\Gamma_h\big(\frac{14}{15}\omega+\tilde{x}\big)\,\Gamma_h\big(\frac{2}{15}\omega-2\tilde{x}\big). \label{eq:3dZinInftyXiFibBar2} \end{split}
\end{equation}
Here $\tilde{k}_{fR}=43/300$ and $\tilde{k}_{ff}=17/200$, while $\tilde{k}'_{fR}=\tilde{k}_{fR}-q^{}_f\tilde{k}_{gR}=1/30$ and $\tilde{k}'_{ff}=\tilde{k}_{ff}-2q^{}_f\tilde{k}_{gf}+q^2_f{k}_{gg}=1/10$.

\section{Effective CS coupling calculations}\label{app:effective_CS}

For rank-one theories a practical way of computing $k_{gg}$ and $k_{gR}$ is via $\partial^2 Q$ and $\partial L$ as in \eqref{eq:kij&kjR_from_Q&L} for outer patches, and then finding the inner-patch values by averaging the two sides (see footnote~\ref{fn:inner_average}). Below we present the more direct calculations via the formulas \eqref{eq:EFT_CS_couplings}.

\subsection*{Second sheet of $(A_1,A_2)$}

The formulas \eqref{eq:EFT_CS_couplings} applied to the saddle at $x^\ast=.2$ on the 2nd sheet of $(A_1,A_2)$ give:
\begin{equation}
    \begin{split}
        -k_{gg}&=\overline{B}_1(x^\ast+\frac{2}{5})+\overline{B}_1(-x^\ast+\frac{2}{5})+\overline{B}_1(-x^\ast-\frac{1}{5})+4\overline{B}_1(2x^\ast+\frac{1}{5})+4\overline{B}_1(-2x^\ast+\frac{1}{5})\\
        &=\frac{3}{2},\\
    -{k}_{gR}&=\frac{-1}{15}\big(\overline{B}_1(x^\ast+\frac{2}{5})-\overline{B}_1(-x^\ast+\frac{2}{5})\big)+\frac{7}{15}\overline{B}_1(-x^\ast-\frac{1}{5})+2\frac{-13}{15}\big[\,\overline{B}_1(2x^\ast+\frac{1}{5})\\
    &\quad -\overline{B}_1(-2x^\ast+\frac{1}{5})\big]+1\big(2\overline{B}_1(2x^\ast)-2\overline{B}_1(-2x^\ast)\big)=-\frac{1}{30},\\
    -{k}_{RR}&=\big(\frac{-1}{15}\big)^2\big(\overline{B}_1(x^\ast+\frac{2}{5})+\overline{B}_1(-x^\ast+\frac{2}{5})\big)+\big(\frac{-7}{15}\big)^2\overline{B}_1(-x^\ast-\frac{1}{5})+\big(\frac{-13}{15}\big)^2\big[\,\overline{B}_1(2x^\ast+\frac{1}{5})\\
    &\quad +\overline{B}_1(-2x^\ast+\frac{1}{5})\big]+\big(\frac{-1}{5}\big)^2\overline{B}_1(\frac{6}{5})+\big(\frac{-13}{15}\big)^2\overline{B}_1(\frac{1}{5})-\big(\frac{-11}{15}\big)^2\overline{B}_1(\frac{2}{5})=\frac{31}{225},\\
    -k_{\text{grav}}&=2\big(\overline{B}_1(x^\ast+\frac{2}{5})+\overline{B}_1(-x^\ast+\frac{2}{5})+\overline{B}_1(-x^\ast-\frac{1}{5})+\overline{B}_1(2x^\ast+\frac{1}{5})+\overline{B}_1(-2x^\ast+\frac{1}{5})\\
    &\quad +\overline{B}_1(\frac{6}{5})+\overline{B}_1(\frac{1}{5})-\overline{B}_1(\frac{2}{5})\big)=-\frac{2}{5}.
    \end{split}\label{eq:CSlevels n=1}
\end{equation}

\subsection*{Third sheet of $(A_1,A_2)$}

For the patch in$_1$ around $x^\ast=.2$ on the 3rd sheet of $(A_1,A_2)$, the equations \eqref{eq:EFT_CS_couplings} give:
\begin{equation}
    \begin{split}
        -k_{gg}&=\overline{B}_1(-\frac{1}{5}+\frac{4}{5})+\overline{B}_1(\frac{1}{5}-\frac{2}{5})+\overline{B}_1(-\frac{1}{5}-\frac{2}{5})+4\overline{B}_1\big(2\cdot\frac{1}{5}+\frac{2}{5}\big)=\frac{3}{2},\\
    -{k}_{gR}&=\frac{-1}{15}\big(-\overline{B}_1(-\frac{1}{5}+\frac{4}{5})\big)+\frac{-7}{15}\big(\overline{B}_1(\frac{1}{5}-\frac{2}{5})-\overline{B}_1(-\frac{1}{5}-\frac{2}{5})\big)+\frac{-13}{15}\big(2\overline{B}_1(2\cdot\frac{1}{5}+\frac{2}{5})\big)\\
    &\ +1\big(2\overline{B}_1(2\cdot\frac{1}{5})-\overline{B}_1(-2\cdot\frac{1}{5})\big)=-\frac{11}{10}.
    \end{split}
\end{equation}
We also get $k_{RR}=-13/450$ and $k_{\text{grav}}=-1/5.$

\subsection*{Fourth and fifth sheets of $(A_1,A_2)$}

The fourth sheet of $(A_1,A_2)$ is conjugate to its third sheet. It thus has the opposite $k_{gg}$, $k_{RR}$, and $k_\text{grav},$ while having the same $k_{gR}.$

Similarly, the fifth sheet has the opposite $k_{gg}$, $k_{RR}$, and $k_\text{grav}$ compared to the second sheet, while having the same $k_{gR}.$

\subsection*{Second sheet of $(A_1,A_4)$}

For the saddle at $(x_1^\ast,x_2^\ast)=(\frac{1}{7},\frac{2}{7})$ on the 2nd sheet of $(A_1,A_4)$, we get from \eqref{eq:EFT_CS_couplings}:
\begin{equation}
    \begin{split}
        -k_{11}&=\overline{B}_1(x^\ast_1 + \frac{3}{7}) + 
   \overline{B}_1(-x^\ast_1 + \frac{3}{7}) + \overline{B}_1(x^\ast_1 - \frac{2}{7}) + 
   \overline{B}_1(-x^\ast_1 - \frac{2}{7}) + 4 \overline{B}_1(2 x^\ast_1 + \frac{1}{7})\\
   &\ \ + 
   4 \overline{B}_1(-2 x^\ast_1 + \frac{1}{7}) + \overline{B}_1(x^\ast_1 + x^\ast_2 + \frac{1}{7}) + 
   \overline{B}_1(-x^\ast_1 + x^\ast_2 + \frac{1}{7}) + \overline{B}_1(-x^\ast_1 - x^\ast_2 + \frac{1}{7})\\
   &=\frac{3}{2},\\
   -k_{22}&=\overline{B}_1(x^\ast_2 + \frac{3}{7}) + 
   \overline{B}_1(-x^\ast_2 + \frac{3}{7})  +\overline{B}_1(-x^\ast_2 - 
     \frac{2}{7}) + 4 \overline{B}_1(2 x^\ast_2 + \frac{1}{7})+ 
   4 \overline{B}_1(-2 x^\ast_2 + \frac{1}{7})\\
     & \ \ + \overline{B}_1(x^\ast_1 + x^\ast_2 + \frac{1}{7}) + 
   \overline{B}_1(-x^\ast_1 + x^\ast_2 + \frac{1}{7}) + \overline{B}_1(-x^\ast_1 - x^\ast_2 + \frac{1}{7})=1,\\
   -k_{12}&=-k_{21}=\overline{B}_1(x^\ast_1 + x^\ast_2 + \frac{1}{7}) - \overline{B}_1(-x^\ast_1 + x^\ast_2 + \frac{1}{7}) + 
 \overline{B}_1(-x^\ast_1 - x^\ast_2 + \frac{1}{7})=\frac{1}{2}.
    \end{split}
\end{equation}
We also get $k_{1R}=-11/42$ and $k_{2R}=4/7.$

\section{Analytic toolkit for 3d $\mathcal{N}=2$ gauge theories}\label{app:3d_toolkit}

\subsection{Superconformal index}\label{app:3d_index}

The 3d superconformal index is defined for any 3d $\mathcal{N}=2$ theory with a $U(1)_R$ symmetry as \cite{Bhattacharya:2008zy}
\begin{equation}
    {I}^{}(q):=\mathrm{Tr}^{}_{S^2}(-1)^{2j_3} q^{R/2+j_3},\label{eq:3dIndexDefApp}
\end{equation}
where $R$ is the $U(1)_R$ charge, while $j_3$ is the spin quantum number associated with the $\mathrm{SO}(3)$ rotations of the Euclidean three-space, and the trace is over the Hilbert space on $S^2$ (or alternatively, the space of local operators in case the 3d $\mathcal{N}=2$ theory is an SCFT).

We refer to $I(q)$ as the \emph{first-sheet} 3d index. For a gauge theory it can be computed from the formula \cite{Aharony:2013kma,Kim:2009wb,Imamura:2011su,Kapustin:2011jm,Dimofte:2011py}
\begin{equation}
\begin{split}
{I}(q)&=\sum_{\boldsymbol{m}}\frac{(-1)^{c_j(\boldsymbol{m})m_j}}{|W_{\boldsymbol{m}}|} q^{\frac{r(\boldsymbol{m})}{2}}\oint \prod_{j=1}^{r_G}\frac{\mathrm{d}z_j}{2\pi i z_j}\, z_j^{c_j(\boldsymbol{m})} \prod_{\alpha_+} \left(1-z^{\pm\alpha_+}q^{|\alpha_+(\boldsymbol{m})|/2}\right)\\
&\ \ \ \ \prod_{\Phi}\prod_{\rho\in R_\Phi}\frac{(z^{-\rho}q^{|\rho(\boldsymbol{m})|/2+1-r_\Phi/2};q)}{(z^{\rho}q^{|\rho(\boldsymbol{m})|/2+r_\Phi/2};q)},
\end{split}
\end{equation}
with $c(\boldsymbol{m}),r(\boldsymbol{m})$ as in \eqref{eq:c(m)},\eqref{eq:r(m)}. The contour is on the unit circles, assuming that via suitable gauge-$R$ mixing one has ensured that the $R$-charges of all chiral multiplets are strictly between $0$ and $2.$

The 2nd sheet index can be found via sending $q\to q\, e^{2\pi i}$, and simplifying via $a_j\to a_j+m_j\,\pi$ (that is $z_j\to z_j\,(-1)^{m_j}$):
\begin{equation}
\begin{split}
\tilde{I}(q)={I}(q\, e^{2\pi i})&=\sum_{\boldsymbol{m}}\frac{e^{i\pi r(\boldsymbol{m})}}{|W_{\boldsymbol{m}}|}q^{\frac{r(\boldsymbol{m})}{2}}\oint \prod_{j=1}^{r_G}\frac{\mathrm{d}z_j}{2\pi i z_j} z_j^{c^{}_{j}(\boldsymbol{m})} \prod_{\alpha_+} (1-q^{|\alpha_+(\boldsymbol{m})|/2}z^{\pm\alpha_+})\\
&\ \ \ \ \prod_{\Phi}\prod_{\rho\in R_\Phi}\frac{(z^{-\rho}q^{|\rho(\boldsymbol{m})|/2+1-r_\Phi/2}\, e^{-i\pi{r_\Phi}};q)}{(z^{\rho}q^{|\rho(\boldsymbol{m})|/2+r_\Phi/2}\,e^{i\pi{r_\Phi}};q)}.
\end{split}\label{eq:2ndSheetGenConv}
\end{equation}\\

We use the 3d superconformal index in this work mainly as a tool to diagnose whether the 3d $\mathcal{N}=2$ gauge theories that we obtain are in the topological phase, and without local operators, in which case we should find:
\begin{equation}
    I(q)=\tilde{I}(q)=1.
\end{equation}

\subsection{Squashed three-sphere partition function}\label{app:S3_ptn}

Consider a 3d $\mathcal{N}=2$ gauge theory with a $U(1)_R$ as well as a $U(1)_f$ flavor symmetry.

The SUSY partition function on the squashed three-sphere $S_b^3$, with unit radius and squashing parameter $b,$ can be found from the following formula\footnote{Our orientation appears to be opposite to the one in \cite{Closset:2019hyt}. Alternatively, our THF modulus is complex conjugate to that of \cite{Closset:2019hyt}. This complex conjugation should be taken into account when comparing with Eqs.~(5.22)--(5.24) in that work.} \cite{Kapustin:2009kz,Imamura:2011wg,Hama:2011ea,Aharony:2013dha}:
\begin{equation}
\begin{split}
    Z(b,m_f)&=\int_{-\infty}^\infty \frac{\mathrm{d}^{r_G}\sigma}{|W|}\ e^{-2\pi i\, k_{ij}\, \frac{\sigma_i\sigma_j}{2}\, -2\pi i\omega\, k_{jR}\, \sigma_j\,-2\pi im_f\, k_{jf}\, \sigma_j}\\
    &\hspace{3cm}\times\frac{\prod_\chi\prod_{\rho^\chi}\Gamma_h\big(r_\chi\, \omega+\rho^\chi\cdot {\boldsymbol{\sigma}}+q_\chi\,m_f\big)}{\prod_{\alpha_+}\Gamma_h\big(\alpha_+\cdot{\boldsymbol{\sigma}}\big)\Gamma_h\big(-\alpha_+\cdot{\boldsymbol{\sigma}}\big)}\,,
    \end{split}\label{eq:S3_Z_gen}
\end{equation}
where $\omega=i(b+b^{-1})/2$. For simplicity, we have suppressed the contributions from $k_{RR},\,k_{\text{grav}},\,k_{fR},$ and $k_{ff}$ \cite{Closset:2019hyt}.

Above, we have suppressed the dependence of the hyperbolic gamma function $\Gamma_h(\,\cdot\,;\omega_1,\omega_2)$ on $\omega_1=ib$ and $\omega_2=ib^{-1}$. For $b=1$, corresponding to the round $S^3,$ we have
\begin{equation}
    \Gamma_h(x)=(1-e^{-2\pi\,x})^{-i\,x-1}\,e^{\frac{i}{2\pi}\mathrm{Li}_2(e^{-2\pi\,x})+\frac{i\pi}{2}(-i\,x-1)^2-\frac{i\pi}{12}}.
\end{equation}

We denote the partition function $Z(b=1,m^{}_f=0)$ simply as $Z^{}_{S^3}$.

\subsection{Bethe roots, BPS surgery, and the modular data}\label{app:Bethe_techniques}

The effective twisted superpotential and the effective dilaton of a 3d $\mathcal{N}=2$ gauge theory on $S^1$ are given by \cite{Nekrasov:2014xaa,Closset:2019hyt}:
\begin{equation}
    \begin{split}
        \widetilde{W}(u)&=\sum_{j,l}\frac{1}{2}k^{+}_{jl}u_j u_l+\sum_j\frac{1}{2}k^{+}_{jj}u_j+\sum_{\Phi}\sum_{\rho\in R_\Phi}\frac{1}{-4\pi^2}\mathrm{Li}_2\big(z^{ \rho}\big)
        \,,\\
        \widetilde{\Omega}(u)&=\sum_j k^{+}_{jR}u_j-\sum_{\Phi}\sum_{\rho\in R_\Phi}\frac{r_\Phi-1}{2\pi i}\log\big(1-z^{ \rho}\big)-\sum_{\alpha_+}\frac{1}{2\pi i}\log\big(1-z^{\pm  \alpha_+}\big)
        \,,
    \end{split}\label{eq:WandOmega}
\end{equation}
where $z=e^{2\pi i u},$ and the ambiguous sign in the exponent of $z^{\pm\alpha}$ means that every $\alpha_+$ contributes two terms to $\sum_{\alpha_+}$, once with each sign. We have dropped the contributions from $k_{RR}$ and $k_\text{grav}$ for simplicity, and used:
\begin{equation}
    \begin{split}
    k^+_{jl}&:=k_{jl}+\frac{1}{2}\sum_\Phi\sum_{\rho\in R_\Phi}\rho_j\rho^{}_l\,,\\
    k^+_{jR}&:=k_{jR}+\frac{1}{2}\sum_\Phi\sum_{\rho\in R_\Phi}\rho_j(r_\Phi-1)\,.
    \end{split}\label{eq:k+}
\end{equation}
To compare with \cite{Closset:2019hyt}, note that we are \emph{not} using an asymmetric quantization scheme (such as $U(1)_{-1/2}$): we do not implicitly augment our chiral multiplets with uv CS couplings.

To simplify the expressions, we often work with ($Z:=2\pi i u$):
\begin{equation}
    \begin{split}
        W(Z)&:=-4\pi^2\,\widetilde{W}(u)=\frac{1}{2}k^{+}_{jl}Z_j Z_l+\frac{1}{2}k^{+}_{jj}\cdot2\pi i Z_j+\sum_{\Phi}\sum_{\rho\in R_\Phi}\mathrm{Li}_2\big(e^{ \rho Z}\big),\\
        \Omega(Z)&:=2\pi i\,\widetilde{\Omega}(u)=k^{+}_{jR}Z_j+\sum_{\Phi}(1-r_\Phi)\sum_{\rho\in R_\Phi}\log\big(1-e^{\rho Z}\big)-\sum_{\alpha_+}\log\big(1-e^{\pm  \alpha_+Z}\big).
    \end{split}
\end{equation}

The Bethe roots are at
\begin{equation}
    \exp\big(\partial_{Z_i} W(Z_\alpha^\ast)\big)=1,
\end{equation}
with each expected to map to a module of the boundary VOA as in \cite{Dedushenko:2018bpp,Cho:2020ljj,Gang:2023rei}.

The handle-gluing and fibering operators are given by (see \emph{e.g.}~\cite{Closset:2018ghr}):
\begin{equation}
    \mathcal{H}(Z)=e^{\Omega(Z)}\,\det \partial^{}_{Z_i}\!\partial^{}_{Z_j}\! W(Z),
\end{equation}
\begin{equation}
    \mathcal{F}(Z)=e^{[W(Z)-Z_i\,\partial^{}_{Z_i}\! W(Z)]/(2\pi i)}.
\end{equation}
Assuming a one-to-one correspondence between the Bethe roots and the TQFT simple objects, we can identify the components of the modular $S$ and $T$ matrices via the map (\emph{cf.}~\cite{Gang:2023rei,Gang:2021hrd,Baek:2024tuo}):
\begin{equation}
    \{S^2_{0\alpha},T^{2}_{\alpha\alpha}\}\longleftrightarrow\{\mathcal{H}(Z^\ast_\alpha)^{-1},\mathcal{F}(Z^\ast_\alpha)^{2}\}.\label{eq:HandF_vs_SandT}
\end{equation}
Since we have dropped the contributions from $k_{RR}$ and $k_\text{grav}$ in \eqref{eq:WandOmega}, these identifications are accurate up to overall phases for the $n$-tuples $S_{0\alpha}$ and $T_{\alpha\alpha}$. Moreover, since our formulas concern $S^2_{0\alpha},T^{2}_{\alpha\alpha}$ (and not $S_{0\alpha},T_{\alpha\alpha}$), our identifications are additionally plagued with sign ambiguities for each ${\alpha}$. In the case of non-spin TQFTs, we identify the $T$ entries through $T_{\alpha\alpha}\longleftrightarrow \mathcal{F}(Z^\ast_\alpha)$, thus removing the additional sign ambiguities of $T_{\alpha\alpha}.$ In the unitary cases (our gapped scenarios), the sign and phase ambiguities in $S_{0\alpha}$ completely disappear due to the requirement $S_{0\alpha}>0$.

Note that so far we have not distinguished between the Bethe roots $Z^\ast_\alpha$. In the unitary cases, using $S_{00}\leq S_{0\alpha}$ we can narrow down the possibilities of (and in some cases even single out) the Bethe root corresponding to the vacuum module.

The $S^3$ partition function can be found via the BPS surgery formula (\emph{e.g.} \cite{Closset:2018ghr}):
\begin{equation}
    Z_{S^3}=\sum_{\alpha}\mathcal{H}(Z^\ast_\alpha)^{-1}\,\mathcal{F}(Z^\ast_\alpha)\,.\label{eq:Z_surgery}
\end{equation}
For a TQFT, it should match with $S_{00}.$ In this work we verify such matchings up to an overall phase. That is, we only verify
\begin{equation}
    |Z_{S^3}|=\big|\sum_{\alpha}\mathcal{H}(Z^\ast_\alpha)^{-1}\,\mathcal{F}(Z^\ast_\alpha)\big| \overset{?}{=} |\mathcal{H}(Z^\ast_0)^{-1/2}|=|S_{00}|\,.
\end{equation}
In the non-unitary cases, this relation can aid in the identification of the Bethe root corresponding to the vacuum module.

For identification of the Wilson line $L_\alpha$ corresponding to a simple object $\alpha$, the surgery formula of \cite{Witten:1988hf} for the expectation value of $L_\alpha$ on a loop in $S^3$ can help:
\begin{equation}
    \langle L_\alpha\rangle^{}_{S^3}=\frac{S_{\alpha0}}{S_{00}}\,,\label{eq:WL_surgery0}
\end{equation}
which together with \eqref{eq:HandF_vs_SandT} yields:
\begin{equation}
    L_\alpha(Z^\ast_0)=\frac{S_{\alpha0}}{S_{00}}=\pm\frac{\mathcal{H}(Z^\ast_\alpha)^{-1/2}}{\mathcal{H}(Z^\ast_0)^{-1/2}}\,.\label{eq:identifyLines_app}
\end{equation}
Further insight into the identification of the simple lines arises from imposing the expected fusion rules on $L_\alpha(Z^\ast_\beta)$, as well as the index constraints spelled out in Section~3.3 of \cite{Gang:2024loa}. Once the identification is made, we can use the more general surgery formula for BPS Wilson lines (see \emph{e.g.}~Eq.~(A.52) in \cite{Gang:2021hrd}):
\begin{equation}
    L_\alpha(Z^\ast_\beta)=\frac{S_{\alpha\beta}}{S_{0\beta}}\,,\label{eq:WL_surgery}
\end{equation}
to obtain $S_{\alpha\beta}$ for nonzero $\alpha,\beta$.

Note that through this procedure, assuming the Wilson lines $L_\alpha$ can be identified, we get the whole $S$ matrix up to an overall phase, as well as the sign ambiguities in each entry. Demanding that $S$ be symmetric removes half of the off-diagonal sign ambiguities of course. As alluded to above, in the unitary cases (our gapped scenarios), the overall phase ambiguity and the sign ambiguities in the first row/column disappear as well. In the non-unitary cases (our SUSY enhancement scenarios), demanding that $S^2=C$ is a permutation matrix can be used to remove the overall phase ambiguity of $S.$ Further reduction of the ambiguities is possible by demanding that one and only one row of $S$ is strictly positive in the present rational context \cite{Gannon:2003de}.

It would be nice to determine precisely how much of the remaining ambiguities can be removed by further imposing the $\mathrm{SL}(2,\mathbb{Z})$ relations and the positivity of the fusion coefficients arising from the Verlinde formula.

\subsection{Half-index calculations}\label{app:half-index}

The 3d half-indices used in this work \cite{Gadde:2013wq,Yoshida:2014ssa,Dimofte:2017tpi} are defined as:
\begin{equation}
    I\!\!I_{\mathcal{B}}:=\mathrm{Tr}_{\text{Ops}_\mathcal{B}}(-1)^R\,q^{\frac{R}{2}+j_3},
\end{equation}
with the trace taken over the hemisphere Hilbert space with boundary conditions $\mathcal{B}.$

\subsubsection*{Neumann}

With the $\mathcal{N}=(0,2)$ Neumann boundary conditions on all fields, the various contributions to the boundary anomaly \cite{Dimofte:2017tpi} are:
\begin{equation}
    -\frac{1}{2}\rho_i\rho_j\,\mathbf{f}_i\cdot\mathbf{f}_j- \rho_j(r_\Phi-1)\,\mathbf{f}_j\cdot\mathbf{r}+\dots\,,
\end{equation}
from any 3d $\mathcal{N}=2$ chiral multiplet of gauge charge $\rho_j$ and $R$-charge $r_\Phi$,
and
\begin{equation}
    h\mathrm{Tr}\big(\mathbf{f}^2\big)+\dots,
\end{equation}
with $h$ the dual Coxeter number, from any 3d $\mathcal{N}=2$ vector multiplet. The CS levels $k_{ij}$ and $k_{jR}$ contribute
\begin{equation}
    \pm\big(k_{ij}\,\mathbf{f}_i\cdot\mathbf{f}_j+ 2k_{jR}\,\mathbf{f}_j\cdot\mathbf{r}\big)\,,
\end{equation}
with the positive sign for the left and negative sign for the right boundary in our conventions.

Moreover, if there is a $U(1)_J$ in the problem associated with an abelian gauge field $A$, we see from
\begin{equation}
    J^\text{top}_\mu=\frac{i}{2\pi}\varepsilon_{\mu\nu\rho}\partial^\nu A^\rho\Longrightarrow\int J^\text{top}_\mu A^\mu_\text{top}=2\times\frac{i}{4\pi}\int A_\text{top}\wedge\mathrm{d}A,
\end{equation}
that there will be a term
\begin{equation}
    2\,\mathbf{f}_x\cdot \mathbf{f},
\end{equation}
in the boundary anomaly, where $\mathbf{f}$ is the curvature of the $U(1)$ gauge field and $\mathbf{f}_x$ that of the background connection for the $U(1)_J$.

To cancel the bulk-induced boundary anomalies, we often add boundary Fermi and/or chiral multiplets. A 2d Fermi/chiral multiplet of gauge charge $g$ and fermion $R$-charge $r_\text{fermion}$ would contribute to the boundary anomaly via
\begin{equation}
    \pm(g^2\mathbf{f}^2+2g\,r_\text{fermion}\mathbf{f}\cdot\mathbf{r}+\dots)\,,
\end{equation}
where in the notation of \cite{Dimofte:2017tpi}, in the Fermi case, $r_\text{fermion}$ is the $R$-charge of $\gamma_-$ and we take the plus sign, while in the chiral case, $r_\text{fermion}$ is that of $\psi_+$ (hence $r_\phi-1$), and we take the minus sign.

Recalling
\begin{equation}
    \theta_0(z;q):=(z;q)(z^{-1}q;q),
\end{equation}
the boundary degrees of freedom contribute to the Neumann half-index via:
\begin{equation}
    \begin{split}
    Z_{\partial\ \text{fermi}_{r}}&\overset{}{=}\theta_0\big((-q^{1/2})^{1-r}z^{-\rho};q\big),\\
     Z_{\partial\ \text{chiral}_r}&=\frac{1}{\theta_0\big((-q^{1/2})^{r}z^\rho;q\big)},
    \end{split}
\end{equation}
where both are assumed to have charge $\rho$ under the $U(1)$ symmetry associated with the fugacity $z$. In the Fermi case, $r$ stands for $r_{\gamma_-}$, and in the chiral case, for $r_\phi\,.$

The bulk vector and chirals contribute via:
\begin{align}
     Z^{N}_\text{vector} &= (q;q)^{\mathrm{rk}(G)}\,\prod_{\alpha_+ } \left( z^{\pm\alpha};q\right), \\
     Z_\text{chiral}^N &= \prod_{\rho \in R_\Phi}  \left(z^\rho  (-q^{1/2})^{r_\Phi} ;q\right)^{-1} .\label{eq:NeumannChiralDGP}
\end{align}

The full Neumann half-index is given by:
\begin{equation}
    I\!\!I_{\mathcal{N},N,\dots}(q)=\frac{1}{|W_G|}\oint \prod_{j=1}^{\mathrm{rk}(G)}\frac{\mathrm{d}z_j}{2\pi i z_j}\ {Z}^{N}_{\text{vector}}\,{Z}^{N}_{\text{chirals}}\,Z_{\partial\ \text{fermis}}\,Z_{\partial\ \text{chirals}}\,.\label{eq:DGPhalf-indexN}
\end{equation}

\vspace{.2cm}

\noindent\textbf{Difficulty in presence of boundary chirals.} In absence of boundary chiral multiplets, the integration contour in \eqref{eq:DGPhalf-indexN} can be taken to be the unit circle, possibly after suitable gauge-$R$ mixing such that all chiral multiplet $R$-charges are strictly between $0$ and $2.$

With boundary chirals present, however, the correct integration contour in \eqref{eq:DGPhalf-indexN} is not clear \cite{Dimofte:2017tpi}. It was suggested in \cite{Dimofte:2017tpi} that in such cases, one takes a two-step approach: first compute the half-index for the boundary condition $\mathcal{D},N,N,\dots,N$ (namely Dirichlet on the vector multiplet and Neumann on the chirals), and then gauge the boundary global symmetry descending from the bulk gauge symmetry via a boundary gauge multiplet, together with various anomaly canceling boundary chiral and Fermi multiplets. The boundary gauging then follows a well-understood JK-residue prescription \cite{Benini:2013xpa,Benini:2016qnm}.

When considering the left boundary, the first step gives \cite{Dimofte:2017tpi}:
\begin{equation}
\begin{split}
I\!\!I_{\mathcal{D},N,\dots}(q,z)&=\frac{1}{(q;q)^{r_G}}\sum_{\boldsymbol{m}}q^{\frac{\sum_{i,j}k^{-}_{ij}m_i m_j}{2}}\,(-q^{1/2})^{k^{-}_{jR}{m_j}}\,\prod_{j=1}^{r_G} z_j^{\sum_l k^{-}_{jl}m_l}\prod_{\alpha_+} \frac{1}{\left(q^{1\pm\alpha_+(\boldsymbol{m})} z^{\pm\alpha_+};q\right)}\\
    &\qquad\qquad\qquad\qquad\prod_{\Phi}\prod_{\rho\in R_\Phi}\left(z^{\rho}  q^{\rho(\boldsymbol{m})}(-q^{1/2})^{r_\Phi} ;q\right)^{-1},
    \end{split}\label{eq:gen-half-index-DN}
\end{equation}
where $k^-_{ij}$ and $k^-_{jR}$ are defined as
\begin{equation}
    \begin{split}
    k^-_{jl}&:=k_{jl}-\frac{1}{2}\sum_\Phi\sum_{\rho\in R_\Phi}\rho_j\rho^{}_l,\\
    k^-_{jR}&:=k_{jR}-\frac{1}{2}\sum_\Phi\sum_{\rho\in R_\Phi}\rho_j(r_\Phi-1).
    \end{split}
\end{equation}
On the right boundary, instead of $k^-_{ij}$ and $k^-_{jR}$ in \eqref{eq:gen-half-index-DN}, one has to use
\begin{equation}
    \begin{split}
    k^-_{jl}\to -k^+_{jl}&\,,\\
    k^-_{jR}\to-k^+_{jR}&\,,
    \end{split}\label{eq:-k+}
\end{equation}
which are the coefficients arising from the gauge anomaly on the right boundary.

The second step then yields
\begin{equation}
    I\!\!I_{\mathcal{N},N,\dots}(q)=\frac{(q;q)^{2r_G}}{|W_G|}\oint_{\text{JK }} \prod_{j=1}^{r_G}\frac{\mathrm{d}z_j}{2\pi i z_j}\ \prod_{\alpha_+}\theta_0(z^{\pm\alpha_+};q)\,I\!\!I_{\mathcal{D},N,\dots}(q,z)\,Z_{\partial\ \text{fermis}}\,Z_{\partial\ \text{chirals}}\,.\label{eq:DGP2ndStepN}
\end{equation}

\subsubsection*{Dirichlet}

The general formula for the 3d half-index with $\mathcal{N}=(0,2)$ Dirichlet boundary conditions on all fields reads \cite{Dimofte:2017tpi}:
\begin{equation}
\begin{split}
I\!\!I_{\mathcal{D},D,\dots}(q,z)&=\frac{1}{(q;q)^{r_G}}\sum_{\boldsymbol{m}}q^{\frac{\sum_{i,j}k^{+}_{ij}m_i m_j}{2}}\,(-q^{1/2})^{k^{+}_{jR}{m_j}}\,\prod_{j=1}^{r_G} z_j^{\sum_l k^{+}_{jl}m_l}\prod_{\alpha_+} \frac{1}{\left(q^{1\pm\alpha_+(\boldsymbol{m})} z^{\pm\alpha_+};q\right)}\\
    &\qquad\qquad\qquad\qquad\prod_{\Phi}\prod_{\rho\in R_\Phi}\left(z^{-\rho}  q^{1-\rho(\boldsymbol{m})} (-q^{1/2})^{-r_\Phi};q\right),
    \end{split}\label{eq:gen-half-index-D}
\end{equation}
for the left boundary, in our conventions.
See \eqref{eq:k+} for the definition of $k^+_{ij}$ and $k^+_{jR}$.

On the right Dirichlet boundary, instead of $k^+_{ij}$ and $k^+_{jR}$ in the above formula, one has to use
\begin{equation}
    \begin{split}
    k^+_{jl}\to -k^-_{jl}&:=-k_{jl}+\frac{1}{2}\sum_\Phi\sum_{\rho\in R_\Phi}\rho_j\rho^{}_l\,,\\
    k^+_{jR}\to-k^-_{jR}&:=-k_{jR}+\frac{1}{2}\sum_\Phi\sum_{\rho\in R_\Phi}\rho_j(r_\Phi-1)\,,
    \end{split}\label{eq:-k-}
\end{equation}
which are the coefficients arising from the gauge anomaly on the right boundary.

\vspace{.5cm}
\paragraph{Conflict of interest statement.} The authors declare no known competing financial interests or personal relationships that could have appeared to influence the work reported in this paper.

\paragraph{Data availability statement.} No new data were created or analyzed during this study. Data sharing is not applicable to this article.

\bibliographystyle{JHEP}
\bibliography{biblio}

\end{document}